\documentclass[journal]{IEEEtran}
\pdfoutput=1
\usepackage{subfigure}
\usepackage{epsfig,graphicx,rotating}
\usepackage{amssymb,amsmath}
\usepackage{epstopdf}
\usepackage{cite}      
\usepackage{psfrag}
\usepackage{longtable}

\newtheorem{theorem}{Theorem}[section]

\newcommand{\norm}[1]{\left\lVert#1\right\rVert}

\newcommand{\arrayantenna}[2]{
\draw[-,thick] (11,#1-.4,#2) -- (11,#1+1+.4,#2);

\draw[-,thick] (11,#1-.6,#2) -- (11,#1-.6,#2+.3);
\draw[-,thick] (11,#1-.3,#2) -- (11,#1-.3,#2+.3);
\draw[-,thick] (11,#1,#2) -- (11,#1,#2+.3);
\draw[-,thick] (11,#1+.3,#2) -- (11,#1+.3,#2+.3);
\draw[-,thick] (11,#1+.6,#2) -- (11,#1+.6,#2+.3);
\draw[-,thick] (11,#1+.9,#2) -- (11,#1+.9,#2+.3);
\draw[-,thick] (11,#1+1.2,#2) -- (11,#1+1.2,#2+.3);
\draw[-,thick] (11,#1+1.5,#2) -- (11,#1+1.5,#2+.3);
\draw[-,thick] (11,#1+1.8,#2) -- (11,#1+1.8,#2+.3);
\draw[-,thick] (11,#1+2.1,#2) -- (11,#1+2.1,#2+.3);

\draw[-] (11,#1.25,#2-.5) -- (11,#1+.5,#2);
\draw[-] (11,#1.75,#2-.5) -- (11,#1+.5,#2);
\draw[-,-latex] (11,#1+2.1,#2) -- (11,#1+2.1,#2+2) node[anchor=north east]{\scriptsize $z$};
\draw[-,-latex] (11,#1+2.1,#2) -- (11,#1-3+1.2,#2) node[anchor=north east]{\scriptsize $x$};
\draw[-,-latex] (11,#1+2.1,#2) -- (11-2,#1+2.1,#2) node[anchor=north east]{\scriptsize $y$};
}

\makeatletter
\newcommand{\vast}{\bBigg@{4}}
\newcommand{\Vast}{\bBigg@{5}}
\makeatother

\usepackage{tikz} 
\usetikzlibrary{decorations.markings,decorations.pathreplacing,decorations.pathmorphing,shapes,backgrounds,patterns,decorations.text}
\usepackage{tikz-3dplot}
\usepackage{color}
 \usepackage{pgfplots}

\makeatother	
	
\begin{document}
\title{	
Massive MIMO for Communications with Drone Swarms\\
}
\author{Prabhu~Chandhar,~\IEEEmembership{Member,~IEEE,} Danyo~Danev,~\IEEEmembership{Member,~IEEE,} and Erik~G.~Larsson,~\IEEEmembership{Fellow,~IEEE}
  \thanks{The authors are with the Division of Communication Systems,
    Dept. of Electrical Engineering (ISY), Link\"{o}ping University,
    Sweden (email: \{prabhu.c, danyo.danev,
    erik.g.larsson\}@liu.se). Portions of this work were presented at
    ICUAS 2016 \cite{prabhu2016_uav} and at IEEE SPAWC 2016
    \cite{prabhu2016_spawc}.
		
		This work was funded in part by the Swedish Research Council (VR) and ELLIIT.}
}

\maketitle
\thispagestyle{empty}
\pagestyle{headings}

\begin{abstract}
We illustrate the potential of Massive MIMO for communication with
unmanned aerial vehicles (UAVs). We consider a scenario where multiple
single-antenna UAVs simultaneously communicate with a ground station
(GS) equipped with a large number of antennas. Specifically, we
discuss the achievable uplink (UAV to GS) capacity performance in the
case of line-of-sight (LoS) conditions. We develop a realistic
geometric model which incorporates an arbitrary orientation of the GS
and UAV antenna elements to characterize the polarization mismatch
loss which occurs due to the movement and orientation of the UAVs. A
closed-form expression for a lower bound on the ergodic rate for a
maximum-ratio combining receiver with estimated channel state
information is derived. The optimal antenna spacing that maximizes the
ergodic rate achieved by an UAV is also determined for uniform linear
and rectangular arrays. It is shown that when the UAVs are spherically
uniformly distributed around the GS, the ergodic rate per UAV is
maximized for an antenna spacing equal to an integer multiple of
one-half wavelength.
\end{abstract}


\IEEEpeerreviewmaketitle

\section{Introduction} In recent years, the use of
unmanned aerial vehicles (UAVs), also known as drones, for both
civilian and military applications is increasing
worldwide. {There are different types of UAVs, with
  varying sizes and capabilities that are used in multitude
  applications. Depending on the power source, their connectivity
  range varies from a few meters to several kilometers and their
  flight time varies from a few minutes to tens of hours.  For a
  comprehensive survey of different type of UAVs, their capabilities,
  and issues related to communication, readers are referred to
  \cite{hayat2016_1,gupta2016,asadpour2014,andre2014} and references
  therein.  The communication between a ground station (GS) and the
  UAVs involves many challenges.  First, UAVs are often equipped with
  cameras that deliver high-resolution images and videos to the GS,
  requiring high-speed communication in the ranges of tens of Mbps
  \cite{hayat2016_1}}. The main challenge here is to maintain reliable
communication as the link conditions are affected by variations in
signal propagation due to the movement of the UAVs in
{three-dimensional (3D) space}. Particularly, the
antenna characteristics (radiation pattern and polarization) and
orientation can have strong impact on the link performance
\cite{yanmaz2013,asadpour2014,Asadpour2016}. Second, many applications
also require that the information should be delivered with low
latency, {down to the order of $10$ milliseconds}
\cite{asadpour2013LDD}. Third, power consumption may be a limitation
for certain UAV networks.

Currently, existing wireless technologies, such as Wireless Fidelity
(WiFi), ZigBee, and XBee-Pro are being used for communication with
UAVs. Since these technologies were originally
  designed for provision of wireless access in indoor scenarios, their
  use is limited to very short range, low throughput, and low-mobility
  applications. Experimental studies have shown that under line-of-sight (LoS) conditions, IEEE $802.11$n can provide single link data rates of $10$ Mbps up to $500$ m (mobility: $5$ m/s, latency: 100s of ms) and
  XBee-Pro can provide $250$ Kbps up to a $1$ km range
  \cite{andre2014,hayat2016}. These technologies cannot be used for
  long-range, high-throughput, high-mobility UAV applications, where
  the flying speed is in the order of $20$--$50$ m/s
  \cite{yanmaz2013,hayat2016_1}. Moreover, these technologies are not
  suitable for applications where a swarm of UAVs needs simultaneous
  high-throughput communication with the GS. Consider, for example,
  $20$--$30$ UAVs streaming high-resolution videos to the GS, each UAV
  requiring tens of Mbps data rate. Some of the potential applications
  that require such high-throughput link include border surveillance,
  crowd management, crop monitoring, 3D cartography, and search and
  rescue missions after natural disasters such as earthquakes and
  massive flooding. The list of civilian and military applications for
  UAV swarms keeps growing
  \cite{abatti2005,olsson2010,andre2014,hayat2016_1}. Therefore, a new
  breakthrough technology is required in order to support UAV
  applications that need reliable long-range connectivity, high
  throughput, low power consumption, and low latency.

Massive multiple input multiple output (MIMO) is an emerging technique
for 5G cellular wireless access
\cite{Marzetta16Book,marzetta2010,larsson2014}. It is characterized by
its scalability and potential to deliver very high and stable
throughputs. In a massive MIMO cellular system, base stations equipped
with a very large number of antennas simultaneously serve multiple
single-antenna terminals. By coherent closed-loop beamforming, the
power is focused into a small region of space, thus reducing
interference. It also provides significant improvement in energy
efficiency and reduced latency. To avoid channel state information
feedback, Massive MIMO uses time-division multiplexing (TDD),
exploiting channel reciprocity. In this paper, we argue that a solution for
  communication with UAV swarms based on Massive MIMO can offer orders
  of magnitude higher sum-throughput and reliability compared to the
  direct use of existing standards.

\subsection{Contributions} We consider an
uplink communication scenario with LoS and no
multipath. In this setup, multiple single-antenna UAVs simultaneously
communicate with a GS which is equipped with an uniform rectangular
array (URA). We develop a geometric model which captures the
polarization characteristics of the GS and the UAV antennas. Using
this model, we answer the following questions:
\begin{itemize}
\item What is the achievable uplink capacity per UAV, when  a
  swarm of single antenna UAVs simultaneously communicate with a
  GS equipped with a large number of antennas?

\item What is the optimal antenna spacing in the GS antenna array?

\item How does the antenna configuration (i.e. antenna orientation and
  polarization) at the GS and the UAV affect the link reliability, and
  what is the appropriate antenna polarization that should be used in
  order to maintain a reliable communication link?
\end{itemize}

The performance gain of Massive MIMO is achieved by the orthogonality
between the spatial signatures (channel response vectors) of the
terminals \cite{Marzetta16Book}. Unlike in Rayleigh fading channels,
in LoS propagation conditions, the spatial signatures are determined
by the position of the terminals. Hence in the UAV application, the
interference power is determined by the spatial correlation between
the spatial signatures of the different UAVs. This interference power
will be continuously changing as the positions of the UAVs change when
the UAVs move. For example, in micro UAV networks \cite{asadpour2014},
the UAVs typically move at high speed ($10$ m/s to $30$ m/s) in random
directions. {Even if the UAVs move along a
  deterministic trajectory, the interference power will fluctuate due
  to varying elevation and azimuth angles. Once can then expect
  multiple independent realizations of the interference power within a
  short period of time (i.e. in a few milliseconds). Effectively, the
  UAVs then experience many possible interference realizations within
  the transmission duration of a codeword. This fact motivates us to
  analyze the ergodic capacity by averaging over all possible
  positions of the UAVs.  For analytical tractability we assume
  inverse-SNR power control, leading to max-min fairness in terms of
  received power. First, we derive a closed-form lower bound on the
  achievable uplink rate for a maximum-ratio combining (MRC) receiver
  with estimated channel state information (CSI). Then, we analyze the
  optimal GS antenna geometry that maximizes the ergodic rate. To the
  best of our knowledge, this analysis is entirely novel and very
  different, quantitatively and qualitatively, from the analysis in
  cellular communications and Rayleigh fading \cite{Marzetta16Book}. We 
	also study the ergodic rate performance for the zero-forcing (ZF)
receiver with perfect CSI.}

We consider that the elements of the GS antenna array
  as well as the UAV antenna are composed of two orthogonally crossed
  dipoles. The advantage of a cross-dipole antenna is its
  quasi-isotropic antenna pattern, and it can be used to transmit and
  receive electromagnetic waves with different polarizations (linear,
  circular, and elliptical). We develop an analytically tractable
  polarization loss model to characterize the channel between the
  cross-dipole antenna elements at the GS and at the UAV. This model
  could, in principle, at some effort, be extended to the case of a
  tripole antennas, which have a closer to isotropic antenna pattern
  compared to dipoles.  However, we show that in the Massive MIMO
  setup, the use of cross-dipole antennas is sufficient to obtain very
  good performance (in terms of rates and reliability) of the
  communication link, by appropriately orienting the elements of the
  GS array. The reason is the ``polarization diversity'' effect that
  arises when the array comprises many antennas with different
  orientations.

The proposed Massive MIMO based communication framework could be used for wide range of altitudes and different types of UAVs. In this paper, we interchangably use the terms drone and UAV.

\subsection{Related Works} \textit{MIMO for UAV communications:} A simulation-based study of multi-user MIMO
communications for air traffic management for airplanes flying at
altitudes ranging from $5$ km to $10$ km was presented in
\cite{rasool2009}. The authors studied the impact of antenna spacing
on the sum-capacity performance in the uplink. However, they neither
used a detailed geometric model nor studied the impact of the number
of antennas on the achievable capacity.  { MIMO for
  point-to-point aerial communication was studied in
  \cite{jiang2014,su2013,michailidis2010}. The authors used small
  numbers of antennas ($2\times 2$ and $4\times 4$) which is different
  from Massive MIMO where a very large antenna array (with hundreds or
  thousands of elements) serve many single antenna terminals. Further,
  they did not study the impact of polarization mismatch losses due to
  fluctuations of the UAV antenna orientations.

\textit{MIMO performance in LoS conditions:} The
  impact of antenna spacing on the capacity of fixed point-to-point
  MIMO ($20\times 20$) in LoS channels was studied in
  \cite{sarris2005,bohagen2007,halsig2015}. It was shown in
  \cite[Ch.~7]{Marzetta16Book} that in two-dimensional (2D) Massive
  MIMO systems, the LoS channels are asymptotically orthogonal as the
  number of antennas increases. The impact of different array
  geometries on the asymptotic channel orthogonality in Massive MIMO
  systems was studied in \cite{jinhui2013}. The author showed that in
  LoS channels, asymptotic orthogonality holds for uniform linear
  arrays and uniform planar arrays, but not for uniform circular
  arrays. In \cite{yeqing2015}, the authors showed that in 2D LoS
  channels, the mainlobe distribution of the interference can be
  approximated as a Beta-mixture. Assuming that perfect CSI is
  available, a sum-rate analysis for LoS Massive MIMO systems with
  different array configurations was studied in \cite{tan2015}. The
  performance of Massive MIMO in LoS conditions with max-min fairness
  signal-to-interference-plus-noise ratio (SINR) power control was studied in \cite{yang2017}. However, the
  authors did not consider ergodic rate performance. All the above
  mentioned works \cite{jinhui2013,yeqing2015,tan2015} considered a
  fixed half-wavelength antenna spacing and did not consider
  polarization mismatch losses. Moreover, the above-mentioned works
  assume that perfect CSI is available at the BS and did not consider
  mobility of the terminals in their analysis. In contrast, we derive
  a lower bound on the uplink ergodic rate with estimated CSI, and
  optimize the antenna array geometry. Our analysis also takes into
  account the (pseudo-)random orientations of the UAV antennas.

\begin{figure*}[htb]
\centering 
\input{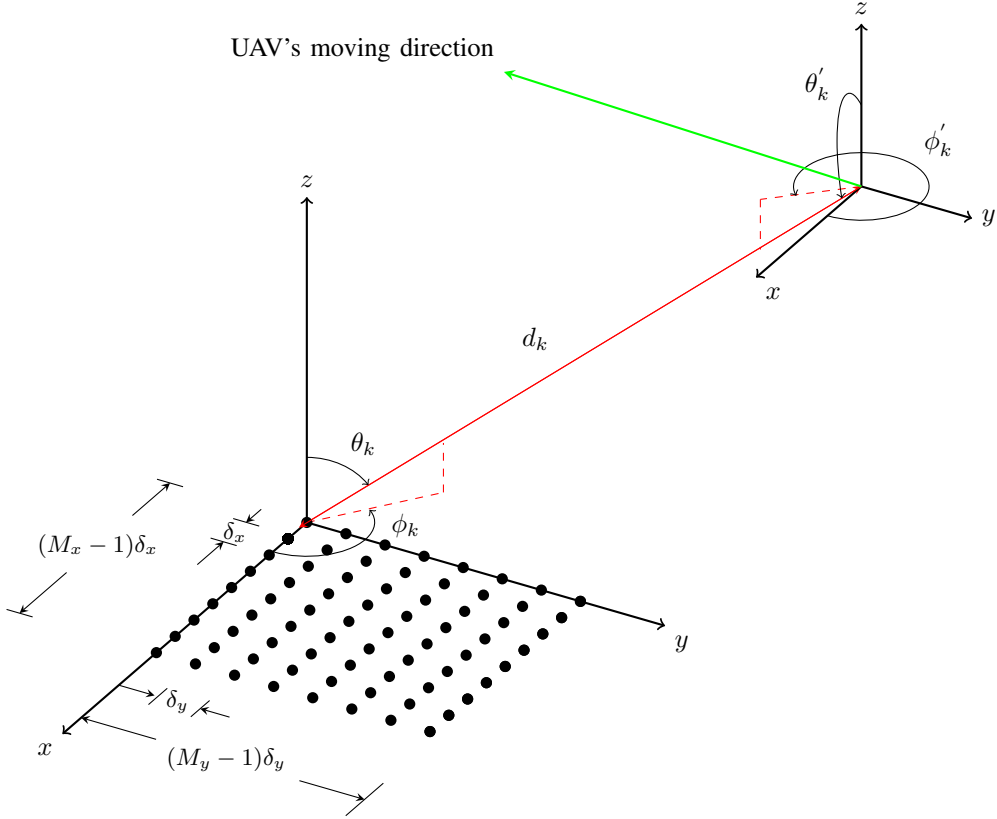}
\caption{Illustration of 3D geometric model with a rectangular array 
	at the GS.}
\label{geometric_model_}
\end{figure*}

\textit{Polarization modeling:} A 3D polarization
  model for MIMO channels with linearly polarized antennas in cellular
  environments was developed in \cite{shafi2006}. In
  \cite{manh_tuan_dao2011}, the authors developed a 3D
  polarization channel model for a $2\times 2$ MIMO configuration in
  cellular environments with vertical (V) and horizontal (H)
  polarizations (i.e., V/V, V/H, and $\pm 45^{\circ}$
  slanted). However, these models cannot be used for UAV
  communications as the propagation conditions are different from
  cellular communications\footnotemark. A 3D polarization rotation
  model for both LoS and NLoS conditions was developed in
  \cite{jaeckel2012}. The authors  used a sequence of complicated
  coordinate system transformations to find the elements of the
  polarization rotation matrix. An experimental study of IEEE $802.11$
  networks with 3D mobility was studied in \cite{yanmaz2013}. It was
  shown that $5$ dB to $15$ dB gains in received signal strength
  is possible using three linearly polarized dipole antennas. The
  authors  analyzed the impact of azimuth and elevation angles but
  they did not analyze the impact of antenna orientations due to
  flight dynamics (pitch, yaw, roll). In this work, we develop a
  simpler method, using rotation matrices for the GS and the UAV
  antennas to incorporate azimuth, elevation, and flight dynamics.

\footnotetext[1]{The effect of polarization mismatch
    is not a major problem in cellular communications, because
    irrespective of the type of transmit antenna polarization, the
    large number of multipath components (MPCs) arriving at the
    receive antenna will comprise a combination of all
    polarizations.}

\section{System Model}
\subsection{Geometric Model}

We consider an uplink of a Massive MIMO system, with LoS and no
multipath. The geometric model of the system is shown in Figure
\ref{geometric_model_}. We fix an orthonormal coordinate system with
unit basis vectors $\hat{\boldsymbol{x}}$,
$\hat{\boldsymbol{y}}$, and
$\hat{\boldsymbol{z}}$ and an origin at some point $O$. We refer to this system as a ``reference
coordinate system". We consider a rectangular antenna array with
$M_x$ and $M_y$ antennas on $x$-axis and
$y$-axis, respectively. The total number of antenna elements is
denoted by $M=M_x M_y$. The spacing between the antenna
elements on $x$-axis and $y$-axis is denoted by $\delta_x$ and
$\delta_y$, respectively. The array elements are described by index
$l=(q-1)M_x+p$, where
$p\in\{1,2,...,M_x\}$ denotes the index on $x$-axis and
$q\in\{1,2,...,M_y\}$ denotes the index on
$y$-axis. The $l$-th antenna position $P^l$ is denoted
by $(x^{l},y^{l},z^{l})=((p-1)\delta_x,(q-1)\delta_y,0)$.

\begin{figure*}[t]
\centering
\subfigure[Relative error with the second term in \eqref{dist_l_final}]
{\includegraphics[scale=.6]{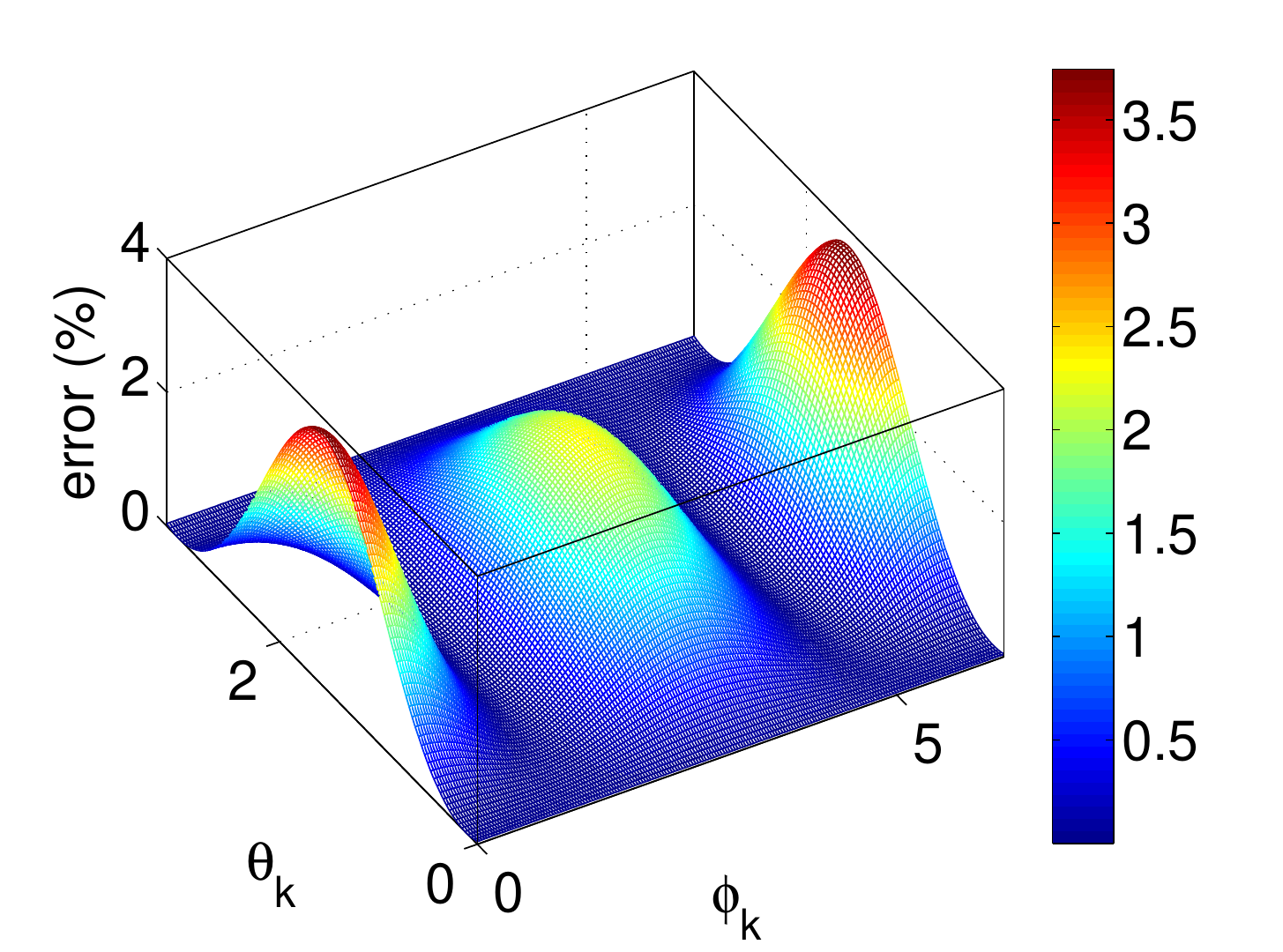}
} 
\subfigure[Relative error without the second term in \eqref{dist_l_final}]
{\includegraphics[scale=.6]{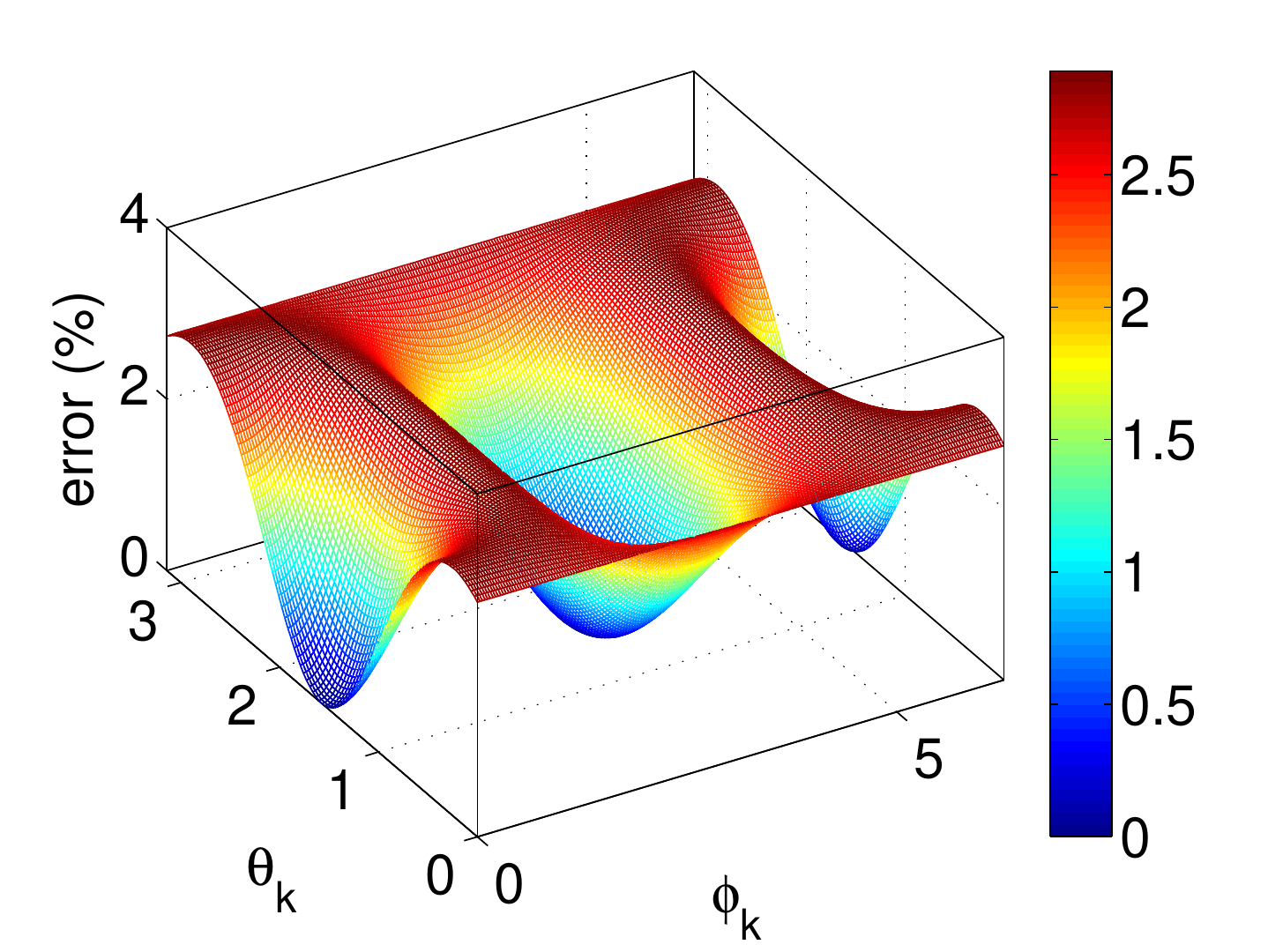} 			
}
\caption{The approximation error in \eqref{dist_l_final} as a function of the 
elevation and azimuth angles for $M_x=100, M_y = 1, d_k =25$ m, and 
$\delta_x=6.25$ cm.}
\label{Dist_Approximation}
\end{figure*}

There are $K$ single-antenna UAVs simultaneously
transmitting data to the GS in the same time-frequency resource. Let
the position $P_{k}$ of the $k$-th UAV has the
coordinates $(x_k,y_k,z_k)$. The direction vector $\boldsymbol{p}_k$
from the origin $O$ towards the $k$-th UAV
at position $P_k$ can be expressed as %
\begin{equation*}
\begin{array}{r@{}l}
  \overrightarrow{OP_k} &{}=
  {\boldsymbol{p}}_k =
  \left(\begin{matrix}
    \hat{\boldsymbol{x}} & \hat{\boldsymbol{y}} & \hat{\boldsymbol{z}}
  \end{matrix}\right)
  \left(\begin{matrix}
    x_k \\ y_k \\z_k
  \end{matrix}\right)\\ &{}=
  \left(\begin{matrix}
    \hat{\boldsymbol{x}} & \hat{\boldsymbol{y}} & \hat{\boldsymbol{z}}
  \end{matrix}\right)
  \left(\begin{matrix}
    d_k\cos{\phi_k}\sin{\theta_k}\\
    d_k\sin{\phi_k} \sin{\theta_k}\\
    d_k\cos{\theta_k}
  \end{matrix}\right),%
\end{array}
\end{equation*}
where $d_k$ is the radial distance between the GS and the $k$-th UAV, 
$\phi_k\in[0,2\pi]$ is the azimuth angle (i.e. the
angle from the positive direction of the $x$-axis towards the positive
$y$-axis, to the vector's (i.e. ${\boldsymbol{p}}_k$'s) orthogonal projection onto the $x$-$y$
plane), and $\theta_k\in[0,\pi]$ is the elevation angle
(i.e. the angle from the positive direction of the $z$-axis towards
the direction vector $ {\boldsymbol{p}}_k$).

The distance between the $l$-th GS antenna and the $k$-th UAV's antenna is then given by 
\begin{equation}\label{dist}
d_{kl}=\sqrt{(x_{k}-(p-1)\delta_x)^2+(y_{k}-(q-1)\delta_y)^2+z_{k}^2}.%
\end{equation}
\normalsize

By expanding \eqref{dist}, we get
\begin{align}
\label{dist_before_approx}
d_{kl}=d_{k}\Bigg[ &
1+\frac{1}{d_k^2}[(p-1)^2\delta_x^2+(q-1)^2\delta_y^2]\\
&-\!\frac{2}{d_k}\sin\theta_k[(p\!-\!1)\delta_x 
\cos\phi_k\!+\!(q\!-\!1)\delta_y\sin\phi_k]
\Bigg]^{\frac{1}{2}}\!.\nonumber
\end{align}

When the distance between the GS antenna and the UAV position is greater than the aperture size of the array i.e. $d_k>\sqrt{(M_x-1)^2\delta_x^2+(M_y-1)^2\delta_y^2}$, by using the approximation $\sqrt{1+t}\approx 1+\frac{t}{2}$, for $|t|<1$, the distance in \eqref{dist_before_approx} can be simplified to
\begin{equation}
\begin{array}{r@{}l}
\label{dist_l_final}
d_{kl}\approx &{} \displaystyle d_{k} + 
\frac{1}{2d_k}[(p-1)^2\delta_x^2+(q-1)^2\delta_y^2]\\
&{} \displaystyle -\sin\theta_k[(p-1)\delta_x \cos\phi_k+(q-1)\delta_y 
\sin\phi_k\big].
\end{array}
\end{equation}

Note that in our previous works \cite{prabhu2016_uav,prabhu2016_spawc}, by assuming $d_k$ to be very large when compared to the aperture size, we neglected the second term in \eqref{dist_l_final}. However, since the micro UAVs typically fly at very low altitudes in the range from $30$ m to $200$ m, the distance $d_k$ can be comparable to the aperture size. In this case, the second term in \eqref{dist_l_final} should not be neglected as it will introduce expressive errors in the analysis. For example, Figure \ref{Dist_Approximation} shows the error (in \%) with and without including the term $\frac{1}{2d_k}\big((p-1)^2\delta_x^2+(q-1)^2\delta_y^2\big)$ for $M_x=100, M_y=1, \delta_x = 6.25$ cm, and $d_k = 25$ m. It can be seen that without the term, the error is significant for most of the elevation and azimuth angles. In contrast, with including the term, the error is small for most of the elevation and azimuth angles. Therefore, in this work, we include the second term in \eqref{dist_l_final} in our analysis.


\subsection{Polarization model with single cross-dipole antenna at the transmitter and the receiver}
Let us assume that the $l$-th GS antenna transmits the signal 
   \[
  u(t) = \cos(2\pi f_0t) = \Re \{e^{i2\pi f_0t}\},
  \]\normalsize
	where $\Re$ denotes the real part.
  In practice, this signal can be realized by creating an alternating
  electrical current of unit amplitude and frequency $f_0$ in the
  transmitter's electrical circuit. The actual transmission takes
  place when we apply electrical current at the transmit
  antenna. In our model we consider amplification and phase shifting
  of the original signal $u(t)$ before applying it to the antenna.
  The signal's wavelength is $\lambda= c/f_0$, where $c\approx 3\times 10^8$ m/s is the speed of
  light.

If the position and the orientation of the $k$-th UAV is fixed, the signal at its receiver can be expressed as 
\begin{align}
\label{received_voltage_vt}
&v(t) =\Re\left\{ \sqrt{\beta_{kl} }\  
h_{kl}(f_0)\ e^{-i\frac{2\pi f_0}{c}d_{kl}}\ u(t)\right\}\\
&= \sqrt{\beta_{kl} }\ |h_{kl}(f_0)|\ \cos\left(2\pi f_0 
\left(t-\frac{d_{kl}}{c}\right)+\mathrm{arg}(h_{kl}(f_0))\right),\nonumber
\end{align}
where $\beta_{kl} = \left(\frac{\lambda}{4\pi d_{kl}}\right)^2$ is the free-space pathloss \cite[Sec: 2.17.1]{Balanis2005} and the complex number $h_{kl}(f_0)$ represents the combined effect of polarization mismatch and antenna gain. In this subsection we detail the calculation of the factor $h_{kl}(f_0)$ for $l$-th GS antenna the antenna at the UAV.

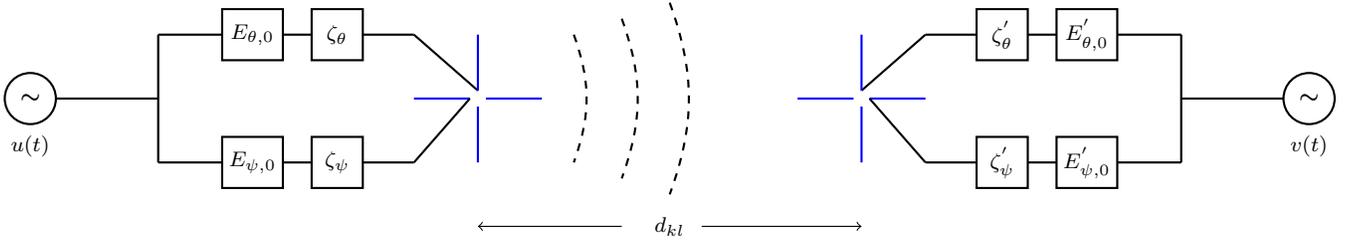
\begin{figure*}
\centering
\begin{tikzpicture}[scale=.85]
\draw[thick] (-2,-1) circle (.4cm);
\node at (-2,-1) {\footnotesize $\boldsymbol{\sim}$};
\draw[thick] (-1.6,-1) -- (0,-1);

\node at (-2,-1.75) {\footnotesize $u(t)$};

\draw[thick,-] (0,0) -- (1,0);
\draw[thick] (1,.4) rectangle (1.95,-.4) node[pos=.5] {\footnotesize $E_{\theta,0}$};			
\draw[thick,-] (1.95,0) -- (2.4,0);
\draw[thick] (2.4,.4) rectangle (3.2,-.4) node[pos=.5] {\footnotesize $\zeta_{\theta}$};			
\draw[thick,-] (3.2,0) -- (4,0);

\draw[thick,-] (4,0) -- (5,-.875);

\draw[thick,-,color=blue] (5,0) -- (5,-.875);
\draw[thick,-,color=blue] (5,-1.125) -- (5,-2);
\draw[thick,-,color=blue] (4,-1) -- (4.875,-1);
\draw[thick,-,color=blue] (5.125,-1) -- (6,-1);




\draw[<-] (5,-3) -- (7.25,-3);
\node at (8,-3) {\footnotesize $d_{kl}$};
\draw[->] (8.5,-3) -- (11,-3);

\draw[thick,-] (12,0) -- (11,-.875);

\draw[thick,-] (12,0) -- (12.8,0);
\draw[thick] (12.8,.4) rectangle (13.6,-.4) node[pos=.5] {\footnotesize $\zeta_{\theta}^{'}$};			
\draw[thick,-] (13.6,0) -- (14.05,0);
\draw[thick] (14.05,.4) rectangle (15,-.4) node[pos=.5] {\footnotesize $E_{\theta,0}^{'}$};			
\draw[thick,-] (15,0) -- (16,0);

\draw[thick,-] (0,0) -- (0,-2);
\draw[thick,-] (16,0) -- (16,-2);

\draw[thick,-] (0,-2) -- (1,-2);
\draw[thick] (1,-2.4) rectangle (1.95,-1.6) node[pos=.5] {\footnotesize $E_{\psi,0}$};			
\draw[thick,-] (1.95,-2) -- (2.4,-2);
\draw[thick] (2.4,-2.4) rectangle (3.2,-1.6) node[pos=.5] {\footnotesize $\zeta_{\psi}$};			
\draw[thick,-] (3.2,-2) -- (4,-2);

\draw[thick,-] (4,-2) -- (4.875,-1);

\draw[thick,-,color=blue] (11,0) -- (11,-.875);
\draw[thick,-,color=blue] (11,-1.125) -- (11,-2);
\draw[thick,-,color=blue] (10,-1) -- (10.875,-1);
\draw[thick,-,color=blue] (11.125,-1) -- (12,-1);


\draw [thick,dashed] (6.5,-.5+.5)  to [out=290,in=70] (6.5,-2);
\draw [thick,dashed] (7.25,-.25+.5)  to [out=290,in=70] (7.25,-2.25);
\draw [thick,dashed] (8,0+.5)  to [out=290,in=70] (8,-2.5);


\draw[thick,-] (12,-2) -- (11.125,-1);

\draw[thick,-] (12,-2) -- (12.8,-2);
\draw[thick] (12.8,-2.4) rectangle (13.6,-1.6) node[pos=.5] {\footnotesize $\zeta_{\psi}^{'}$};			
\draw[thick,-] (13.6,-2) -- (14.05,-2);
\draw[thick] (14.05,-2.4) rectangle (15,-1.6) node[pos=.5] {\footnotesize $E_{\psi,0}^{'}$};			
\draw[thick,-] (15,-2) -- (16,-2);

\draw[thick] (16,-1) -- (17.6,-1);
\draw[thick] (18,-1) circle (.4cm);
\node at (18,-1) {\footnotesize $\boldsymbol{\sim}$};
\node at (18,-1.75) {\footnotesize $v(t)$};
\end{tikzpicture}
\caption{Communication channel model with crossed dipoles at the transmitter 
and at the receiver.}
\label{crossed_dipole_graph_journal}
\end{figure*}

We fix a translated coordinate system parallel to
  the reference coordinate system with origin at the $l$-th GS
  antenna's position $P^l$,
  i.e. $(x^{l},y^{l},z^{l})$. For the first GS antenna,
  we get exactly the reference coordinate system.  The position
  $P_{kl}$ of the $k$-th UAV in the translated
  coordinate system has coordinates $(x_{kl},y_{kl},z_{kl})=(x_{k}-(p-1)\delta_x,y_{k}-(q-1)\delta_y,z_{k})$.
	Each element of the GS array is composed of two orthogonally crossed dipoles
  (one dipole is oriented parallel to the $z$-axis and
  the other to the $y$-axis). As it will be detailed later, the crossed dipoles are
  fed with the same signal but with different magnitude and phase. The
  UAV antenna is also composed of two crossed dipoles oriented along
  the $y$- and $z$-axes  (refer Figure \ref{crossed_dipole_graph_journal}). Here we consider the downlink communication from the $l$-th GS antenna to the $k$-th UAV. For the uplink communication the results
  are similar due to the antenna reciprocity principle.

\paragraph*{Polarization of the wave transmitted by the dipole placed along the $z$-axis}
  Polarization of the electromagnetic wave is usually described by the
  direction of the transmitted wave's electric field vector over time
  at a given point. The orientation of the electric field is always
  in the plane orthogonal to the direction of wave propagation. For
  the dipole placed along the $z$-axis, the wave
  travels in the direction towards the $k$-th UAV
  \begin{equation*}
  \begin{array}{r@{}l}
  \overrightarrow{P^lP_k} =
  \boldsymbol{p}_{kl}&{}=
  \left(\begin{matrix}\hat{\boldsymbol{x}} &
    \hat{\boldsymbol{y}}&\hat{\boldsymbol{z}}\end{matrix}\right)
  \left(\begin{matrix}x_{kl}\\ y_{kl}\\ z_{kl}\end{matrix}\right)\\
  &{}=
  \left(\begin{matrix}\hat{\boldsymbol{x}}&\hat{\boldsymbol{y}} &
    \hat{\boldsymbol{z}}\end{matrix}\right)
  \left(\begin{matrix}x_{k}-(p-1)\delta_x\\ y_{k}-(q-1)\delta_y\ \\
    z_{k}\end{matrix}\right).
  \end{array}
  \end{equation*}
  We denote by $\hat{\boldsymbol{p}}_{kl}=
      \frac{\boldsymbol{p}_{kl}}{\norm{\boldsymbol{p}_{kl}}}=
      \frac{\boldsymbol{p}_{kl}}{d_{kl}}$ the unit vector in the
  direction of the radio wave propagation. The induced electric field
  will be oriented in the direction of the vector orthogonal to both
  $\hat{\boldsymbol{p}}_{kl}$ and
  $\hat{\boldsymbol{z}}\times\hat{\boldsymbol{p}}_{kl}$ \cite{Balanis2005}. The
  unit length vector determining this direction and having
  non-negative scalar product with
  $\hat{\boldsymbol{z}}$ is
  \begin{equation}
  \hat{\boldsymbol{\theta}}_{kl}=
  \frac{\left(\begin{matrix}\hat{\boldsymbol{x}}&\hat{\boldsymbol{y}}&
      \hat{\boldsymbol{z}}\end{matrix}\right)}{d_{kl}\sqrt{x_{kl}^2+y_{kl}^2}}
  \left(\begin{matrix} -x_{kl}z_{kl} \\-y_{kl}z_{kl}
    \\x_{kl}^2+y_{kl}^2
  \end{matrix}\right).
  \label{theta_hat}
  \end{equation}

  The elevation angle between the $z$-axis directed dipole and the
  propagation direction is given by
  \begin{equation*}
   \theta_{kl}=\cos^{-1}(\hat{\boldsymbol{p}}_{kl}\cdot
  \hat{\boldsymbol{z}} )=\cos^{-1}\left(\frac{z_{kl}}{d_{kl}}\right).
  \end{equation*}
  Then, for a plane wave propagating in the direction $\hat{\boldsymbol{p}}_{kl}$, by neglecting the distance factor and phase shift due to propagation delay, the electrical field solution of the wave equation at the direction of the receiver can be expressed in a simplified form as \cite{Balanis2005}
  \begin{equation}  \label{elec_theta}
  \boldsymbol{E}_{\theta}^{l}(f_0,t)
  =\hat{\boldsymbol{\theta}}_{kl}
  \Re\left\{E^{l}_{\theta}F^{l}_{\theta}(\theta_{kl},f_0)\
  e^{i2\pi f_0t}\right\}.
  \end{equation}
In \eqref{elec_theta}, $E^{l}_{\theta}$ is a complex scalar, i.e. $E^{l}_{\theta}=E^{l}_{\theta,0}\ e^{i\zeta_{\theta}^{l}}$,
  where $E^{l}_{\theta,0}$ is the amplitude and $\zeta_{\theta}^{l}$ is the phase delay of the signal
  fed to the dipole along the $z$-axis. The function $F^{l}_{\theta}(\theta,f)$ gives the field 
	pattern for the elevation angle $\theta$ and the
  frequency $f_0$. If $d_{\mathrm{len}}$ is the length of a dipole, the normalized field 
	pattern (to have a unity maximum gain) of this dipole is given by the function
  \cite[Sec: 4.52]{Balanis2005}
  \begin{equation}
  \begin{array}{r@{}l}
  \label{ant_gain_theta}
      F(\theta,f_0)&{}=\displaystyle
      \frac{\cos\left(\frac{\pi
      d_{\mathrm{len}}}{\lambda}\cos \theta\right)-\cos\left(\frac{\pi
      d_{\mathrm{len}}}{\lambda}\right)}{\sin \theta}\\
      &{}= \displaystyle\frac{\cos\left(\pi f_0\cos \theta \frac{
      d_{\mathrm{len}}}{c}\right)-\cos\left(\pi f_0\frac{
      d_{\mathrm{len}}}{c}\right)}{\sin \theta}.
  \end{array}
  \end{equation}
  It can be observed from \eqref{ant_gain_theta} that
  the dipole antenna has an omni-directional radiation pattern only in
  the azimuth direction but not in elevation. Further, in order to achieve maximum signal reception, the receiving antenna should be aligned along the direction of the incoming wave $\hat{\boldsymbol{\theta}}_{kl}$. If the receiving antenna is aligned along the direction orthogonal to $\hat{\boldsymbol{\theta}}_{kl}$, due to polarization mismatch the received signal strength will be very low. Therefore, we consider
  another dipole oriented along the $y$-axis to
  compensate for it\footnotemark. The total electric field received
at a distant point is the superposition of the field components
emitted from the dipole antennas oriented along the
$y$- and $z$-axes.
\footnotetext[2]{{By placing third dipole along the
    $x$-axis, one can achieve a radiation pattern that
    is even closer to an ideal, isotropic antenna pattern. Further, the polarization mismatch loss also can be significantly reduced.}}

\paragraph*{Polarization of the wave transmitted by the dipole placed along the $y$-axis}
   Let $\psi_{kl}=\cos^{-1}(\hat{\boldsymbol{y}}\cdot
      \hat{\boldsymbol{p}}_{kl})=\cos^{-1}
      \left(\frac{y_{kl}}{d_{kl}}\right)\ $ be the elevation angle
  between the dipole oriented along the $y$-axis and
  the propagation direction. Then, similarly to \eqref{elec_theta}, the
  electric field component emitted by the $y$-axis directed dipole at the direction of the receiver is
\begin{align*}
  \boldsymbol{E}_{\psi}^{l}(f_0,t)
  & = \hat{\boldsymbol{\psi}}_{kl}
  \Re\bigg\{E^{l}_{\psi}F^{l}_{\psi}(\psi_{kl},f_0)\
  e^{i2\pi f_0t}\bigg\},
\end{align*}
  where $E^{l}_{\psi}=E^{l}_{\psi,0}\ e^{i\zeta^{l}_{\psi}}$. 
  The unit direction vector $\hat{\boldsymbol{\psi}}_{kl}$ represents the orientation of the electric
  field emitted by the $y$-axis directed dipole. It is in the direction of the 
  vector orthogonal to both $\hat{\boldsymbol{p}}_{kl}$ and 
  $\hat{\boldsymbol{y}}\times\hat{\boldsymbol{p}}_{kl}$, i.e.
	\begin{align} \hat{\boldsymbol{\psi}}_{kl}=
  \frac{\left(\begin{matrix}\hat{\boldsymbol{x}}&\hat{\boldsymbol{y}}&\hat{\boldsymbol{z}}
    \end{matrix}\right)}{d_{kl}\sqrt{x_{kl}^2+z_{kl}^2}}\left(\begin{matrix}
    -x_{kl}y_{kl} \\x_{kl}^2+z_{kl}^2 \\-y_{kl}z_{kl}
  \end{matrix}\right).
  \label{psi_hat}
  \end{align}\normalsize
  
	\paragraph*{Polarization of the wave transmitted by the cross-dipole}
	The total electric field at position $P_{kl}$ is
  \begin{align}
  \label{elec_tot}
   &\boldsymbol{E}_{\mathrm{Tot}}^{l}(f_0,t)=
  \boldsymbol{E}_{\theta}^{l}(f_0,t)+\boldsymbol{E}_{\psi}^{l}(f_0,t)\\
  & = \Re\left\{\bigg(\hat{\boldsymbol{\theta}}_{kl}
  E^{l}_{\theta}F^{l}_{\theta}(\theta_{kl},f_0)+
  \hat{\boldsymbol{\psi}}_{kl}E^{l}_{\psi}F^{l}_{\psi}(\psi_{kl},f_0)\bigg)\
  e^{i2\pi f_0t}\right\}.\nonumber
  \end{align}
  At various time instances, the orientation of the electric field
  will be in the $\hat{\boldsymbol{\theta}}_{kl}-\hat{\boldsymbol{\psi}}_{kl}$ plane which is orthogonal to the wave travel direction\footnotemark. Let the vector function $\boldsymbol{\mathrm{E}}^{l}(\theta_{kl},\psi_{kl},f_0)
      $ be defined as
  \begin{equation}
    \label{GS_response_vec}
    \boldsymbol{\mathrm{E}}^{l}(\theta_{kl},\psi_{kl},f_0)
      =\left(\begin{matrix}E^{l}_{\theta}
        F^{l}_{\theta}(\theta_{kl},f_0) \\ E^{l}_{\psi}
        F^{l}_{\psi}(\psi_{kl},f_0)\end{matrix}\right).
  \end{equation}
  The total electric field vector in \eqref{elec_tot} can be rewritten as $\boldsymbol{E}_{\mathrm{Tot}}^{l}(P_{kl},f_0,t) = \Re\big\{ \boldsymbol{\mathcal{E}}_l
      e^{i2\pi f_0t}\big\}$, where $\boldsymbol{\mathcal{E}}_l$ is the response vector
  of the transmit antenna defined as
  \begin{equation}
  \begin{array}{r@{}l}
  \label{inc_response_vector}
  \boldsymbol{\mathcal{E}}_{kl}&{}=
  \left(\hat{\boldsymbol{\theta}}_{kl}
  \ \ \hat{\boldsymbol{\psi}}_{kl}\right) \cdot
  \boldsymbol{\mathrm{E}}^{l}(\theta_{kl},\psi_{kl},f_0)\\
  &{}=
  \hat{\boldsymbol{\theta}}_{kl}E^{l}_{\theta}F^{l}_{\theta}(\theta_{kl},f_0)+
  \hat{\boldsymbol{\psi}}_{kl}E^{l}_{\psi}F^{l}_{\psi}(\psi_{kl},f_0).
  \end{array}
  \end{equation}
  The polarization of the electric field can be expressed by the unit
  vector $\hat{\boldsymbol{\mathcal{E}}}_l=
      \frac{{\boldsymbol{\mathcal{E}}_l}}{\norm{\boldsymbol{\mathcal{E}}_l}}$.

  \footnotetext[3]{The polarization (i.e. linear, circular or
    elliptical) of the resultant wave is determined by the magnitudes
    and phase difference between the quantities
    $E^{l}_{\theta}$ and
    $E^{l}_{\psi}$ and the direction of wave
    propagation. For example, if the two dipoles were fed with equal magnitude
    (i.e. $E^{l}_{\theta,0}=E^{l}_{\psi,0}=\frac{1}{\sqrt{2}}$) and
    a $\frac{\pi}{2}$ phase difference
    (i.e. $\zeta^{l}_{\theta}-\zeta^{l}_{\psi}=\frac{\pi}{2}$)
    then the polarization along the direction perpendicular to both
    dipole axes (the $x$-axis) would be circular and
    elliptical (or linear) in other directions (for more details see
    \cite[Sec: 2.12]{Balanis2005}).  }

	\paragraph*{Polarization of the receive cross-dipole}

  Let $\hat{\boldsymbol{p}}_{kl}^{'}$ be the
  propagation direction unit vector measured from the receiver, i.e. $\hat{\boldsymbol{p}}_{kl}^{'} = -\hat{\boldsymbol{p}}_{kl}$. When the transmitted electromagnetic wave is illuminated on the
  antenna at the receiver side, the induced field strength depends on
  the field patterns of the receive dipoles in the incoming
  propagation direction. The elevation angles for the receive
  dipoles oriented along the $z$- and
  $y$-axes are obtained as $\theta_{{kl}}^{'}
      =\cos^{-1}(\hat{\boldsymbol{z}}\cdot\hat{\boldsymbol{p}}_{kl}^{'})=
      \cos^{-1}\left(-\frac{z_{kl}}{d_{kl}}\right)=\pi-\theta_{kl}$
  and $\psi_{{kl}}^{'}
      =\cos^{-1}(\hat{\boldsymbol{y}}\cdot\hat{\boldsymbol{p}}_{kl}^{'})=
      \cos^{-1}\left(-\frac{y_{kl}}{d_{kl}}\right)=\pi-\psi_{kl}$,
  respectively. Let $F^{'k}_{\theta}(\theta_{kl}^{'},f)$ and
  $F^{'k}_{\psi}(\psi_{kl}^{'},f)$ be the field patterns 
	of $z$ and $y$ directed
  receiving dipoles, respectively. The outputs from the dipoles are
  appropriately amplified and phase shifted in order to match with the
  polarization of the wave. Let the complex magnitudes of these
  amplifications be $E^{'k}_{\theta}=E^{'k}_{\theta,0}\ e^{i\zeta^{'k}_{\theta}}$
  and $E^{'k}_{\psi}=E^{'k}_{\psi,0}\ e^{i\zeta^{'k}_{\psi}}$. 
	Similarly to \eqref{GS_response_vec}, we can define
  \begin{equation}
  \label{UAV_response_vec}
    \boldsymbol{\mathrm{E}}^{'k}(\theta_{kl}^{'},\psi_{kl}^{'},f_0)=
      \left(\begin{matrix}E^{'k}_{\theta}F^{'k}_{\theta}(\theta_{kl}^{'},f_0)
        \\ E^{'k}_{\psi}F^{'k}_{\psi}(\psi_{kl}^{'},f_0)\end{matrix}\right)
  \end{equation}
  and the response vector of the receive antenna can be written as
  \begin{equation}
  \begin{array}{r@{}l}
  \label{rece_response_vector}
    \boldsymbol{\mathcal{E}}_{kl}^{'}&{}=
    \left(\hat{\boldsymbol{z}}
    \ \ \hat{\boldsymbol{y}}\right)\cdot
    \boldsymbol{\mathrm{E}}^{'k}(\theta_{kl}^{'},\psi_{kl}^{'},f_0)\\
    &{}=
    \hat{\boldsymbol{z}}E^{'k}_{\theta}F^{'k}_{\theta}(\theta_{kl}^{'},f_0)+
    \hat{\boldsymbol{y}}E^{'k}_{\psi}F^{'k}_{\psi}(\psi_{kl}^{'},f_0).
  \end{array}
  \end{equation}
  The polarization of the electric field of the receive antenna can
  be expressed as $\hat{\boldsymbol{\mathcal{E}}}_k^{'}=
      \frac{{\boldsymbol{\mathcal{E}}_k^{'}}}{\norm{\boldsymbol{\mathcal{E}}_k^{'}}}$.
			
				\paragraph*{Polarization loss factor (PLF)}
  
	The quantity $h_{kl}(f_0)$ in \eqref{received_voltage_vt} can be obtained by
  projecting the incident electric field vector ($\boldsymbol{\mathcal{E}}_{kl}$ as given in \eqref{inc_response_vector}) upon the receiving antenna
  response vector ($\boldsymbol{\mathcal{E}}_{kl}^{'}$ as given in \eqref{rece_response_vector}), i.e.
  \begin{align*}
    h_{kl}(f_0)& 
    =  \boldsymbol{\mathcal{E}}_{kl}\cdot\boldsymbol{\mathcal{E}}_{kl}^{'}
    =  \boldsymbol{\mathcal{E}}_{kl}^{H}\boldsymbol{\mathcal{E}}_{kl}^{'}\nonumber\\& =
    \left(\boldsymbol{\mathrm{E}}^{l}(\theta_{kl},\psi_{kl},f_0)\!\right)^{H}\!\!
    \left(\!\begin{matrix}
      \hat{\boldsymbol{\theta}}_{kl}^T \\
      \hat{\boldsymbol{\psi}}_{kl}^T
    \end{matrix}\right)\!\!
    \left(\!\begin{matrix}
      \hat{\boldsymbol{z}} & \!\!
      \hat{\boldsymbol{y}}
    \end{matrix}\!\right)
    \boldsymbol{\mathrm{E}}^{'k}(\theta_{kl}^{'},\psi_{kl}^{'},f_0)\nonumber \\
    &= \left(\boldsymbol{\mathrm{E}}^{l}(\theta_{kl},\psi_{kl},f_0)\!\right)^{H}\!
    \boldsymbol{\mathrm{T}}_{kl}
    \boldsymbol{\mathrm{E}}^{'k}(\theta_{kl}^{'},\psi_{kl}^{'},f_0). 
  \end{align*}
	We can calculate the $2\times 2$ matrix $\boldsymbol{\mathrm{T}}_{kl}$ as 
  \begin{equation}
  \begin{array}{r@{}l}
  \label{twobytwomatrix}
  \boldsymbol{\mathrm{T}}_{kl} &{} \displaystyle =
  \left(\begin{matrix}
    \hat{\boldsymbol{\theta}}_{kl}^T \\
    \hat{\boldsymbol{\psi}}_{kl} ^T    
  \end{matrix}\right)\left(
  \begin{matrix}
    \hat{\boldsymbol{z}} &  \hat{\boldsymbol{y}}
  \end{matrix}\right)=
  \left(\begin{matrix}
    \hat{\boldsymbol{\theta}}_{kl}^T \hat{\boldsymbol{z}}   &
    \hat{\boldsymbol{\theta}}_{kl}^T \hat{\boldsymbol{y}}  \\
    \hat{\boldsymbol{\psi}}_{kl}^T \hat{\boldsymbol{z}} &
    \hat{\boldsymbol{\psi}}_{kl}^T \hat{\boldsymbol{y}}
  \end{matrix}\right)\\
  &{} \displaystyle =
  \frac{1}{d_{kl}} \left(\begin{matrix}
    \sqrt{x_{kl}^2+y_{kl}^2}  & -\frac{y_{kl}z_{kl}}{\sqrt{x_{kl}^2+y_{kl}^2}}     \\
    -\frac{y_{kl}z_{kl}}{\sqrt{x_{kl}^2+z_{kl}^2}} &  \sqrt{x_{kl}^2+z_{kl}^2} 
  \end{matrix}\right),
  \end{array}
  \end{equation}
  where the matrix entries denote the polarization
  mismatch factors between the orientations of electric field
  components and the dipoles at the UAV. Using this result we
  obtain that 
  \begin{align}
  \label{CTF}
   & h_{kl}(f_0) = 
   \left(\boldsymbol{\mathrm{E}}^{l}(\theta_{kl},\psi_{kl},f_0)\right)^{H}
    \boldsymbol{\mathrm{T}}_{kl} 
    \boldsymbol{\mathrm{E}}^{'k}(\theta_{kl}^{'},\psi_{kl}^{'},f_0) \nonumber\\
   & \displaystyle \!=\!\frac{1}{d_{kl}}\!
    \left(\begin{matrix}
      E^{l}_{\theta,0}F^{l}_{\theta}(\theta_{kl},f_0) 
      e^{-i\zeta^{l}_{\theta}}\! \\
      E^{l}_{\psi,0}F^{l}_{\psi}(\psi_{kl},f_0) e^{-i\zeta^{l}_{\psi}}\!
    \end{matrix}\right)^T\!\!\!\!
    \left(\begin{matrix}
      \!\!\sqrt{x_{kl}^2+y_{kl}^2}\!  & 
      \!-\frac{y_{kl}z_{kl}}{\sqrt{x_{kl}^2+y_{kl}^2}}\!\!     \\
      \!\!-\frac{y_{kl}z_{kl}}{\sqrt{x_{kl}^2+z_{kl}^2}}\! &  
      \!\sqrt{x_{kl}^2+z_{kl}^2}\!\!
    \end{matrix}\right)\nonumber\\
    &\hspace{1cm}\times
    \left(\begin{matrix}
      E^{'k}_{\theta,0}F^{'k}_{\theta}(\theta_{kl}^{'},f_0) e^{i\zeta^{'k}_{\theta}} \\
      E^{'k}_{\psi,0}F^{'k}_{\psi}(\psi_{kl}^{'},f_0) e^{i\zeta^{'k}_{\psi}}
    \end{matrix}\right)\\
    &\!= \frac{\sqrt{x_{kl}^2+y_{kl}^2}}{d_{kl}}
    E^{l}_{\theta,0}E^{'k}_{\theta,0}F^{l}_{\theta}(\theta_{kl},f_0)
    F^{'k}_{\theta}(\theta_{kl}^{'},f_0)
    e^{i\left(\zeta^{'k}_{\theta}-\zeta^{l}_{\theta} \right)}\nonumber\\
    & -\frac{y_{kl}z_{kl}}{d_{kl}\sqrt{x_{kl}^2+y_{kl}^2}}     
    E^{l}_{\theta,0}E^{'k}_{\psi,0}F^{l}_{\theta}(\theta_{kl},f_0)
    F^{'k}_{\psi}(\psi_{kl}^{'},f_0)
    e^{i\left(\zeta^{'k}_{\psi}-\zeta^{l}_{\theta} \right)}\nonumber\\
    & -\frac{y_{kl}z_{kl}}{d_{kl}\sqrt{x_{kl}^2\!+\!z_{kl}^2}}     
    E^{l}_{\psi,0}E^{'k}_{\theta,0}F^{l}_{\psi}(\psi_{kl},f_0)
    F^{'k}_{\theta}(\theta_{kl}^{'},f_0)
    e^{i\left(\zeta^{'k}_{\theta}-\zeta^{l}_{\psi} \right)}\nonumber\\
    & + \frac{\sqrt{x_{kl}^2+z_{kl}^2}}{d_{kl}}
    E^{l}_{\psi,0}E^{'k}_{\psi,0}F^{l}_{\psi}(\psi_{kl},f_0)
    F^{'k}_{\psi}(\psi_{kl}^{'},f_0)
    e^{i\left(\zeta^{'k}_{\psi}-\zeta^{l}_{\psi} \right)}.\nonumber
\end{align}
	
	The polarization loss factor between the transmitted electromagnetic wave and the
  receive antennas is given by \cite{Balanis2005}
  \begin{equation}
  \begin{array}{r@{}l}
      \mathrm{PLF}_{kl} &{} =|h_{kl}(f_0)|^2  
      =|\hat{\boldsymbol{\mathcal{E}}}_{kl}\cdot
      \hat{\boldsymbol{\mathcal{E}}}_{kl}^{'}|^2=
      |\hat{\boldsymbol{\mathcal{E}}}_{kl}^H\hat{\boldsymbol{\mathcal{E}}}_{kl}^{'}|^2
      \\ &{} =
      \hat{\boldsymbol{\mathcal{E}}}_{kl}^H
      \hat{\boldsymbol{\mathcal{E}}}_{kl}^{'}
      (\hat{\boldsymbol{\mathcal{E}}}_{kl}^{'})^H
      \hat{\boldsymbol{\mathcal{E}}}_{kl},
  \end{array}
  \end{equation}
  where $(\cdot)^H$ denotes Hermitian
  transpose. {Note that even with fixed orientations of
  the transmit and receive antennas, if we move the receive antenna's
  position around the transmit antenna, the PLF can be high for
  certain elevation and azimuth angles irrespective of the type of
  polarization of transmit and receive antennas.}

In this subsection we considered the situation when the dipoles at both
transmitting and receiving end are aligned with the $y$- and
$z$-axes. In the following subsection, we detail the calculation of $h_{kl}(f_0)$ due to the rotation of
antennas at the transmitter and the receiver.

\begin{figure*}[t]
\centering
\subfigure[]{\includegraphics[scale=.85,trim={.1cm .15cm .1cm .1cm},clip]{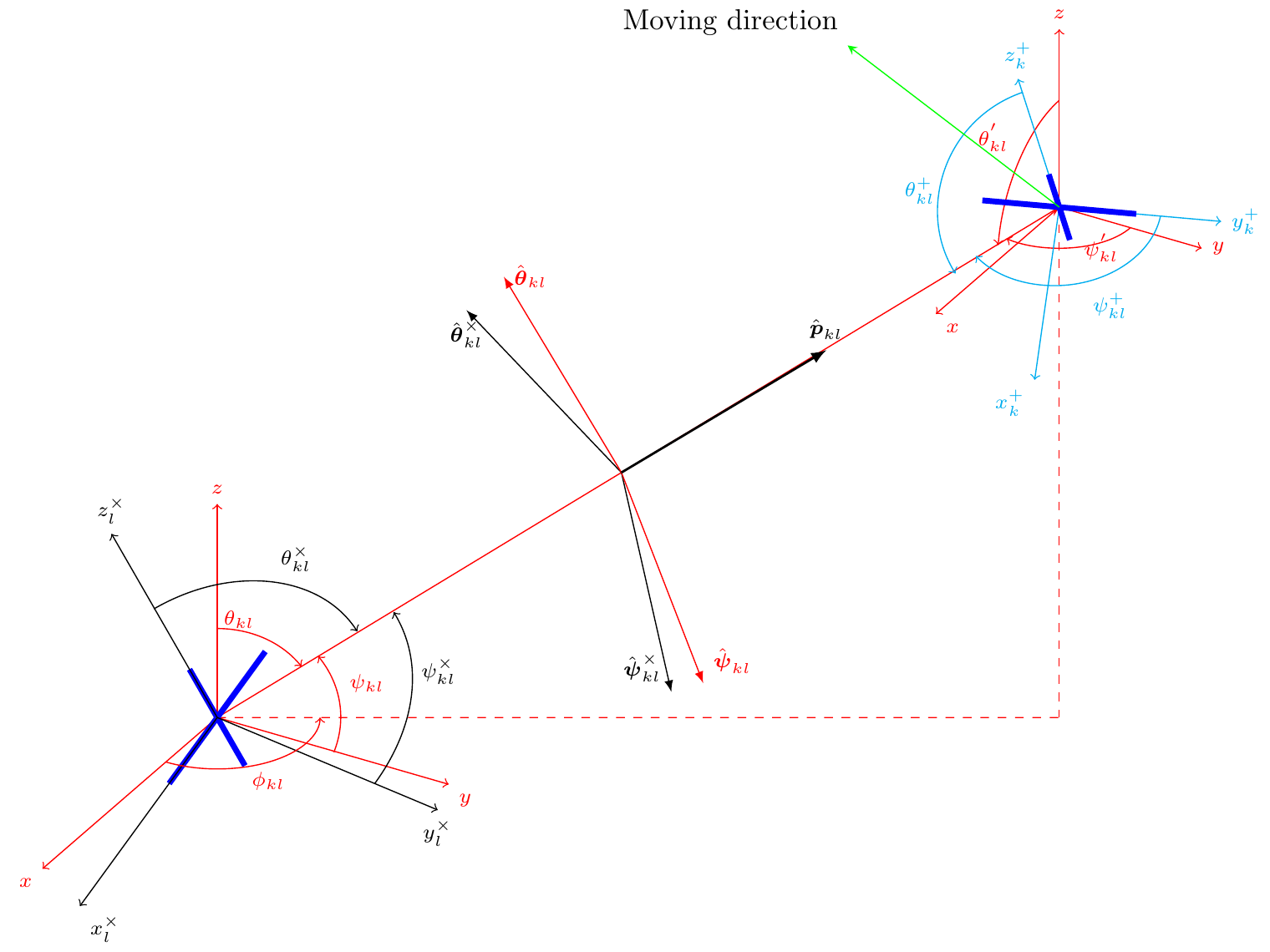}\label{geometric_model_ula_final_journal}}\hspace{-.025cm}
\subfigure[]{\includegraphics[scale=.82,trim={-1.01cm -3.01cm .1cm
      .1cm},clip]{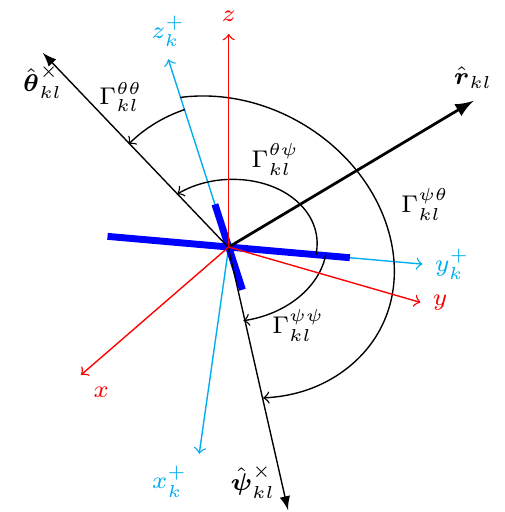}\label{geometric_model_ula_pol_maf_journal}}
\caption{(a) Schematic representation of the elevation angles measured with 
respect to the coordinated axes at the transmitting and receiving end. This 
figure also illustrates the orientation of the electric field vectors. (b) 
Schematic representation of the projections of the electric field vectors on 
UAV's dipoles.}
\end{figure*}

\subsection{Single antenna channel transfer function for arbitrary orientation of GS and UAV antennas}
Since arbitrary rotation of transmit and receive antennas change
the orientation of the transmitted electric field vectors and the
elevation angles, they have to be calculated with respect to the
rotated coordinate axes as shown in Figure
\ref{geometric_model_ula_final_journal}. We denote the components of
rotated coordinated systems of the transmit and receive antennas
using the superscripts $^{\times}$ and $^{+}$, respectively. The unit
direction vectors of the transmit antenna's (i.e. $l$-th GS
antenna's) rotated coordinate system are obtained as 
\begin{equation*} 
\left(\begin{matrix}
  \hat{\boldsymbol{x}}_l^{\times} & \hat{\boldsymbol{y}}_l^{\times} &
  \hat{\boldsymbol{z}}_l^{\times}
\end{matrix}\right) =
\left(\begin{matrix}
  \hat{\boldsymbol{x}} & \hat{\boldsymbol{y}} & \hat{\boldsymbol{z}}
\end{matrix}\right)
\boldsymbol{\mathrm{R}}_{l}^\times\left(\alpha_{l,x}^{\times},\alpha_{l,y}^{\times},\alpha_{l,z}^{\times}\right),
\end{equation*}
\normalsize where
 $\boldsymbol{\mathrm{R}}_{l}^\times(\alpha_{l,x}^{\times},\alpha_{l,y}^{\times},\alpha_{l,z}^{\times})$ 
is the $3\times 3$ rotation matrix obtained as a function of the roll
($\alpha_{l,x}^{\times}\in [-\frac{\pi}{2},\frac{\pi}{2}]$), pitch
($\alpha_{l,y}^{\times}\in  [-\frac{\pi}{2},\frac{\pi}{2}]$), and yaw
($\alpha_{l,z}^{\times}\in [0,2\pi]$) angles. For the calculation
of this rotation matrix, see Appendix \ref{section_uav_orientation}.
In this coordinate system, the receiver's (i.e. $k$-th UAV's) position
has coordinates
\begin{equation*}
\begin{array}{r@{}l}
\left(\begin{matrix}
  x_{kl}^{\times} \\ y_{kl}^{\times} \\ z_{kl}^{\times}
\end{matrix}\right) &{} =
\left(\boldsymbol{\mathrm{R}}_{l}^{\times}\left(\alpha_{k,x}^{\times},
	  \alpha_{k,y}^{\times},\alpha_{k,z}^{\times}\right)\right)^T
\left(\begin{matrix}
  x_{kl} \\ y_{kl}\\ z_{kl}
\end{matrix}\right) \\
   &{}=
\left(\begin{matrix}
  \boldsymbol{\mathrm{R}}_{l,11}^{\times}x_{kl}+\boldsymbol{\mathrm{R}}_{l,21}^{\times}y_{kl}+\boldsymbol{\mathrm{R}}_{l,31}^{\times}z_{kl}\\
  \boldsymbol{\mathrm{R}}_{l,12}^{\times}x_{kl}+\boldsymbol{\mathrm{R}}_{l,22}^{\times}y_{kl}+\boldsymbol{\mathrm{R}}_{l,32}^{\times}z_{kl}\\
  \boldsymbol{\mathrm{R}}_{l,13}^{\times}x_{kl}+\boldsymbol{\mathrm{R}}_{l,23}^{\times}y_{kl}+\boldsymbol{\mathrm{R}}_{l,33}^{\times}z_{kl}
\end{matrix}\right).
\end{array}
\end{equation*}
The elevation angles with respect to the rotated coordinate axes at the transmitter are calculated as $\theta_{kl}^\times =
    \cos^{-1}(\hat{\boldsymbol{z}}_l^\times\cdot\hat{\boldsymbol{p}}_{kl})=\cos^{-1}\left(\frac{z_{kl}^\times}{d_{kl}}\right)$ 
and $\psi_{kl}^\times =
    \cos^{-1}(\hat{\boldsymbol{y}}_l^\times\cdot\hat{\boldsymbol{p}}_{kl})=\cos^{-1}\left(\frac{y_{kl}^\times}{d_{kl}}\right)$.

Similarly, let $\boldsymbol{\mathrm{R}}_{k}^+(\alpha_{k,x}^{+},\alpha_{k,y}^{+},\alpha_{k,z}^{+})$ 
be the rotation matrix of the receive antenna. The unit direction
vectors of the receive antenna's rotated coordinate system are
obtained as
\begin{equation*}
 \left(\begin{matrix}
  \hat{\boldsymbol{x}}_k^{+} &
  \hat{\boldsymbol{y}}_k^{+} &
  \hat{\boldsymbol{z}}_k^{+}
\end{matrix}\right) =
\left(\begin{matrix}
  \hat{\boldsymbol{x}} & \hat{\boldsymbol{y}} & \hat{\boldsymbol{z}}
\end{matrix}\right)
\boldsymbol{\mathrm{R}}_{k}^+\left(\alpha_{k,x}^{+},\alpha_{k,y}^{+},\alpha_{k,z}^{+}\right).
\end{equation*}
The elevation angles with respect to the rotated coordinate axes at
the receiving end are calculated as $\theta_{{kl}}^{+}
    =\cos^{-1}(\hat{\boldsymbol{z}}^{+}\cdot\hat{\boldsymbol{p}}_{kl}^{'})=\cos^{-1}\left(-\frac{z_{kl}^+}{d_{kl}}\right)$
and $\psi_{{kl}}^{+}
    =\cos^{-1}(\hat{\boldsymbol{y}}^{+}\cdot\hat{\boldsymbol{p}}_{kl}^{'})=\cos^{-1}\left(-\frac{y_{kl}^+}{d_{kl}}\right)$,
where
\begin{equation*}
\begin{array}{r@{}l}
\left(\begin{matrix}
  x_{kl}^{+} \\ y_{kl}^{+} \\ z_{kl}^{+}
\end{matrix}\right) &{} =
\left(\boldsymbol{\mathrm{R}}_{k}^{+}\left(\alpha_{k,x}^{+},\alpha_{k,y}^{+},\alpha_{k,z}^{+}\right)\right)^T
\left(\begin{matrix}
  x_{kl} \\ y_{kl}\\ z_{kl}
\end{matrix}\right)\\ &{}=
\left(\begin{matrix}
  \boldsymbol{\mathrm{R}}_{k,11}^{+}x_{kl}+\boldsymbol{\mathrm{R}}_{k,21}^{+}y_{kl}+\boldsymbol{\mathrm{R}}_{k,31}^{+}z_{kl}\\
  \boldsymbol{\mathrm{R}}_{k,12}^{+}x_{kl}+\boldsymbol{\mathrm{R}}_{k,22}^{+}y_{kl}+\boldsymbol{\mathrm{R}}_{k,32}^{+}z_{kl}\\
  \boldsymbol{\mathrm{R}}_{k,13}^{+}x_{kl}+\boldsymbol{\mathrm{R}}_{k,23}^{+}y_{kl}+\boldsymbol{\mathrm{R}}_{k,33}^{+}z_{kl}
\end{matrix}\right).
\end{array}
\end{equation*}

The quantity $h_{kl}(f_0)$ when we consider the rotations of the
transmit and receive antennas can be calculated
as in \eqref{CTF} to be 
\begin{equation}\label{CTF_rot}
  h_{kl}^{\times,+}(f_0)=
    \left(\boldsymbol{\mathrm{E}}^{l}(\theta_{kl}^\times,\psi_{kl}^\times,f_0)\right)^H\
    \boldsymbol{\mathrm{T}}_{kl}^{\times,+}\
    \boldsymbol{\mathrm{E}}^{'k}(\theta_{kl}^{+},\psi_{kl}^{+},f_0),
\end{equation}
where $\boldsymbol{\mathrm{E}}^{l}(\theta_{kl}^\times,\psi_{kl}^\times,f_0)$
and $\boldsymbol{\mathrm{E}}^{'k}(\theta_{kl}^{+},\psi_{kl}^{+},f_0)$
are defined similarly to \eqref{GS_response_vec} and
\eqref{UAV_response_vec} as
\begin{align*}
  \boldsymbol{\mathrm{E}}^{l}(\theta^{\times}_{kl},\psi^{\times}_{kl},f_0)
   & =\left(\begin{matrix}
      E^{l}_{\theta}F^{l}_{\theta}(\theta^{\times}_{kl},f_0) \\
      E^{l}_{\psi}  F^{l}_{\psi} (\psi^{\times}_{kl},f_0)
    \end{matrix}\right)\nonumber\\
  &\text{ and }\nonumber\\
   \boldsymbol{\mathrm{E}}^{'k}(\theta^{+}_{kl},\psi^{+}_{kl},f_0)
    &=\left(\begin{matrix}
      E^{'k}_{\theta}F^{'k}_{\theta}(\theta^{+}_{kl},f_0) \\
      E^{'k}_{\psi}  F^{'k}_{\psi} (\psi^{+}_{kl},f_0)
    \end{matrix}\right),
\end{align*}
and the matrix
$\boldsymbol{\mathrm{T}}_{kl}^{\times,+}$  is
\begin{align*}
  \boldsymbol{\mathrm{T}}_{kl}^{\times,+} & =
  \left(\begin{matrix}
    (\hat{\boldsymbol{\theta}}_{kl}^{\times})^T \\
    (\hat{\boldsymbol{\psi}}_{kl}^{\times})^T
  \end{matrix}
  \right)
  \left(\begin{matrix}
    \hat{\boldsymbol{z}}^{+} & \!\hat{\boldsymbol{y}}^{+}
  \end{matrix}\right) \! = \! 
  \left(\begin{matrix}
    (\hat{\boldsymbol{\theta}}_{kl}^{\times})^T \hat{\boldsymbol{z}}^{+}   &
    (\hat{\boldsymbol{\theta}}_{kl}^{\times})^T \hat{\boldsymbol{y}}^{+}  \\
    (\hat{\boldsymbol{\psi}}_{kl}^{\times})^T \hat{\boldsymbol{z}}^{+} &
    (\hat{\boldsymbol{\psi}}_{kl}^{\times})^T \hat{\boldsymbol{y}}^{+}
  \end{matrix}\right). 
\end{align*}

Similarly to the calculation of $\hat{\boldsymbol{\theta}}_{kl}$ and $\hat{\boldsymbol{\psi}}_{kl}$ in \eqref{theta_hat} and
\eqref{psi_hat}, the unit directions of the electric fields,
$\hat{\boldsymbol{\theta}}_{kl}^\times$ and
$\hat{\boldsymbol{\psi}}_{kl}^\times$ can be calculated
in the rotated transmit antenna coordinate system as 
\begin{align*}
\hat{\boldsymbol{\theta}}_{kl}^\times &=
  \frac{\left(\begin{matrix} \hat{\boldsymbol{x}}_l^{\times} &
      \hat{\boldsymbol{y}}_l^{\times} &
      \hat{\boldsymbol{z}}_l^{\times}
    \end{matrix}\right)
  }{d_{kl}\sqrt{(x_{kl}^{\times})^2+(y_{kl}^{\times})^2}}
  \left(\begin{matrix}
    -x_{kl}^{\times}z_{kl}^{\times} \\
    -y_{kl}^{\times}z_{kl}^{\times} \\
    (x_{kl}^{\times})^2+(y_{kl}^{\times})^2
  \end{matrix}\right)\nonumber\\&= 
  \left(\begin{matrix}
    \hat{\boldsymbol{x}} &
    \hat{\boldsymbol{y}} &
    \hat{\boldsymbol{z}}
    \end{matrix}\right)
  \frac{\boldsymbol{\mathrm{R}}_{l}^{\times}\left(\alpha_{l,x}^{\times},\alpha_{l,y}^{\times},\alpha_{l,z}^{\times}\right)}
       {d_{kl}\sqrt{(x_{kl}^{\times})^2+(y_{kl}^{\times})^2}}
  \left(\begin{matrix}
    -x_{kl}^{\times}z_{kl}^{\times} \\
    -y_{kl}^{\times}z_{kl}^{\times} \\
    (x_{kl}^{\times})^2+(y_{kl}^{\times})^2
  \end{matrix}\right)
\end{align*}
and
\begin{align*}
  \hat{\boldsymbol{\psi}}_{kl}^\times  & =
  \frac{\left(\begin{matrix}
      \hat{\boldsymbol{x}}_l^{\times} &
      \hat{\boldsymbol{y}}_l^{\times} &
      \hat{\boldsymbol{z}}_l^{\times}
    \end{matrix}\right)
  }{d_{kl}\sqrt{(x_{kl}^{\times})^2+(z_{kl}^{\times})^2}}
  \left(\begin{matrix}
    -x_{kl}^{\times}y_{kl}^{\times} \\
    (x_{kl}^{\times})^2+(z_{kl}^{\times})^2\\
    -y_{kl}^{\times}z_{kl}^{\times}
  \end{matrix}\right) \nonumber\\&= 
  \left(\begin{matrix}
    \hat{\boldsymbol{x}} &
    \hat{\boldsymbol{y}} &
    \hat{\boldsymbol{z}}
  \end{matrix}\right)
  \frac{\boldsymbol{\mathrm{R}}_{l}^{\times}\left(\alpha_{l,x}^{\times},\alpha_{l,y}^{\times},\alpha_{l,z}^{\times}\right)}
       {d_{kl}\sqrt{(x_{kl}^{\times})^2+(z_{kl}^{\times})^2}}
  \left(\begin{matrix}
    -x_{kl}^{\times}y_{kl}^{\times} \\
    (x_{kl}^{\times})^2+(z_{kl}^{\times})^2\\
    -y_{kl}^{\times}z_{kl}^{\times}
  \end{matrix}\right)\!.
\end{align*}
Obviously, for the receive antenna dipole directions we have 
\begin{equation*}
\begin{array}{r@{}l}
    \hat{\boldsymbol{z}}_{k}^{+} &{} =
  \left(\begin{matrix}
    \hat{\boldsymbol{x}}_k^{+} &
    \hat{\boldsymbol{y}}_k^{+} &
    \hat{\boldsymbol{z}}_k^{+}
  \end{matrix}\right)
  \left(\begin{matrix}
    0 & 0 & 1
  \end{matrix}\right)^T \\ &{}=
  \left(\begin{matrix}
    \hat{\boldsymbol{x}} &
    \hat{\boldsymbol{y}} &
    \hat{\boldsymbol{z}}
  \end{matrix}\right)
  \boldsymbol{\mathrm{R}}_{k}^{+}\left(\alpha_{k,x}^{+},\alpha_{k,y}^{+},\alpha_{k,z}^{+}\right)
  \left(\begin{matrix}
    0 & 0 & 1
  \end{matrix}\right)^T
\end{array}
\end{equation*}
and 
\begin{equation*}
\begin{array}{r@{}l}
\hat{\boldsymbol{y}}_{k}^{+} &{} =
\left(\begin{matrix}
\hat{\boldsymbol{x}}_k^{+} &
\hat{\boldsymbol{y}}_k^{+} &
\hat{\boldsymbol{z}}_k^{+}
\end{matrix}\right)
\left(\begin{matrix}
0 & 1 & 0
\end{matrix}\right)^T \\ &{}=
\left(\begin{matrix}
\hat{\boldsymbol{x}} &
\hat{\boldsymbol{y}} &
\hat{\boldsymbol{z}}
\end{matrix}\right)
\boldsymbol{\mathrm{R}}_{k}^{+}\left(\alpha_{k,x}^{+},\alpha_{k,y}^{+},\alpha_{k,z}^{+}\right)
\left(\begin{matrix}
0 & 1 & 0
\end{matrix}\right)^T
\end{array}
\end{equation*}
Since the reference coordinate system is orthonormal, i.e.
$\left(\begin{matrix}
\hat{\boldsymbol{x}} & \hat{\boldsymbol{y}} & \hat{\boldsymbol{z}}
\end{matrix}\right)^T\left(\begin{matrix}
\hat{\boldsymbol{x}} & \hat{\boldsymbol{y}} & \hat{\boldsymbol{z}}
\end{matrix}\right)= \boldsymbol{\mathrm{I}}_3$, we calculate the matrix
 $\boldsymbol{\mathrm{T}}_{kl}^{\times,+}$ as 
\begin{align}
\label{twobytwomatrix_final}
& \boldsymbol{\mathrm{T}}_{kl}^{\times,+}  =
  \left(\begin{matrix}
    (\hat{\boldsymbol{\theta}}_{kl}^{\times})^T \\
    (\hat{\boldsymbol{\psi}}_{kl}^{\times})^T
  \end{matrix}
  \right)
  \left(\begin{matrix}
    \hat{\boldsymbol{z}}^{+} & \hat{\boldsymbol{y}}^{+}
  \end{matrix}\right) \nonumber\\
  & \! = \!
  \frac{1}{d_{kl}}\!\!
  \small
  \left(\begin{matrix}
         -\frac{x_{kl}^{\times}z_{kl}^{\times}}{\sqrt{(x_{kl}^{\times})^2+(y_{kl}^{\times})^2}}
          \! & \!
         -\frac{y_{kl}^{\times}z_{kl}^{\times}}{\sqrt{(x_{kl}^{\times})^2+(y_{kl}^{\times})^2}}
          \! & \! \sqrt{(x_{kl}^{\times})^2+(y_{kl}^{\times})^2} \\
         -\frac{x_{kl}^{\times}y_{kl}^{\times}}{\sqrt{(x_{kl}^{\times})^2+(z_{kl}^{\times})^2}}
          \! & \! \sqrt{(x_{kl}^{\times})^2+(z_{kl}^{\times})^2} \! & \!
         -\frac{z_{kl}^{\times}y_{kl}^{\times}}{\sqrt{(x_{kl}^{\times})^2+(z_{kl}^{\times})^2}}
  \end{matrix}\right)
  \nonumber\\
  & \normalsize
  \hspace{2cm}\times
  \boldsymbol{\mathrm{R}}_{kl}^{\times,+}
  \left(\begin{matrix}
    0 & 0 \\ 0 & 1 \\ 1 & 0
  \end{matrix}\right), 
\end{align}
where $\boldsymbol{\mathrm{R}}_{kl}^{\times,+}$ is
the combined rotation matrix given by 
\begin{align*}
  \boldsymbol{\mathrm{R}}_{kl}^{\times,+}  &=
  \left(\boldsymbol{\mathrm{R}}_{l}^{\times}\left(\alpha_{l,x}^{\times},\alpha_{l,y}^{\times},\alpha_{l,z}^{\times}\right)\right)^T
  \boldsymbol{\mathrm{R}}_{k}^{+}\left(\alpha_{k,x}^{+},\alpha_{k,y}^{+},\alpha_{k,z}^{+}\right)\nonumber\\& =
  \left(\begin{matrix}
  \boldsymbol{\mathrm{R}}_{kl,11}^{\times,+} & \boldsymbol{\mathrm{R}}_{kl,12}^{\times,+} & \boldsymbol{\mathrm{R}}_{kl,13}^{\times,+}\\
  \boldsymbol{\mathrm{R}}_{kl,21}^{\times,+} & \boldsymbol{\mathrm{R}}_{kl,22}^{\times,+} & \boldsymbol{\mathrm{R}}_{kl,23}^{\times,+}\\
  \boldsymbol{\mathrm{R}}_{kl,31}^{\times,+} & \boldsymbol{\mathrm{R}}_{kl,32}^{\times,+} & \boldsymbol{\mathrm{R}}_{kl,33}^{\times,+}
\end{matrix}\right).
\end{align*}

Note that the elements of $\boldsymbol{\mathrm{T}}_{kl}^{\times,+}$ are the
projections of the unit electric field vectors components $\hat{\boldsymbol{\theta}}_{kl}^\times$ and
$\hat{\boldsymbol{\psi}}_{kl}^\times$ on the receive
dipoles unit vectors $\hat{\boldsymbol{z}}_{k}^{+}$ and $\hat{\boldsymbol{y}}_{k}^{+}$ as shown in Figure
\ref{geometric_model_ula_pol_maf_journal}, i.e.
\begin{equation}
    \boldsymbol{\mathrm{T}}_{kl}^{\times,+} =
    \left(\begin{matrix}
      \cos \Gamma_{kl}^{\theta\theta}   &  \cos \Gamma_{kl}^{\theta\psi}   \\
      \cos \Gamma_{kl}^{\psi\theta}   &  \cos \Gamma_{kl}^{\psi\psi} 
  \end{matrix}\right).
\end{equation}
The magnitudes of these projections are
\begin{align*}
    & \cos\Gamma_{kl}^{\theta\theta}  = 
    \hat{\boldsymbol{\theta}}_{kl}^{\times}\cdot
    \hat{\boldsymbol{z}}_k^{+}\\
    & \ = 
    \frac{-\mathrm{R}_{kl,13}^{\times,+}x_{kl}^{\times}z_{kl}^{\times}
      -\mathrm{R}_{kl,23}^{\times,+}y_{kl}^{\times}z_{kl}^{\times}+
      \mathrm{R}_{kl,33}^{\times,+}((x_{kl}^{\times})^2+(y_{kl}^{\times})^2)}
         {d_{kl}\sqrt{(x_{kl}^{\times})^2+(y_{kl}^{\times})^2}},\\   
    & \cos\Gamma_{kl}^{\theta\psi}  = 
    \hat{\boldsymbol{\theta}}_{kl}^{\times}\cdot\hat{\boldsymbol{y}}_k^{+}\\
    & \ = 
    \frac{-\mathrm{R}_{kl,12}^{\times,+}x_{kl}^{\times}z_{kl}^{\times}
      -\mathrm{R}_{kl,22}^{\times,+}y_{kl}^{\times}z_{kl}^{\times}+
      \mathrm{R}_{kl,32}^{\times,+}((x_{kl}^{\times})^2+(y_{kl}^{\times})^2)}
         {d_{kl}\sqrt{(x_{kl}^{\times})^2+(y_{kl}^{\times})^2}},\\
    & \cos\Gamma_{kl}^{\psi\theta}  = 
    \hat{\boldsymbol{\psi}}_{kl}^{\times}\cdot\hat{\boldsymbol{z}}_k^{+}\\
    & \ = 
    \frac{-\mathrm{R}_{kl,13}^{\times,+}x_{kl}^{\times}y_{kl}^{\times}
      -\mathrm{R}_{kl,33}^{\times,+}z_{kl}^{\times}y_{kl}^{\times}+
      \mathrm{R}_{kl,23}^{\times,+}((x_{kl}^{\times})^2+(z_{kl}^{\times})^2)}
         {d_{kl}\sqrt{(x_{kl}^{\times})^2+(z_{kl}^{\times})^2}},\\
    & \cos\Gamma_{kl}^{\psi\psi}  = 
    \hat{\boldsymbol{\psi}}_{kl}^{\times}\cdot\hat{\boldsymbol{y}}_k^{+}\\
    & \ = 
    \frac{-\mathrm{R}_{kl,12}^{\times,+}x_{kl}^{\times}y_{kl}^{\times}
      -\mathrm{R}_{kl,32}^{\times,+}z_{kl}^{\times}y_{kl}^{\times}+
      \mathrm{R}_{kl,22}^{\times,+}((x_{kl}^{\times})^2+(z_{kl}^{\times})^2)}
         {d_{kl}\sqrt{(x_{kl}^{\times})^2+(z_{kl}^{\times})^2}}.
\end{align*}
\normalsize
It can be verified that if the transmit and receive antennas are not rotated, i.e. when $\boldsymbol{\mathrm{R}}_{kl}^{\times,+}=\boldsymbol{\mathrm{I}}_3$, the $2 \times 2 $ matrix in \eqref{twobytwomatrix_final} is equal to \eqref{twobytwomatrix}.


\subsection{Antenna array channel model}
From \eqref{received_voltage_vt}, the $M\times 1$ channel vector from the $k$-th UAV at position $P_k$ to the GS array with carrier frequency $f_0=f_c$ is given by 
\begin{equation}
\label{channel_gain_vector}
\boldsymbol{g}_k=[g_{k1} \ g_{k2} \ ....\ g_{kM}]^T,
\end{equation}
where the elements are obtained from \eqref{received_voltage_vt} as
\begin{equation*}
g_{kl} = \sqrt{\beta_{kl}}\ h_{kl}\ e^{-i\frac{2\pi }{\lambda}d_{kl}}, \ \ l=1,2,...,M,
\end{equation*}
where the complex quantity $h_{kl}$ incorporates the transmit and receive dipoles' antenna gain and field patterns, and the polarization mismatch loss factors, i.e. $h_{kl}=\sqrt{\mathcal{G}_t \mathcal{G}_r} h_{kl}^{\times,+}(f_c)$, where $\mathcal{G}_t$ and $\mathcal{G}_r$ are the gains of transmit and receive dipole, respectively, and $h_{kl}^{\times,+}(f_c)$ is given in \eqref{CTF_rot}. Note that we introduced $\mathcal{G}_t$ and $\mathcal{G}_r$ here, because the field patterns are normalized in \eqref{ant_gain_theta}. For half-wavelength dipole, the antenna gain is approximately equal to $1.643$ ( $\approx 2.15$ dB). Irrespective of the type of UAV, depending on the placement of the antenna, the UAV's body may also introduce additional loss. For example, measurement results show that the body blockage loss in fixed-wing aircraft is between $5$ to $35$ dB \cite{sun2017_airframe} and in quadcopters it is between $5$ to $15$ dB \cite{asadpour2014}. For simplicity of the analysis, in our model, we do not include the body blockage loss. However, it could be incorporated in the model as a function of the azimuth, elevation, roll, pitch, and yaw angles. 

\subsection{Uplink pilot signaling and channel estimation}
For the purpose of receiver processing, the GS need to know the CSI. Since the UAVs move in a 3D space with high speed, the signal received at the GS will experience Doppler shift. Therefore, the channel vectors at the GS have to be re-estimated after certain time duration, i.e. coherence time. The coherence time is defined as a time interval over which the impact of Doppler shift on the received signal is insignificant. We adopt an over-conservative design to re-estimate the channel after every coherence time $T_{\mathrm{coh}}$ calculated as follows. Given the maximum speed of the UAVs $v_{\mathrm{max}}$, the coherence time can be calculated as $T_{\mathrm{coh}}\approx \frac{1}{2 f_\mathrm{max}}$, where the Doppler frequency $f_\mathrm{max}=\frac{v_{\mathrm{max}}}{\lambda}$. For example, if the maximum speed of UAV is $30$ m/s, at a carrier frequency of $2.4$ GHz, the coherence time is $T_{\mathrm{coh}}\approx 2$ ms.

As the massive MIMO systems operate in TDD mode, downlink transmission, uplink pilot transmission, and uplink data transmission happen within the coherence interval $T_\mathrm{len}$, i.e. $\tau_{\mathrm{dl}}+\tau_{\mathrm{ul,p}}+\tau_{\mathrm{ul,d}} \le T_\mathrm{len}$. The parameters $\tau_{\mathrm{dl}}$, $\tau_{\mathrm{ul,p}}$, and $\tau_{\mathrm{ul,d}}$ denote the number of symbols used for downlink, uplink pilot, and uplink data transmission, respectively. The coherence interval $T_\mathrm{len}$ is defined as the product of coherence time and coherence bandwidth. In LoS, since there is no multipath, the coherence bandwidth is infinite. However, in over-water and mountainous settings, due to a few multipath components, the coherence bandwidth is finite \cite{matolak2017_water,sun2017_hilly,matolak2017_urban}\footnotemark. Therefore, we define 
\begin{equation}\label{coh_length}
T_\mathrm{len}=T_{\mathrm{coh}}\times B_c=\frac{B_c \cdot c}{2\cdot v_{\mathrm{max}}\cdot f_c}, 
\end{equation}
where $B_c$ is the coherence bandwidth. 

\footnotetext[4]{{Recent measurements performed in the
    C-band ($5.03$--$5.091$ GHz) show that the average
    root-mean-square delay spread is typically very small in
    over-water ($\sim 10$ ns), hilly and mountain ($\sim 10$ ns),
    suburban and near-urban ($10$--$60$ ns) environments (with an average UAV altitude of
$600$ m and link ranges from $860$ m to several kilometers)
    \cite{matolak2017_water,sun2017_hilly,matolak2017_urban}. Hence,
    if the coherence bandwidth is defined as the bandwidth over which
    the frequency correlation function is above $0.5$
    \cite{rappaport96}, depending on the environment the coherence
    bandwidth varies between $3$ MHz and $20$ MHz ($300$ KHz and $2$
    MHz if the frequency correlation is 0.9). Further, the measured
    values of Rician $K$-factors in different
    environments is greater than $25$ dB
    \cite{matolak2017_water,sun2017_hilly,matolak2017_urban}. Therefore,
    it is appropriate to consider LoS propagation between the GS and
    the UAVs.}}

During the training phase, $K$ UAVs are assigned $K$ orthogonal pilot sequences of length $\tau_{\mathrm{ul,p}}$. Let the $M\times K$ channel matrix between the GS and the UAVs be $\boldsymbol{G}=[\boldsymbol{g}_1\ \boldsymbol{g}_2\ ... \ \boldsymbol{g}_K]$. The $M\times \tau_{\mathrm{ul,p}}$ received pilot matrix at the GS is given by
\begin{equation*}
\boldsymbol{Y}_{p} = \sqrt{p_p} \boldsymbol{G}\boldsymbol{\Phi}^{T}+\boldsymbol{N}_p,
\end{equation*}
where $p_p$ is transmit power of each pilot symbol, $\boldsymbol{\Phi}$ is 
$\tau_{\mathrm{ul,p}} \times K$ orthogonal pilot matrix satisfies 
$\boldsymbol{\Phi}^H\boldsymbol{\Phi}=\boldsymbol{\mathrm{I}}_{K}$ and 
$\boldsymbol{N}_p$ is $M\times~\tau_{\mathrm{ul,p}}$ noise matrix with i.i.d 
$\mathcal{CN}(0,1)$ elements. For notational convenience, we take noise 
variance to be $1$. Therefore, $p_p$ can be interpreted as normalized transmit 
SNR.

 In order to obtain reliable channel estimate, the pilot power has to be chosen based on the worst-case values of the distance and effective antenna gain, 
\begin{equation}\label{pilot_power}
p_p = \rho_p \left(\frac{4\pi d_\mathrm{wc}}{\lambda}\right)^2 \frac{1}{\chi_\mathrm{wc}},
 \end{equation}
where $\rho_p$ is the target pilot SNR, $d_\mathrm{wc}$ is the maximum possible distance between the GS and UAV, and $\chi_\mathrm{wc}$ is the lowest possible gain over all possible values of azimuth, elevation, and UAV's rotation angles as discussed in Appendix \ref{section_uav_orientation}, i.e. $\chi_\mathrm{wc} = \min \limits_{\phi,\theta,\alpha_x,\alpha_y,\alpha_z}\frac{1}{M}\sum_{l=1}^M\chi_{kl}$.

The maximum likelihood (ML) estimate of $\boldsymbol{G}$ given $\boldsymbol{Y}_p$ is
\begin{equation}
\hat{\boldsymbol{G}}= \frac{1}{\sqrt{p_p}}\boldsymbol{Y}_p\boldsymbol{\Phi}^{*} = \boldsymbol{G}+\frac{1}{\sqrt{p_p}} \boldsymbol{W}.
\end{equation}
Here, $\boldsymbol{W}=\boldsymbol{N}_p\boldsymbol{\Phi}^{*}$ is the estimation error that is uncorrelated with $\boldsymbol{G}$. The elements of $\boldsymbol{W}$ are i.i.d zero-mean complex Gaussian with unit variance. We use ML as finding the minimum mean square error estimate is nontrivial under the assumed LoS model.

\subsection{Uplink Data Transmission}
The $M\times 1$ received signal vector at the GS is given by
\begin{equation*}
\boldsymbol{y} =  \boldsymbol{G} (\sqrt{\boldsymbol{p}_u}\odot\boldsymbol{q}) +\boldsymbol{n},
\end{equation*}
where $\odot$ denotes element wise multiplication; $\boldsymbol{q}$ is the vector of symbols simultaneously transmitted by the $K$ UAVs, i.e. $\boldsymbol{q}=[q_1,q_2,...,q_K]^T$ (normalized such that $\mathbb{E}\{|q_k|^2\}=1$ for all $k\in \{1,2,...,K\}$); $\boldsymbol{p}_u=[p_{u1},p_{u2},...,p_{uK}]^T$ is the vector of transmit power of symbols of $K$ UAVs; $\boldsymbol{n}$ is a complex AWGN vector, $\boldsymbol{n} \sim \mathcal{CN}(0,\boldsymbol{\mathrm{I}}_M)$.

In order to maintain the same average SNR ($\rho_u$) for all UAVs, we consider channel inversion power control, i.e. the power allocated by the $k$-th UAV to each data symbol is 
\begin{equation}\label{power_ch_inv_kl}
p_{uk}=\mathrm{min}\left(\frac{\rho_u}{\frac{1}{M}\sum_{l=1}^M\beta_{kl}\chi_{kl}}, p_{u}\right),
\end{equation}
where $\chi_{kl}=|h_{kl}|^2$ and $p_{u}$ is the maximum power available at the UAV for data symbol. {Note that for power control, the UAV needs to know the large scale channel gain (i.e. the denominator term in \eqref{power_ch_inv_kl}). This can be accomplished through downlink pilot transmission. Unlike uplink, the downlink pilot transmission requires only one symbol.}

We consider that the pilot symbols are transmitted with fixed power $p_{p}$ according to \eqref{pilot_power}. The value of data power $p_{u}$ is calculated from the total energy constraint $\mathcal{P}$ of each UAV in a coherence interval given by 
\begin{equation}
p_p\tau_{\mathrm{ul,p}}+p_u\tau_{\mathrm{ul,d}}\leq \mathcal{P}.
\end{equation}
Here $\mathcal{P}$ is a design parameter selected based on source of power supply and flying range of the UAVs. Due to the uplink power constraint in \eqref{power_ch_inv_kl}, the combined effect of free-space path loss, polarization mismatch, and directional antenna gains may result in signal outage, i.e. the $k$-th UAV is in outage if $\frac{\rho_u}{\frac{1}{M}\sum_{l=1}^M\beta_{kl}\chi_{kl}}>p_{u}$. The outage probability is defined as 
\begin{equation}\label{outage_prob}
P_{\mathrm{out}} = \mathbb{P}\left(\frac{\rho_u}{\frac{1}{M}\sum_{l=1}^M\beta_{kl}\chi_{kl}}>p_{u}\right).
\end{equation}

When the GS array elements are identically oriented and the distance between 
the GS and the UAV location (i.e. $d_k$) is much larger than the aperture size 
of the GS array, $h_{kl}$ and $\beta_{kl}$ are approximately the same across 
the antenna elements, i.e. for all $l=1,2,...,M$, we have
\begin{equation}
\begin{array}{c}
\label{plane_wave_approximation1}
h_{kl} \approx h_k,\ \beta_{kl}\approx \beta_k, \ 
\boldsymbol{\mathrm{R}}_{kl}^{\times,+}=\boldsymbol{\mathrm{I}}_3,\ 
{\text{and}}\\ 
d_{kl} \gg \sqrt{(M_x-1)^2\delta_x^2+(M_y-1)^2\delta_y^2}.
\end{array}
\end{equation} 
Therefore, from \eqref{power_ch_inv_kl}, we obtain that 
\begin{equation}\label{power_ch_inv}
p_{uk}=\mathrm{min}\left(\frac{\rho_u}{\beta_{k}\chi_{k}}, p_{u}\right).
\end{equation}
For analytical tractability, we use \eqref{plane_wave_approximation1} and \eqref{power_ch_inv} for the  ergodic rate analysis in Section \ref{sec_achievable_analysis}. In Section \ref{sec_link_budget_analysis}, we separately show the impact of arbitrary orientation of GS array elements on the link reliability.

\section{Achievable Rate Analysis}\label{sec_achievable_analysis}
It is known that the linear detectors (MRC and ZF) perform fairly well when $K\ll M$ \cite{Marzetta16Book}. In this section, we derive the uplink achievable rate for MRC receiver considering the estimated CSI. For ZF receiver, we analyze the achievable rate considering perfect CSI.

\subsection{MRC receiver}
By using the MRC detector, the received signal $\boldsymbol{y}$ is separated into $K$ streams by multiplying it with $\hat{\boldsymbol{G}}^{H}$ as follows
\begin{equation*}
\label{received_signal_vectore_2}
\boldsymbol{r} =\hat{\boldsymbol{G}}^{H} \boldsymbol{y}= \hat{\boldsymbol{G}}^{H} \boldsymbol{G}(\sqrt{\boldsymbol{p}_u}\odot\boldsymbol{q})+\hat{\boldsymbol{G}}^{H}\boldsymbol{n}.
\end{equation*}

Let $r_k$ and $q_k$ be the $k$-th elements of the vectors $\boldsymbol{r}$ and $\boldsymbol{q}$, respectively. Then, 
\begin{equation}\label{received_signal_k}
r_k = \sqrt{p_{uk}}\hat{\boldsymbol{g}}_k^H\boldsymbol{g}_k q_k+\sum\nolimits_{j=1,j\neq k}^{K}\sqrt{p_{uj}}\hat{\boldsymbol{g}}_k^H\boldsymbol{g}_j q_j+\hat{\boldsymbol{g}}_k^H\boldsymbol{n},
\end{equation}
where $\hat{\boldsymbol{g}}_{k}$ is the $k$-th column of $\hat{\boldsymbol{G}}$. In \eqref{received_signal_k}, the quantities $\boldsymbol{g}_k$ and $\boldsymbol{g}_j$ will be continuously changing as the positions of the UAVs change due to their movement. Even if the location of the $k$-th UAV is fixed, it is more likely that any of the other $K-1$ UAVs will interfere that UAV. Hence, the quantity $\hat{\boldsymbol{g}}_k^H\boldsymbol{g}_j$ will also be changing as a function of the positions of the UAVs i.e. $d_k$, $\theta_k$, and $\phi_k$ for all $k\in\{1,2,...,K\}$. For example, in micro UAV networks \cite{asadpour2014}, since the UAVs move at high speed ($10$ m/s to $30$ m/s) in random directions, one can expect multiple independent realizations of $\boldsymbol{g}_k^H\boldsymbol{g}_j$ within a short time duration. For example, consider an ULA with $M_x$ antennas on the $x$-axis and $M_y =1$, i.e. $M=M_x$. If $d_k$ and $d_j$ are very large when compared to the aperture size of the array, after some manipulations, the square of the inner product between the channel vectors of the $k$-th and the $j$-th UAV can be obtained from \eqref{channel_gain_vector} and \eqref{plane_wave_approximation1} as  
\begin{align*}
|\boldsymbol{g}_k^H\boldsymbol{g}_j|^2= 
& \beta_k\beta_j\chi_k\chi_j M^2\nonumber\\
& \times\frac{\mathrm{sinc}^2\left(M\frac{\delta_x}{\lambda} 
(\sin{\theta_k}\cos{\phi_k}-\sin{\theta_j}\cos{\phi_j})\right)}
{\mathrm{sinc}^2\left(\frac{\delta_x}{\lambda} 
(\sin{\theta_k}\cos{\phi_k}-\sin{\theta_j}\cos{\phi_j})\right)}.
\end{align*}
Here $\mathrm{sinc}(x)=\frac{\sin(\pi x)}{\pi x}$. The fluctuations in $|\boldsymbol{g}_k^H\boldsymbol{g}_j|^2$ are determined by the following factors: number of antennas, $M$, antenna spacing, $\delta_x$, velocity, and moving direction of the UAVs.  
Therefore, by assuming multiple independent realizations of the interference power within the codeword transmission time, we compute the ergodic rate by averaging over all possible UAV positions.

Next we derive closed form expression for achievable rate using the method from \cite{Marzetta16Book} i.e. we assume that the receiver at the GS uses only statistical knowledge of the channel when performing the detection. The $k$-th element of $\boldsymbol{r}$ in \eqref{received_signal_k}, i.e. $r_k$, can be rewritten in the form 
\begin{equation}
\begin{array}{r@{}l}
\label{received_signal_stm}
\sqrt{p_{uk}}r_k = &{} 
\mathbb{E}\{p_{uk}\hat{\boldsymbol{g}}_k^H\boldsymbol{g}_k\}q_k
  \!+\!\big(p_{uk}\hat{\boldsymbol{g}}_k^H\boldsymbol{g}_k-
 \mathbb{E}\{p_{uk}\hat{\boldsymbol{g}}_k^H\boldsymbol{g}_k\}\big)q_k\\
 &{} \displaystyle +\sum\nolimits_{j=1,j\neq 
 k}^{K}\sqrt{p_{uj}p_{uk}}\hat{\boldsymbol{g}}_k^H\boldsymbol{g}_j 
 q_j+\sqrt{p_{uk}}\hat{\boldsymbol{g}}_k^H\boldsymbol{n}.
\end{array}
\end{equation}

By defining the effective additive noise as 
\begin{equation}
\begin{array}{r@{}l}
\label{effective_noise}
a_k' = &{} \underbrace{\bigg(p_{uk}\hat{\boldsymbol{g}}_k^H\boldsymbol{g}_k-
 \mathbb{E}\{p_{uk}\hat{\boldsymbol{g}}_k^H\boldsymbol{g}_k\}\bigg)q_k}_{a_1}\\
 &{} + \underbrace{\sum\nolimits_{j=1,j\neq 
 k}^{K}\sqrt{p_{uj}p_{uk}}\hat{\boldsymbol{g}}_k^H\boldsymbol{g}_j q_j}_{a_2}+ 
 \underbrace{\sqrt{p_{uk}}\hat{\boldsymbol{g}}_k^H\boldsymbol{n}}_{a_3},
\end{array}
\end{equation} 
the expression in \eqref{received_signal_stm} can be written as
\begin{equation}\label{received_signal_stm_effective_noise}
\sqrt{p_{uk}}r_k = 
\mathbb{E}\left\{p_{uk}\hat{\boldsymbol{g}}_k^H\boldsymbol{g}_k\right\}q_k+a_k'.
\end{equation}\normalsize

Since $\mathbb{E}\{\hat{\boldsymbol{g}}_k^H\boldsymbol{g}_k\}$ is deterministic and $q_k$ is independent of $\hat{\boldsymbol{g}}_k^H\boldsymbol{g}_k$, the first two terms of \eqref{received_signal_stm} are uncorrelated. Similarly, the last two terms of \eqref{received_signal_stm} are uncorrelated with the first term of \eqref{received_signal_stm}. Hence, the desired signal and the effective additive noise in \eqref{received_signal_stm_effective_noise} are uncorrelated. By using the fact that the worst-case uncorrelated additive noise is independent Gaussian noise of same variance \cite{Marzetta16Book}, the ergodic rate achieved by the $k$-th UAV can be lower bounded as
\begin{equation}\label{ergodic_lower_bound_hgo}
S_k^{\mathrm{MRC}} \geq S_k^{\mathrm{lb,MRC}} 
\triangleq \Lambda\log_2 \left(1+ 
\frac{|\mathbb{E}\{p_{uk}\hat{\boldsymbol{g}}_k^H\boldsymbol{g}_k\}|^2}
{\mathrm{var}(a_k')}\right),
\end{equation}
where $\Lambda$ denotes the fraction of symbols used for uplink data transmission within the coherence length and $p_{uk}\leq p_{u}$. If the number of uplink pilot symbols $\tau_{\mathrm{ul,p}}=K$, then from \eqref{coh_length} we can write 
\begin{equation} \label{pre_log_factor}
\Lambda=1-\frac{\tau_{\mathrm{dl}}+\tau_{\mathrm{ul,p}}}{T_\mathrm{len}} = 1-\frac{2\cdot v_{\mathrm{max}} \cdot f_c \cdot (\tau_{\mathrm{dl}}+K)}{B_c\cdot c}.
\end{equation}\normalsize
Since all three terms in \eqref{effective_noise} are independent of each other, 
the variance of  effective noise in \eqref{ergodic_lower_bound_hgo} is 
$\mathrm{var}(a_k')=\mathrm{var}(a_1)~+~\mathrm{var}(a_2)~+~\mathrm{var}(a_3)$. 
After substituting the expectation and variance terms in 
\eqref{ergodic_lower_bound_hgo}, the lower bound $S_k^{\mathrm{lb,MRC}}$ of the 
ergodic rate achieved by the $k$-th UAV is obtained as given in 
\eqref{ergodic_lower_bound_gk} (shown on top of next page).
\begin{figure*}[htb]
\begin{align}\label{ergodic_lower_bound_gk}
&S_k^{\mathrm{lb,MRC}} =\Lambda \log_2\left(1+\frac{M\rho_u}{\frac{1}{M\rho_u}\sum\limits_{j=1,j\neq k}^{K}\mathbb{E}\big\{p_{uj}p_{uk}|\boldsymbol{g}_k^H\boldsymbol{g}_j|^2\big\}+\big(1+K\rho_u\big)\mathbb{E}\big\{\frac{1}{\beta_k\chi_k}\big\}\left(\frac{\lambda}{4\pi d_\mathrm{wc}}\right)^2\frac{\chi_\mathrm{wc}}{\rho_u \rho_p}+1}\right).
\end{align}
\hrule
\end{figure*}
For the proof, see Appendix \ref{ergodic_lower_bound_gk_appendix}.

In \eqref{ergodic_lower_bound_gk}, it can be observed
  that the numerator term inside the logarithm increases
  proportionally with $M$. This is an effect of the
  array gain. For example, with $100$ antennas and the same radiated
  power as in a single-antenna system, the array gain is $20$ dB which
  implies a range extension of $10$ times in LoS. The first term in
  the denominator represents the cumulative interference caused by the
  other $K-1$ UAVs. The second and third terms stem
  from channel estimation errors and noise, respectively.

Equation \eqref{ergodic_lower_bound_gk} can be used to analyze the
achievable rate for any arbitrary distribution and placement of the
drones (i.e. distributions of $d_k$, $\theta_k$, and $\phi_k$ for 
$k\in\{1,2,...,K\}$). For some distributions one may
have to compute the ergodic rate numerically as it is difficult to
obtain a closed form expression. Next we derive a lower bound on the
ergodic rate for the case with uniformly distributed UAV locations
inside a spherical shell and the results follow in closed form in
this case.

\begin{theorem}\label{theorem_1}
By employing MRC receiver at the GS and using the ML estimate of the channel matrix, for the independently and spherically uniformly distributed UAV locations inside the spherical shell with inner radius 
\begin{equation}\label{inner_radius}
R_\mathrm{min}>\sqrt{(M_x-1)^2\delta_x^2+(M_y-1)^2\delta_y^2}
\end{equation}
and outer radius $R$, the lower bound on the achievable ergodic rate for the $k$-th UAV is given in \eqref{ergodic_plane_wave_mrc}  (shown on top of next page).
\begin{figure*}
\begin{equation}\label{ergodic_plane_wave_mrc}
S_k^{\mathrm{lb,MRC}}= \Lambda\log_2\left(1+ \frac{M\rho_u}{\rho_u (K-1)(1+\frac{\Omega}{M})+1+\frac{1}{\rho_u \rho_p}\big(1+K\rho_u\big)\frac{3\kappa  \chi_\mathrm{wc}(R^5-R_\mathrm{min}^5)}{5R^2(R^3-R_\mathrm{min}^3)}}\right),
\end{equation}
where \begin{align}\label{omega}
\Omega & =\sum\limits_{l=1}^{M}\sum\limits_{l'=1,l'\neq l}^{M} \Bigg\{\mathrm{sinc}^2\left(\frac{2}{\lambda}\sqrt{(p'-p)^2\delta_x^2+(q'-q)^2\delta_y^2}\right) \times \left(\mathbb{C}^2(b_{ll'})+\mathbb{D}^2(b_{ll'})\right)\Bigg\},
\end{align}
\begin{align}\label{b_definition}
b_{ll'}=\frac{\pi}{\lambda}\left(((p-1)^2-(p'-1)^2)\delta_x^2+((q-1)^2-(q'-1)^2)\delta_y^2\right),
\end{align}\normalsize
\begin{equation*}
l=(q-1)M_x+p, \ \ \ \ l'=(q'-1)M_x+p',  \ \ \ \
p,p'\in\{1,2,...,M_x\}, \ \ \ q,q'\in\{1,2,...,M_y\},
\end{equation*}
\begin{align}\label{c_b}
\mathbb{C}(b_{ll'})=\frac{1}{2(R^3-{R_{\mathrm{min}}^3)}}\times\Bigg(
&\big(2R^2-b_{ll'}^2\big)R\cos\big(b_{ll'}/R\big)-b_{ll'}R^2
\sin\big(b_{ll'}/R\big)-b_{ll'}^3\mathrm{Si}\big(b_{ll'}/R\big)\\
&\hspace{.5cm}-\big(2R_\mathrm{min}^2-b_{ll'}^2\big)R_\mathrm{min}
\cos\big(b/R_\mathrm{min}\big)+b_{ll'}R_\mathrm{min}^2
\sin\big(b_{ll'}/R_\mathrm{min}\big)+b_{ll'}^3
\mathrm{Si}\big(b_{ll'}/R_\mathrm{min}\big)\Bigg),\nonumber
\end{align}
\begin{align}\label{d_b}
\mathbb{D}(b_{ll'})=\frac{1}{2(R^3-{R_{\mathrm{min}}^3)}} 
\times\Bigg(&\big(2R^2-b_{ll'}^2\big)R\sin\big(b_{ll'}/R\big)+
b_{ll'}R^2\cos\big(b_{ll'}/R\big)+b_{ll'}^3\mathrm{Ci}\big(b/R\big)\\&\hspace{.5cm}
 -\big(2R_\mathrm{min}^2-b_{ll'}^2\big)R_\mathrm{min}\sin\big(b_{ll'}/
R_\mathrm{min}\big)-b_{ll'}R_\mathrm{min}^2\cos\big(b_{ll'}/
R_\mathrm{min}\big)-b_{ll'}^3\mathrm{Ci}\big(b_{ll'}/R_\mathrm{min}\big)\Bigg),\nonumber
\end{align}
\normalsize
and $\kappa=\mathbb{E}\left\{\frac{1}{\chi_k}\right\}$. Here, $\mathrm{Si}(x)=\int_0^x\frac{\sin t}{t} dt$ and $\mathrm{Ci}(x)=-\int_x^\infty\frac{\cos t}{t} dt$.
\hrule
\end{figure*}
\end{theorem}

\begin{IEEEproof}
Consider that the UAV positions are independently and spherically uniformly distributed within a spherical shell with inner radius radius $R_{\mathrm{min}}$ and outer radius $R$. The distribution of the distance $d_j$ (for all $j\in \{1,2,...,K\}$) is given by 
\begin{equation}\label{pdf_dist}
f_{d_j}(r) = \frac{3r^2}{R^3-R_{\mathrm{min}}^3}, \ \ \ \ \ \ \ R_{\mathrm{min}} \leq r \leq R.
\end{equation}
The distributions of the elevation and azimuth angles are given by 
\begin{equation}\label{pdf_el_az}
f_{\theta_j}(\theta)=\frac{\sin\theta}{2},\ 0\leq \theta \leq \pi\ \ \  \text{and} \ \ \ 
f_{\phi_j}(\phi)=\frac{1}{2\pi},\ 0\leq \phi \leq 2\pi, 
\end{equation}
respectively. 

By using \eqref{pdf_dist}, since $\chi_k$ is independent of the distance (in spherical coordinates $\chi_k$ is only a function of $\theta_k$ and $\phi_k$), the expected value of $\frac{1}{\beta_k \chi_k}$ can be obtained as
\begin{align}\label{E_inv_beta_k_text}
  \mathbb{E}\left\{\!\frac{1}{\beta_k\chi_k}\!\right\}
  & = \mathbb{E}\left\{\left(\frac{4\pi d_k}{\lambda}\right)^2\right\} 
  \mathbb{E}\left\{\!\frac{1}{\chi_k}\!\right\}\ \nonumber\\
  & =  \mathbb{E}\left\{\frac{1}{\chi_k}\right\}
  \left(\frac{4\pi}{\lambda}\right)^2\int_{R_{\mathrm{min}}}^R
  r^2 f_{d_k}(r) \ dr \\
  & = \mathbb{E}\left\{\frac{1}{\chi_k}\right\} 
  \left(\frac{4\pi}{\lambda}\right)^2\int_{R_{\mathrm{min}}}^R
  r^2\frac{3r^2}{R^3-{R_{\mathrm{min}}^3}}\ dr\nonumber\\
  & = \left(\frac{4\pi}{\lambda}\right)^2\kappa\
 \frac{3(R^5-R_{\mathrm{min}}^5)}{5(R^3-{R_\mathrm{min}^3})},\nonumber
\end{align}
where $\kappa= \mathbb{E}\left\{\frac{1}{\chi_k}\right\}$. This expectation has to be calculated numerically as $h_{kl}$ is a complicated function of rotation angles, field patterns, and polarization mismatch factors. We will discuss this in detail in Section \ref{sec_link_budget_analysis}. 

The inner product between the channel vectors of $k$-th and $j$-th UAVs can be written as
\begin{align}\label{inner_product_g_k_g_j_first}
\boldsymbol{g}_k^H\boldsymbol{g}_j&=\sum\nolimits_{l=1}^{M}\sqrt{\beta_{kl}\beta_{jl}}h_{kl}h_{jl}\ e^{i\frac{2\pi}{\lambda}(d_{kl}-d_{jl})}.
\end{align}

By applying \eqref{plane_wave_approximation1}, the expression in \eqref{inner_product_g_k_g_j_first} can be rewritten as
\begin{align}
\boldsymbol{g}_k^H\boldsymbol{g}_j&=\sqrt{\beta_{k}\beta_{j}}h_{k}h_{j}\sum\nolimits_{l=1}^{M}e^{i\frac{2\pi}{\lambda}(d_{kl}-d_{jl})}.\nonumber
\end{align} 

Since $|\boldsymbol{g}_k^H\boldsymbol{g}_j|^2=(\boldsymbol{g}_k^H\boldsymbol{g}_j)(\boldsymbol{g}_k^H\boldsymbol{g}_j)^H$, we can write 
\begin{align}
|\boldsymbol{g}_k^H\boldsymbol{g}_j|^2&=\beta_{k}\beta_{j}|h_{k}|^2|h_{j}|^2\nonumber\\&\hspace{.5cm}\times\left(\sum\limits_{l=1}^{M}e^{i\frac{2\pi}{\lambda}(d_{kl}-d_{jl})}\right)\left(\sum\limits_{l'=1}^{M}e^{-i\frac{2\pi}{\lambda}(d_{kl'}-d_{jl'})}\right)\nonumber\\&=\beta_{k}\beta_{j}\chi_{k}\chi_{j}  \sum\limits_{l=1}^{M}\sum\limits_{l'=1}^{M}e^{i\frac{2\pi}{\lambda}(d_{kl}-d_{kl'})}\ e^{-i\frac{2\pi}{\lambda}(d_{jl}-d_{jl'})}.\nonumber
\end{align} 
The expectation of $p_{uj}p_{uk}|\boldsymbol{g}_k^H\boldsymbol{g}_j|^2$ can be written as
\begin{equation}
\begin{array}{r@{}l}
\label{exp_inner_product_g_k_g_j_square_first}
\mathbb{E}&{}\{p_{uj}p_{uk}|\boldsymbol{g}_k^H\boldsymbol{g}_j|^2\}\\
&{} = \mathbb{E}\Bigg\{(p_{uk}\beta_{k}\chi_{k})(p_{uj}\beta_{j}\chi_{j})\\
&{}\displaystyle \hspace{1cm}\times\sum\limits_{l=1}^{M}
\sum\limits_{l'=1}^{M}e^{i\frac{2\pi}{\lambda}(d_{kl}-d_{kl'})}\ 
e^{-i\frac{2\pi}{\lambda}(d_{jl}-d_{jl'})}\Bigg\}.
\end{array}
\end{equation} 

Since $d_{kl}$ and $d_{jl}$ are independent, by applying \eqref{power_ch_inv}, the expectation in \eqref{exp_inner_product_g_k_g_j_square_first} can be written as
\begin{equation*}
\mathbb{E}\big\{p_{uj}p_{uk}|\boldsymbol{g}_k^H\boldsymbol{g}_j|^2\big\}=\rho_u^2\sum\limits_{l=1}^{M}\sum\limits_{l'=1}^{M} N_{kll'} \ N_{jll'}^*,
\end{equation*} 
where 
\[
N_{kll'}=\mathbb{E}\big\{e^{i\frac{2\pi}{\lambda}(d_{kl}-d_{kl'})}\big\} 
\mbox{ and } 
N_{jll'}=\mathbb{E}\big\{e^{-i\frac{2\pi}{\lambda}(d_{jl}-d_{jl'})}\big\}.
\] 
Obviously $N_{kll'}=N_{jll'}$. Therefore,
\begin{align}\label{inner_product_g_k_g_j_square_sec}
\mathbb{E}\{p_{uj}p_{uk}|\boldsymbol{g}_k^H\boldsymbol{g}_j|^2\}
& = \rho_u^2\sum\nolimits_{l=1}^{M}\sum\nolimits_{l'=1}^{M} |N_{kll'}|^2 \\
& = \rho_u^2\bigg(M+\sum\nolimits_{l=1}^{M}\sum\nolimits_{l'=1,l'\neq l}^{M} 
|N_{kll'}|^2\bigg).\nonumber
\end{align}
The distance difference between the $l$-th and $l'$-th elements to the $k$-th UAV is given by 
\begin{align*}
& d_{kl}-d_{kl'}\\
& = \frac{1}{2d_k}\Big[((p-1)^2-(p'-1)^2)\delta_x^2 
+((q-1)^2\!-\!(q'-1)^2)\delta_y^2\Big]\nonumber\\
& \hspace{1cm} -\sin\theta_k\Big[(p-p')\delta_x\cos{\phi_k}+
(q-q')\delta_y\sin{\phi_k}\Big].
\end{align*} 
Since $d_k$, $\theta_k$, and $\phi_k$ are independent for all $k\in\{1,2,...,K\}$,
\begin{align}\label{Nkll}
N_{kll'}
& =\mathbb{E}\Bigg\{e^{i\frac{2\pi}{\lambda}(d_{kl}-d_{kl'})}\Bigg\}\\
& = \mathbb{E}\Bigg\{e^{i\frac{\pi}{\lambda}\frac{1}{d_k}[((p-1)^2-(p'-1)^2)
	\delta_x^2+((q-1)^2-(q'-1)^2)\delta_y^2]}\Bigg\}\nonumber\\
&\hspace{.5cm}\times\mathbb{E}\Bigg\{e^{-i\frac{2\pi}{\lambda}
	\sin\theta_k[(p-p')\delta_x\cos{\phi_k}+(q-q')\delta_y
	\sin{\phi_k}]}\Bigg\}.\nonumber
\end{align}
By using \eqref{pdf_dist}, the first expectation in \eqref{Nkll} can be obtained as
\begin{equation}
\begin{array}{r@{}l}
\label{mean_higher_order_term}
\mathbb{E}\bigg\{
	&{} e^{i\frac{\pi}{\lambda}\frac{1}{d_k}[((p-1)^2-(p'-1)^2)
	\delta_x^2+((q-1)^2-(q'-1)^2)\delta_y^2]}\bigg\}\\
	&{} = \mathbb{C}(b_{ll'})+i\ \mathbb{D}(b_{ll'}),
\end{array}
\end{equation}
where $b_{ll'}$, $\mathbb{C}(b_{ll'})$, and $\mathbb{D}(b_{ll'})$ are defined in \eqref{b_definition}, \eqref{c_b}, \eqref{d_b}, respectively.

By using \eqref{pdf_el_az}, the second expectation in \eqref{Nkll} can be obtained as
\begin{equation}
\begin{array}{r@{}l}
\label{mean_sinc}
\mathbb{E}\bigg\{&{}
   e^{-i\frac{2\pi}{\lambda}\sin\theta_k[(p-p')\delta_x
	\cos{\phi_k}+(q-q')\delta_y\sin{\phi_k}]}\bigg\}\\
   &{} \displaystyle = 
   \mathrm{sinc}\left(\frac{2}{\lambda}\sqrt{(p'-p)^2\delta_x^2
   	+(q'-q)^2\delta_y^2}\right).
\end{array}
\end{equation}
For the proofs of \eqref{mean_higher_order_term} and \eqref{mean_sinc}, see Appendix \ref{proof_mean_sinc_higher_order_term}.

By substituting \eqref{mean_higher_order_term} and \eqref{mean_sinc} into \eqref{Nkll}, we obtain that
\begin{equation}
\begin{array}{r@{}l}
\label{Nkll_final}
N_{kll'} = &{}\left(\mathbb{C}(b_{ll'})+i\ \mathbb{D}(b_{ll'})\right)\\
&{} \displaystyle \hspace{0.3cm}\times 
\mathrm{sinc}\left(\frac{2}{\lambda}\sqrt{(p'-p)^2\delta_x^2+(q'-q)^2\delta_y^2}
\right). 
\end{array}
\end{equation}

By substituting \eqref{Nkll_final} into \eqref{inner_product_g_k_g_j_square_sec}, we get 
\begin{align}\label{mean_inner_product_sq_final}
\mathbb{E}\bigg\{p_{uj}p_{uk}|\boldsymbol{g}_k^H\boldsymbol{g}_j|^2\bigg\} 
&=  \rho_u^2(M+\Omega),
\end{align}
where 
\begin{align*}
\Omega =&\sum\limits_{l=1}^{M}\sum\limits_{l'=1,l'\neq l}^{M} \Bigg\{\mathrm{sinc}^2\bigg(\frac{2}{\lambda}\sqrt{(p'-p)^2\delta_x^2+(q'-q)^2\delta_y^2}\bigg)\nonumber\\&\hspace{2.5cm}\times \left(\mathbb{C}^2(b_{ll'})+\mathbb{D}^2(b_{ll'})\right)\Bigg\}.
\end{align*}
 Finally, by using the fact that $d_\mathrm{wc} = R$ and after substituting \eqref{E_inv_beta_k_text} and \eqref{mean_inner_product_sq_final} into \eqref{ergodic_lower_bound_gk}, we get \eqref{ergodic_plane_wave_mrc}.
\end{IEEEproof}

From the Theorem \ref{theorem_1}, we derive the following result.  When the 
UAVs are located on the surface of the sphere, i.e. when 
$R_{\mathrm{min}}\rightarrow R$, we obtain that
\[\mathbb{C}(b_{ll'})\rightarrow \frac{1}{6R^2}f'(R)=\cos(b_{ll'}/R)\] and \[  \mathbb{D}(b_{ll'})\rightarrow \frac{1}{6R^2}g'(R)=\sin(b_{ll'}/R),\]
where $f'(x)$ and $g'(x)$ are the derivatives of 
\[
f(x) \!=\! 
(2x^2-b_{ll'}^2) x\cos\Big(\frac{b_{ll'}}{x}\Big)
- b_{ll'}x^2\sin\Big(\frac{b_{ll'}}{x}\Big)
\!- b_{ll'}^3\mathrm{Si}\Big(\frac{b_{ll'}}{x}\Big)
\]  
and 
\[
g(x) \!=\! 
(2x^2-b_{ll'}^2) x\sin\Big(\frac{b_{ll'}}{x}\Big)
+ b_{ll'}x^2\cos\Big(\frac{b_{ll'}}{x}\Big)
\!+ b_{ll'}^3\mathrm{Ci}\Big(\frac{b_{ll'}}{x}\Big)\!,
\]  
respectively. 

Since 
$\mathbb{C}^2(b_{ll'})+\mathbb{D}^2(b_{ll'})\rightarrow\cos^2(b_{ll'}/R)+\sin^2(b_{ll'}/R)=1$,
 the lower bound becomes as given in \eqref{limit_R_Rmin} (shown on top of next 
page).
\begin{figure*}[!htb]
\begin{equation}\label{limit_R_Rmin}
S_k^{\mathrm{lb,MRC}} \rightarrow \Lambda\log_2\left(1+ 
\frac{M\rho_u}{\rho_u(K-1)\left(1+\frac{\Omega_1}{M}\right)+1+\frac{\kappa 
\chi_\mathrm{wc} }{\rho_u \rho_p}\big(1+K\rho_u\big)}\right),\ 
R_{\mathrm{min}}\rightarrow R,
\end{equation}
where $\Omega_1=\sum\limits_{l=1}^{M}\sum\limits_{l'=1,l'\neq l}^{M} \mathrm{sinc}^2\bigg(\frac{2}{\lambda}\sqrt{(p'-p)^2\delta_x^2+(q'-q)^2\delta_y^2}\bigg)$. 
\hrule
\end{figure*}
The quantity $\Omega_1$ in \eqref{limit_R_Rmin} depends on the spacing between 
the elements of the GS array. We discuss the impact of antenna spacing on the 
ergodic rate in Section \ref{sec_erg_opt_ant}. Note that since $\kappa = 
\mathbb{E}\left\{\frac{1}{\chi_k}\right\}$, we have that 
$\kappa\chi_\mathrm{wc}<1$. 

\subsection{Zero forcing (ZF) receiver}
In this section by assuming that perfect CSI is available at the GS we derive and analyze a lower bound on ergodic capacity with the ZF receiver. By using the ZF detector, the received signal $\boldsymbol{y}$ is separated into $K$ streams by multiplying it with ${\boldsymbol{G}}^\dagger = \big({\boldsymbol{G}}^H {\boldsymbol{G}}\big)^{-1}{\boldsymbol{G}}^H$ as follows 
\begin{equation*}
\boldsymbol{r} = {\boldsymbol{G}}^{\dagger} \boldsymbol{y}.
\end{equation*}

The output of the ZF detector can be written as
\begin{equation}
\label{received_signal_vectore_2}
\boldsymbol{r} =\boldsymbol{G}^{\dagger} \left(\boldsymbol{G} (\sqrt{\boldsymbol{p}_u}\odot\boldsymbol{q}) +\boldsymbol{n}\right) = (\sqrt{\boldsymbol{p}_u}\odot\boldsymbol{q}) +\boldsymbol{G}^{\dagger}\boldsymbol{n}.
\end{equation}
Since the first and the second terms in \eqref{received_signal_vectore_2} are independent of each other, for a given channel matrix $\boldsymbol{G}$, the $K \times K$ covariance matrix of $\boldsymbol{r}$ can be written as
\begin{align}
\mathbb{E}\big\{&\boldsymbol{r}\boldsymbol{r}^H\big|\boldsymbol{G}\big\}\nonumber\\&= \mathbb{E}\left\{\big((\sqrt{\boldsymbol{p}_u}\odot\boldsymbol{q}) +\boldsymbol{G}^{\dagger}\boldsymbol{n}\big)\big((\sqrt{\boldsymbol{p}_u}\odot\boldsymbol{q}) +\boldsymbol{G}^{\dagger}\boldsymbol{n}\big)^H\right\}\nonumber\\&= \mathbb{E}\left\{(\sqrt{\boldsymbol{p}_u}\odot\boldsymbol{q})(\sqrt{\boldsymbol{p}_u}\odot\boldsymbol{q})^H\right\} +\mathbb{E}\big\{\boldsymbol{G}^{\dagger}\boldsymbol{n}\boldsymbol{n}^H(\boldsymbol{G}^{\dagger})^H\big\}\nonumber \\&=  \mathrm{diag}(p_{u1},...,p_{uK})+\boldsymbol{G}^{\dagger}\mathbb{E}\{\boldsymbol{n}\boldsymbol{n}^H\} (\boldsymbol{G}^{\dagger})^H\nonumber\\& = \mathrm{diag}(p_{u1},...,p_{uK})+\boldsymbol{G}^{\dagger}\boldsymbol{\mathrm{I}}_M(\boldsymbol{G}^{\dagger})^H\nonumber\\&= \mathrm{diag}(p_{u1},...,p_{uK})+(\boldsymbol{G}^H\boldsymbol{G})^{-1}, \nonumber
\end{align}
where we used the facts that  
$\mathbb{E}\big\{(\sqrt{\boldsymbol{p}_u}\odot\boldsymbol{q})(\sqrt{\boldsymbol{p}_u}\odot\boldsymbol{q})^H\big\}
 = \mathrm{diag}(p_{u1},...,p_{uK})$, 
$\mathbb{E}\{\boldsymbol{n}\boldsymbol{n}^H\}=\boldsymbol{\mathrm{I}}_M$, and 
$\boldsymbol{G}^{\dagger}(\boldsymbol{G}^{\dagger})^H = 
(\boldsymbol{G}^H\boldsymbol{G})^{-1}$.

The post processing SINR for the $k$-th UAV is 
$\gamma_k^{\mathrm{ZF}} = 
\frac{p_{uk}}{\big[\big(\boldsymbol{G}^H\boldsymbol{G}\big)^{-1}\big]_{kk}}$.
 Since we know that 
$\big[\big(\boldsymbol{G}^H\boldsymbol{G}\big)^{-1}
\big]_{kk} = \frac{1}{\mathrm{det}\big(\boldsymbol{G}^H
\boldsymbol{G}\big)}\big[\mathrm{adj}
\big(\boldsymbol{G}^H\boldsymbol{G}\big)\big]_{kk}$,
 we can write 
\begin{equation*}
\gamma_k^{\mathrm{ZF}} =\frac{p_{uk}\ \mathrm{det}\big(\boldsymbol{G}^H\boldsymbol{G}\big)}{ \big[\mathrm{adj}\big(\boldsymbol{G}^H\boldsymbol{G}\big)\big]_{kk}}.
\end{equation*}

By the convexity of $\log_2\left(1+\frac{1}{t}\right)$ and Jensen's inequality, a lower bound on the achievable uplink rate using ZF receiver can be obtained as 
\begin{align}
\label{ergodic_rate_lb_zf}
S_k^{\mathrm{ZF}}\ge S_k^{\mathrm{lb,ZF}} 
& = \Lambda\log_2\left(\!1+\left(\!
\mathbb{E}\left\{\frac{1}{\gamma_k^{\mathrm{ZF}}}\right\}\right)^{-1}\right)\\
& = \Lambda\log_2\left(\!1+\left(\mathbb{E}\left\{\frac{ 
\big[\mathrm{adj}\big(\boldsymbol{G}^H\boldsymbol{G}\big)\big]_{kk}}{p_{uk}\ 
\mathrm{det}\big(\boldsymbol{G}^H\boldsymbol{G}\big)}\right\}\right)^{-1}
\right)\!\!,\nonumber
\end{align}
where $p_{uk}\leq p_u$.

The expectation in \eqref{ergodic_rate_lb_zf} has to be taken over the distances $d_k$, the elevation angles $\theta_k$, and the azimuth angles $\phi_k$ for all $k\in\{1,2,...,K\}$. Since it is difficult to find the expectation for general $K$ UAVs scenario, we analyze the lower bound for $K=2$. In this case, the determinant of $\boldsymbol{G}^H\boldsymbol{G}$ can be obtained as 
\begin{align*}
 \mathrm{det}&\big(\boldsymbol{G}^H\boldsymbol{G}\big)= \norm{\boldsymbol{g}_1}^2 \norm{\boldsymbol{g}_2}^2-(\boldsymbol{g}_1^H\boldsymbol{g}_2)(\boldsymbol{g}_2^H\boldsymbol{g}_1)\nonumber \\
 &= M^2 \beta_1 \beta_2\chi_1\chi_2\nonumber \\
 &\hspace{.5cm} -\beta_1 \beta_2\chi_1\chi_2\bigg(M+\sum\limits_{l=1}^{M}\sum\limits_{\substack{l{'}=1,\\l{'}\neq l}}^{M}e^{i\frac{2\pi}{\lambda}(d_{1l}-d_{1l{'}}-d_{2l}+d_{2l{'}})}\bigg).
\end{align*}
We have also 
$\big[\mathrm{adj}\big(\boldsymbol{G}^H\boldsymbol{G}\big)\big]_{kk}=\norm{\boldsymbol{g}_{3-k}}^2=
 M \beta_{3-k}\chi_{3-k}$ for $k = 1,2$. Therefore, by applying 
\eqref{power_ch_inv}, the expectation in \eqref{ergodic_rate_lb_zf} is obtained 
as
\begin{align}
\label{sinr_zf_2}
& \mathbb{E}\left\{\frac{\big[\mathrm{adj}\big(\boldsymbol{G}^H
	\boldsymbol{G}\big)\big]_{kk}}{p_{uk}\ 
	\mathrm{det}\big(\boldsymbol{G}^H\boldsymbol{G}\big)}\right\}  = 
	\mathbb{E}\left\{\frac{M \beta_1\beta_2\chi_1\chi_2}{\rho_u\mathrm{det}
	\big(\boldsymbol{G}^H\boldsymbol{G}\big)}\right\}\\
& \!=\!\frac{1}{\rho_u}\mathbb{E}\!\left\{\!\!\left(\!\!M\! 
-\!\!\bigg(\!1+\!\frac{1}{M}\sum\limits_{l=1}^{M}
\sum\limits_{\substack{l{'}=1,\\l{'}\neq l}}^{M}
e^{i\frac{2\pi}{\lambda}(d_{1l}-d_{1l{'}}-d_{2l}+d_{2l{'}})}\!\bigg)
\!\!\right)^{-1}\right\}\!\!.\nonumber
\end{align}

Finally, by substituting \eqref{sinr_zf_2} into \eqref{ergodic_rate_lb_zf}, we get the lower bound as given in \eqref{ergodic_rate_lower_bound_zf} (shown on top of next page). 
\begin{figure*}
\begin{align}\label{ergodic_rate_lower_bound_zf}
&S_k^{\mathrm{lb,ZF}} =\Lambda\log_2\left(1+\rho_u \left[\mathbb{E}\left\{\left(M-\left(1+\frac{1}{M}\sum\limits_{l=1}^{M}\sum\limits_{l{'}=1,l{'}\neq l}^{M}e^{i\frac{2\pi}{\lambda}(d_{1l}-d_{1l{'}}-d_{2l}+d_{2l{'}})}\right)\right)^{-1}\right\}\right]^{-1}\right), \ k = 1, 2.
\end{align}
\hrule
\end{figure*}

\begin{figure*}[t]
\centering
\subfigure[$\Omega$ (dB) for $M_x = 50$, $M_y=1$]{\includegraphics[scale=.535]{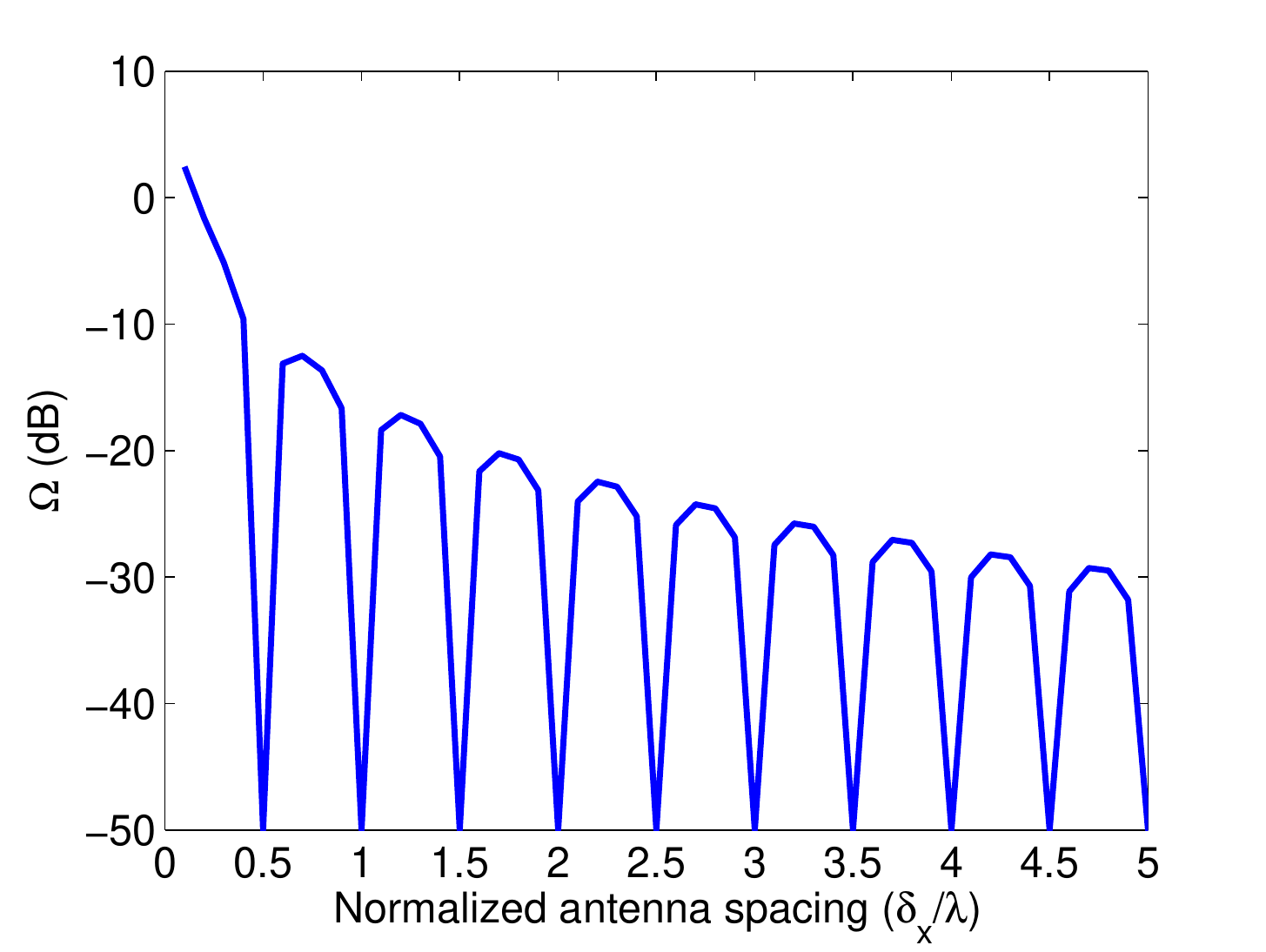} \label{Omega_actual}}\\
\subfigure[$\Omega$ (dB) for $M_x=5$, $M_y=5$]{\includegraphics[scale=.535]{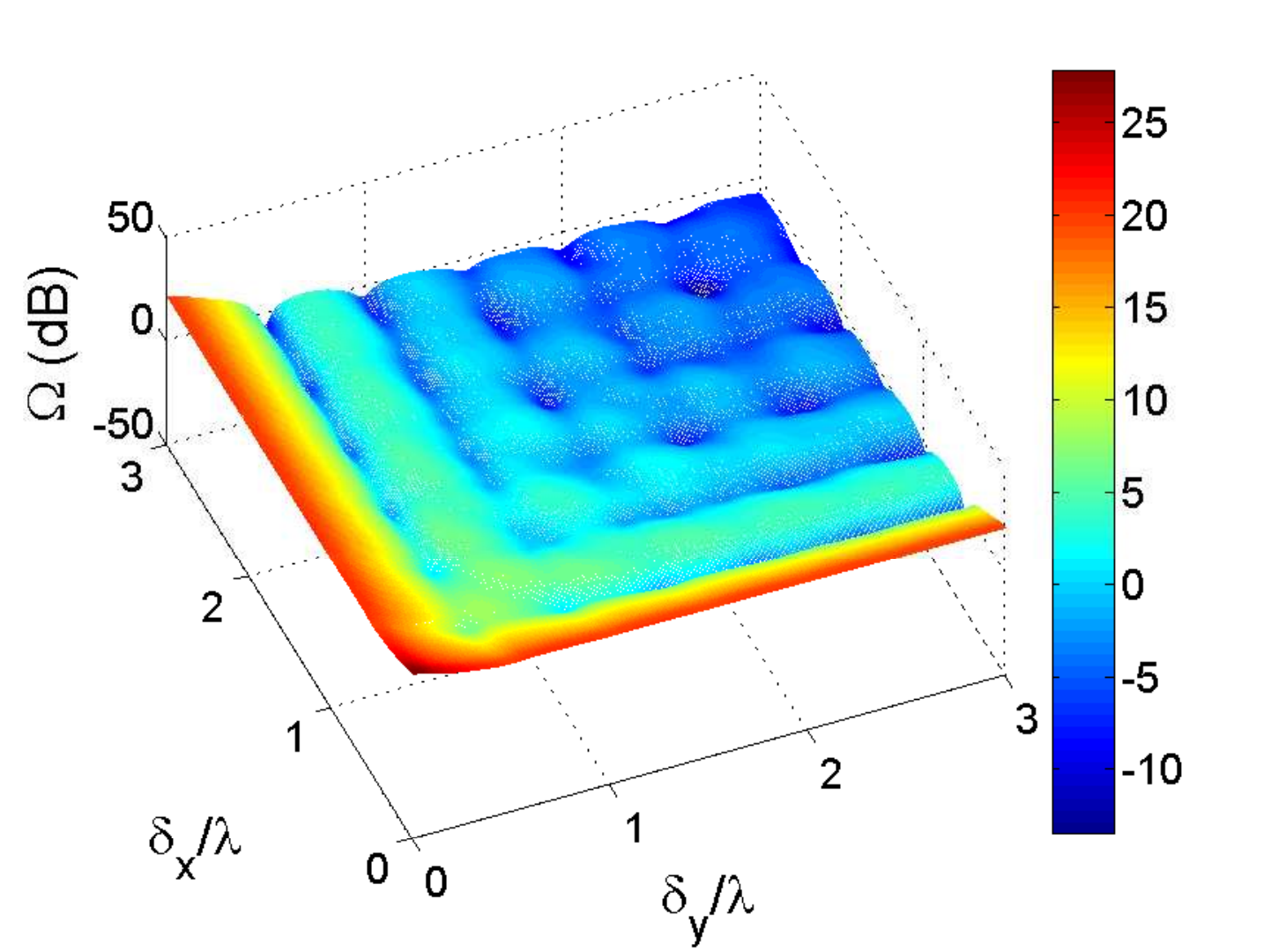} \label{omega_x_y_5_5_ura}}
\subfigure[$\Omega$ (dB) for $M_x=5$, $M_y=5$]{\includegraphics[scale=.535]{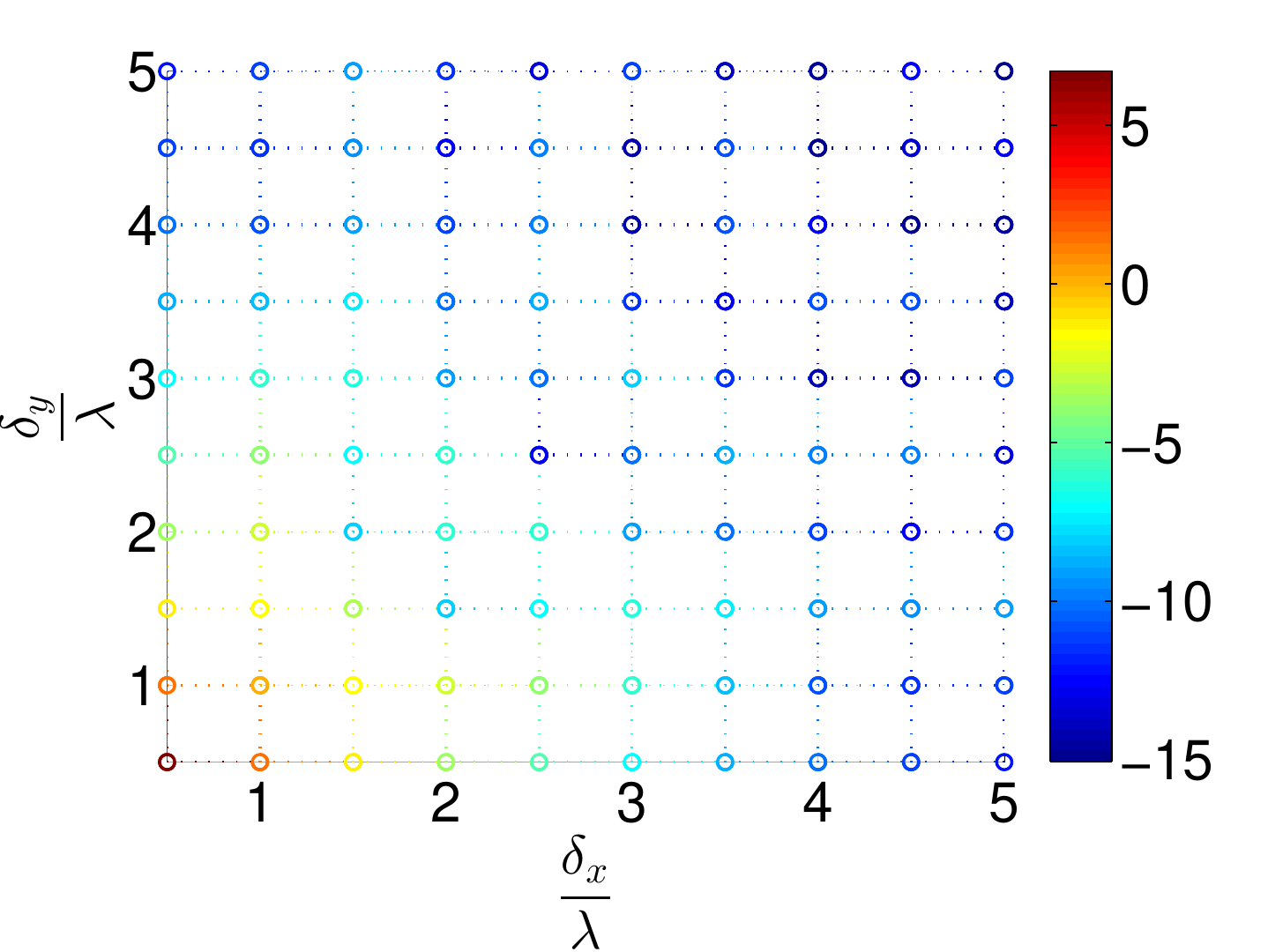} \label{omega_x_y_50_5_ura}}
\caption{$\Omega$ as a function of antenna spacing-wavelength ratio for 
$\lambda = 12.5$ cm and $R=500$ m.}
\end{figure*}

\section{Ergodic rate performance with optimal antenna spacing}\label{sec_erg_opt_ant}
In this section we analyze the ergodic rate performance with optimal antenna spacing for ULA and URA structures. 

\subsection{Optimal antenna spacing}
\paragraph*{Uniform linear array}
For an ULA, i.e. $M_y=1$, since the zero crossings of the $\mathrm{sinc}(u)$ are at non-zero integer multiples of $u$, the quantity $\Omega$ (as given in \eqref{omega}) is zero whenever $\delta_x = n\frac{\lambda}{2},\ n= 1,2,...,\big\lfloor\frac{2R_{\mathrm{min}}}{\lambda (M-1)}\big\rfloor$ (Here the maximum value of $n$ is obtained from \eqref{inner_radius}). Therefore, for an ULA, the optimal antenna spacing for maximizing the ergodic rate is 
\begin{equation}\label{optimal_antenna_spacing_ula}
\delta_x^* = n\frac{\lambda}{2},\ \ n= 1,2,...,\bigg\lfloor\frac{2R_{\mathrm{min}}}{\lambda (M-1)}\bigg\rfloor.
\end{equation}
Interestingly, the result in \eqref{optimal_antenna_spacing_ula} is the same as 
in the case of isotropic scattering \cite{teal2002}. Figure \ref{Omega_actual} 
shows the quantity $\Omega$ as a function of antenna spacing-wavelength ratio 
on $x$-axis (i.e.~$\frac{\delta_x}{\lambda}$) for $M_x = 50$ and $M_y=1$.

\paragraph*{Uniform rectangular array}
For an URA, the $\mathrm{sinc}$ function in \eqref{omega} will not be vanishing 
as the arguments will not be an integer for cross diagonal elements. However, 
we observe that there are local minimas at 
\begin{equation}\label{spacing_ura}
(\delta_x,\delta_y) \approx \left(n\frac{\lambda}{2},m\frac{\lambda}{2}\right),
\end{equation}\normalsize
where $n$ and $m$ are positive integers. This can be seen from Figure 
\ref{omega_x_y_5_5_ura} which shows $\Omega$ as a function of the antenna 
spacing-wavelength ratio on $x$-axis ($\frac{\delta_x}{\lambda}$) and $y$-axis 
($\frac{\delta_y}{\lambda}$). Note that for a given inner radius of the 
spherical shell $R_{\mathrm{min}}$ as in \eqref{inner_radius}, the values of 
$n$ and $m$ are limited by the maximum allowable aperture size of the array, 
i.e. 
\[
(M_x-1)^2n^2+(M_y-1)^2m^2<\frac{4R_{\mathrm{min}}^2}{\lambda^2}.
\] 
Furthermore, numerical observations show that $\Omega$ is close to zero 
whenever $n \ge M_y$ and $m \ge M_x$. For instance, with $n = M_y$ and $m = 
M_x$, we observe that $\Omega\approx 0.053$. This can be seen from Figure 
\ref{omega_x_y_50_5_ura} which shows only the values of $\Omega$ at 
\eqref{spacing_ura}. Therefore, for an URA, an appropriate choice of antenna 
spacing for maximizing the ergodic rate could be   
\begin{equation}
\begin{array}{c}
\label{optimal_antenna_spacing_ura}
\displaystyle
(\delta_x^*,\delta_y^*) = \left(n\frac{\lambda}{2},m\frac{\lambda}{2}\right),\ 
n \ge M_y,\ m \ge M_x,\\ 
\displaystyle
(M_x-1)^2n^2+(M_y-1)^2m^2<\frac{4R_{\mathrm{min}}^2}{\lambda^2}.
\end{array}
\end{equation}

\begin{figure*}[t]
\centering
\subfigure[]{\includegraphics[scale=.6]{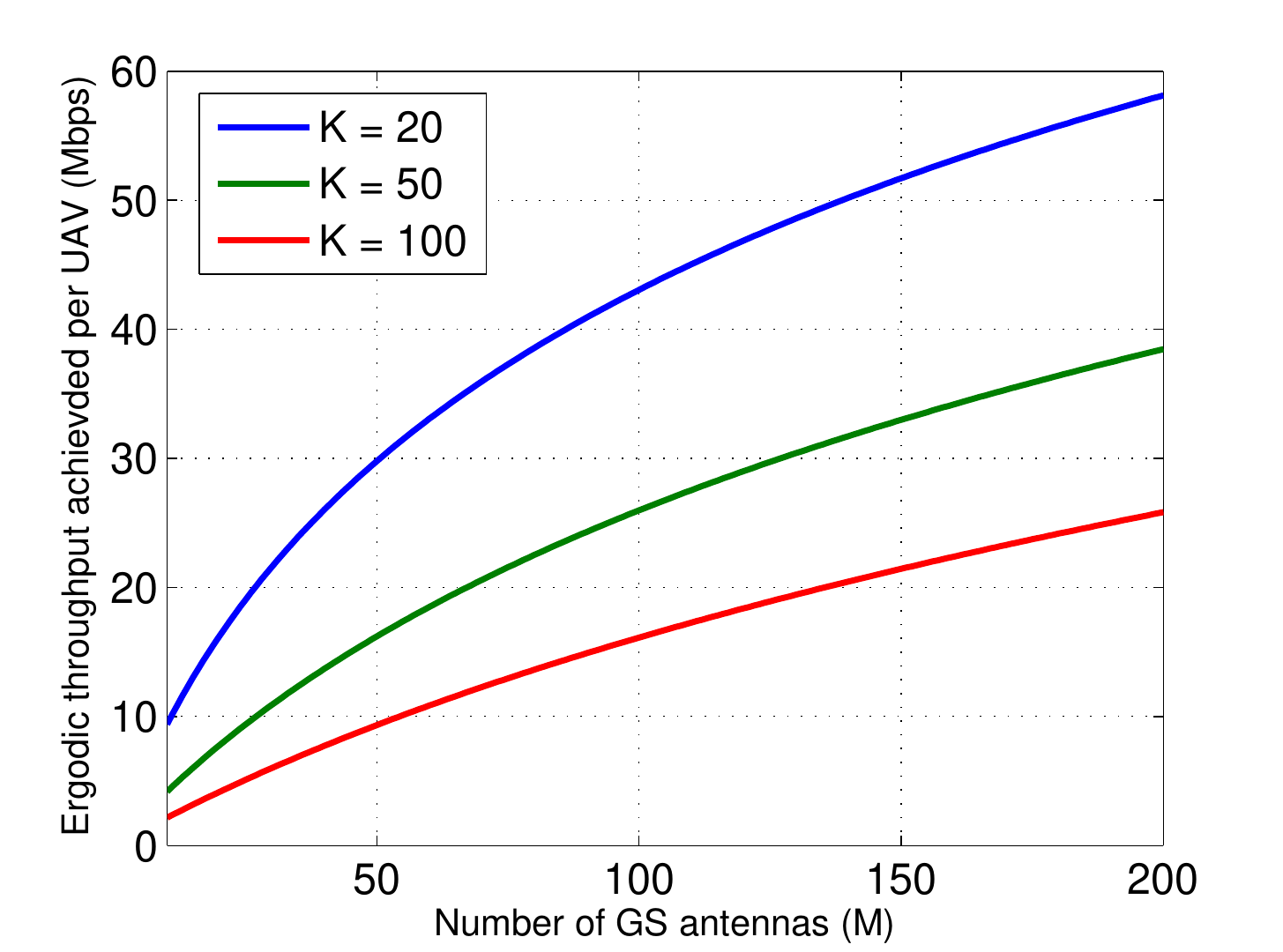}\label{rate_MRC_K}}
\subfigure[]{\includegraphics[scale=.6]{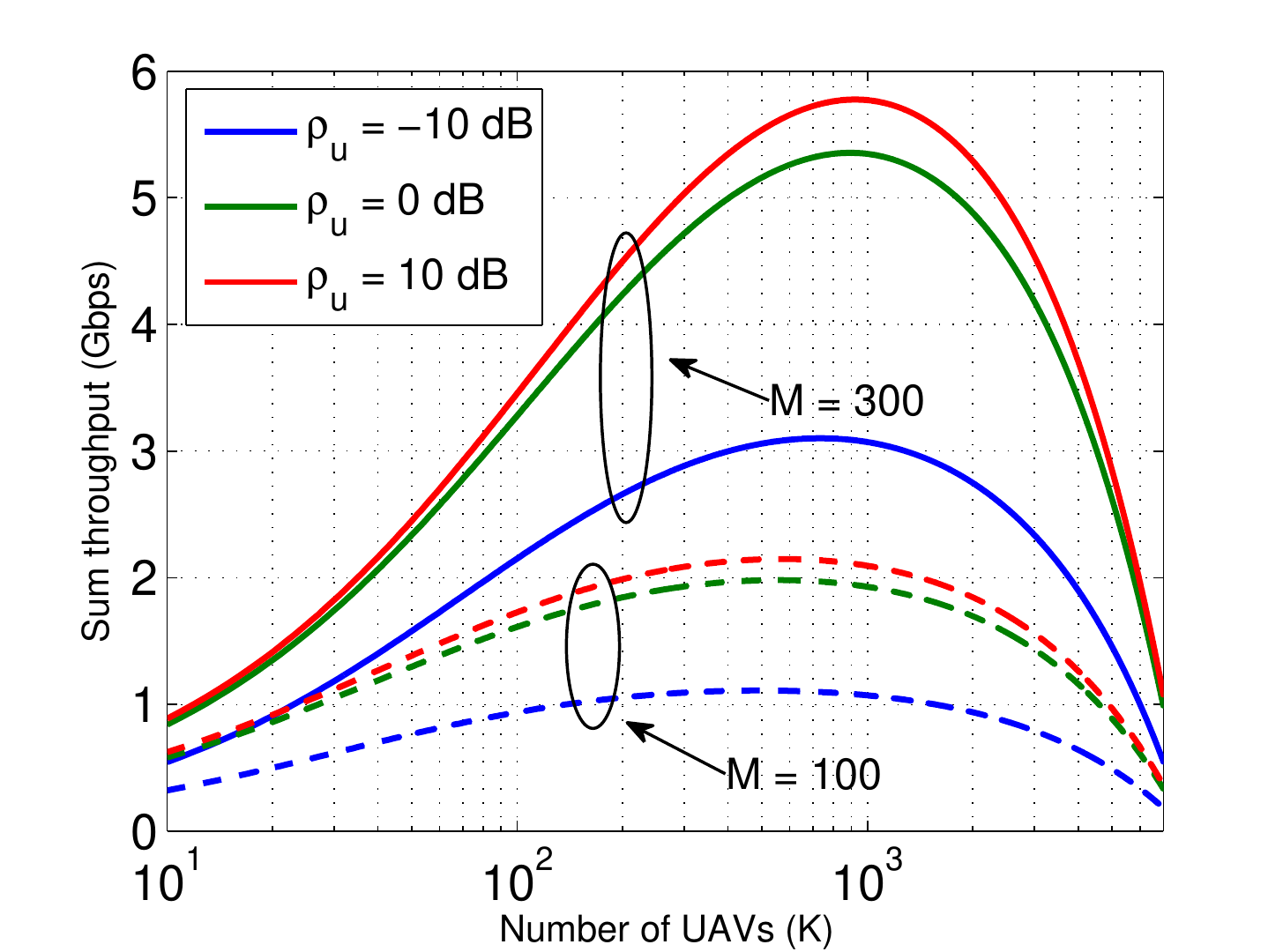} \label{rate_MRC_rhou}}
\caption{(a) Ergodic throughput achieved per UAV vs. Number of GS antennas 
($M$) with the MRC receiver for $\rho_u=0$ dB, $\rho_p=10$ dB, 
$\tau_{\mathrm{dl}} = \frac{T_{\mathrm{len}}}{8}$, $B=20$ MHz, $B_c=3$ MHz, 
$f_c = 2.4$ GHz, and $v_{\mathrm{max}}=20$ m/s. (b) Sum throughput (in Gbps) 
vs. Number of UAVs for $\rho_p=10$ dB and different values of $\rho_u$ and $M$.}
\end{figure*}

\subsection{Achievable rate performance with optimal antenna spacing} \label{rate_opt_ant_spacing}
\paragraph*{MRC receiver} With the optimal antenna spacing as given in 
\eqref{optimal_antenna_spacing_ula} and \eqref{optimal_antenna_spacing_ura}, by 
employing MRC receiver at the GS, when $R_{\mathrm{min}}\rightarrow R$, from 
\eqref{limit_R_Rmin} we obtain that 
\begin{align}
\label{asym_high_snr_mrc}
& \!\! S_k^{\mathrm{lb,MRC}}\\
& \!\! = \!\! \begin{cases}
\!\Lambda\log_2\!\!\bigg(\!1\!+\!\frac{M\rho_u}{\rho_u(K-1)+1+
		\frac{\kappa\chi_\mathrm{wc}}{\rho_u\rho_p}(1+K\rho_u)}\!\bigg),
& \!\!\!\!\!\text{for ULA,} \\
\!\Lambda\log_2\!\!\bigg(\!1\!+\!\frac{M\rho_u}{\rho_u(K-1)
	\left(1+\frac{\Omega_{\mathrm{URA}}}{M}\right)+1+\frac{\kappa 
	\chi_\mathrm{wc}}{\rho_u\rho_p}(1+K\rho_u)}\!\!\bigg),
&\!\!\!\!\!\text{for URA}, \\
\end{cases}\nonumber
\end{align}
where $\Omega_{\mathrm{URA}} \!=\! 
\sum\limits_{l=1}^{M}\sum\limits_{\substack{l'=1,\\ l'\neq l}}^{M}  
\!\!\mathrm{sinc}^2\bigg(\sqrt{(p'-p)^2n^2+(q'-q)^2m^2}\bigg)$, 
$l=(q-1)M_x+p$, $p\in\{1,2,...,M_x\}$, $q\in\{1,2,...,M_y\}$, 
$n \ge M_y$, $m \ge M_x$, and $(M_x-1)^2n^2+(M_y-1)^2m^2 < 
\frac{4R_{\mathrm{min}}^2}{\lambda^2}$.

For an ULA, the sum rate achieved by the $K$ UAVs is then 
\begin{align}
S_{\mathrm{sum},K}^{\mathrm{MRC}} 
& \!=\! \sum\nolimits_{k=1}^K S_k^{\mathrm{lb,MRC}}\\
& \!=\! \Lambda K\log_2\!\left(1+\frac{M\rho_u}{\rho_u(K-1)+1+
	\frac{\kappa\chi_\mathrm{wc}}{\rho_u\rho_p}(1+K\rho_u)}\right)\!\!.\nonumber
\end{align} 

It can be observed from \eqref{asym_high_snr_mrc} that for a given $K$, the rate grows unbounded with $M$. If we define ergodic throughput, $Q=B\cdot S$ bits/sec (where $S$ is the ergodic rate in bits/sec/Hz and $B$ is the system bandwidth in Hz), it can be derived from \eqref{asym_high_snr_mrc} that when $R_{\mathrm{min}}\rightarrow R$, the minimum number of antennas required to support a target data rate of $Q_{\mathrm{tar}}$ (bits/s) is 
\begin{equation}
\label{M_req}
M_{\mathrm{req}}=\left((K-1)+\frac{1}{\rho_u}+\frac{\kappa 
\chi_\mathrm{wc}}{\rho_u^2\rho_p}(1+K\rho_u)\right)
\big(2^{{\frac{Q_\mathrm{tar}}{\Lambda B}}}-1\big).
\end{equation}

Figure \ref{rate_MRC_K} shows the ergodic throughput achieved per UAV ($Q$) 
versus the number of GS antennas ($M$) for varying number of simultaneously 
communicating UAVs when $\kappa\chi_\mathrm{wc}=1$, $\tau_{\mathrm{dl}} = 
\frac{T_{\mathrm{len}}}{8}$, $\rho_u=0$ dB, $\rho_p=10$ dB, $B=20$ MHz, 
$B_c=3$~MHz, $f_c = 2.4$ GHz, and $v_{\mathrm{max}}=20$ m/s. It can be seen 
that, in 
order to support the data rate of $20$ Mbps, for $K=20$, $50$, and $100$ UAVs 
the number of antenna elements required is approximately equal to $27$, $68$, 
and $136$, respectively. With the same parameters as in the previous case, 
Figure \ref{rate_MRC_rhou} shows the sum throughput ($Q_{\mathrm{sum}}=K\cdot B 
\cdot S$) versus the number of UAVs ($K$) for different values of data SNR 
($\rho_u$) and varying number of GS antennas. For a given $\rho_u$, the sum 
throughput increases up to a certain value of $K$ and decreases with further 
increase in $K$. This is because the pre-log term decreases with the number of 
UAVs due to the finite number of symbols in a coherence interval.
\paragraph*{ZF receiver}
With ZF receiver, since it is difficult to calculate the expectation in \eqref{ergodic_rate_lower_bound_zf}, we provide the asymptotic result. For large $M$, we observed that the term $\frac{1}{M}\sum\limits_{l=1}^{M}\sum\limits_{l{'}=1,l{'}\neq l}^{M}e^{i\frac{2\pi}{\lambda}(d_{1l}-d_{1l{'}}-d_{2l}+d_{2l{'}})}$ in \eqref{ergodic_rate_lower_bound_zf} tends to zero, i.e.
\begin{equation}\label{asym_high_snr_zf}
\frac{S_k^{\mathrm{lb,ZF}}}{\Lambda\log_2\left(1+M\rho_u\right)} \rightarrow 1, \ k=1,2.
\end{equation}

Note that when compared to MRC receiver, additional sum-rate performance gains at medium and high SNR regime are possible with the ZF receiver \cite{Marzetta16Book}. This is a topic for future work.

It is interesting to see from \eqref{asym_high_snr_mrc} and \eqref{asym_high_snr_zf} that only by increasing the number of antenna elements at the GS, one can increase the uplink capacity of UAV communication system without increasing the UAV's transmit power. 

\paragraph*{Power scaling law} Consider a case with the perfect CSI (i.e. $\rho_p \rightarrow \infty$). If the UAV's transmit power is scaled down according to $\rho_u=\frac{\varepsilon_u}{M}$, where $\varepsilon_u$ is fixed, then for large $M$, the lower bounds in \eqref{asym_high_snr_mrc} and \eqref{asym_high_snr_zf} (i.e. $S_k^{\mathrm{lb,MRC}}$ and $S_k^{\mathrm{lb,ZF}}$) tend to $\Lambda\log_2(1+\varepsilon_u)$. This implies that, with finite $K$, when $M$ grows large, each UAV obtains the same rate performance as in the single UAV case.

\begin{figure*}[t]
\centering
\subfigure[]{\includegraphics[scale=.6]{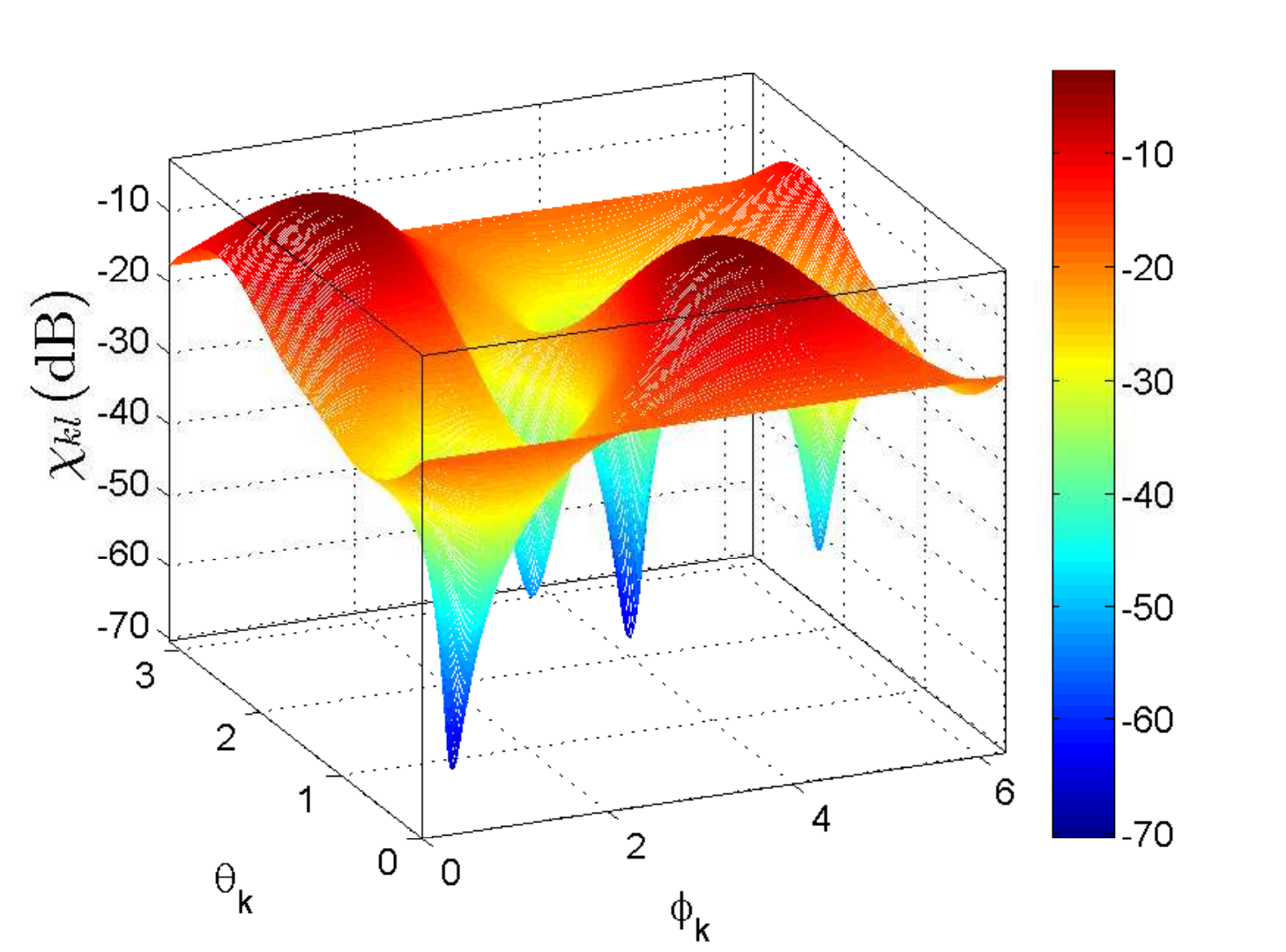}
	\label{Effective_Gain_3D_Journal_}}
\subfigure[]{\includegraphics[scale=.6]{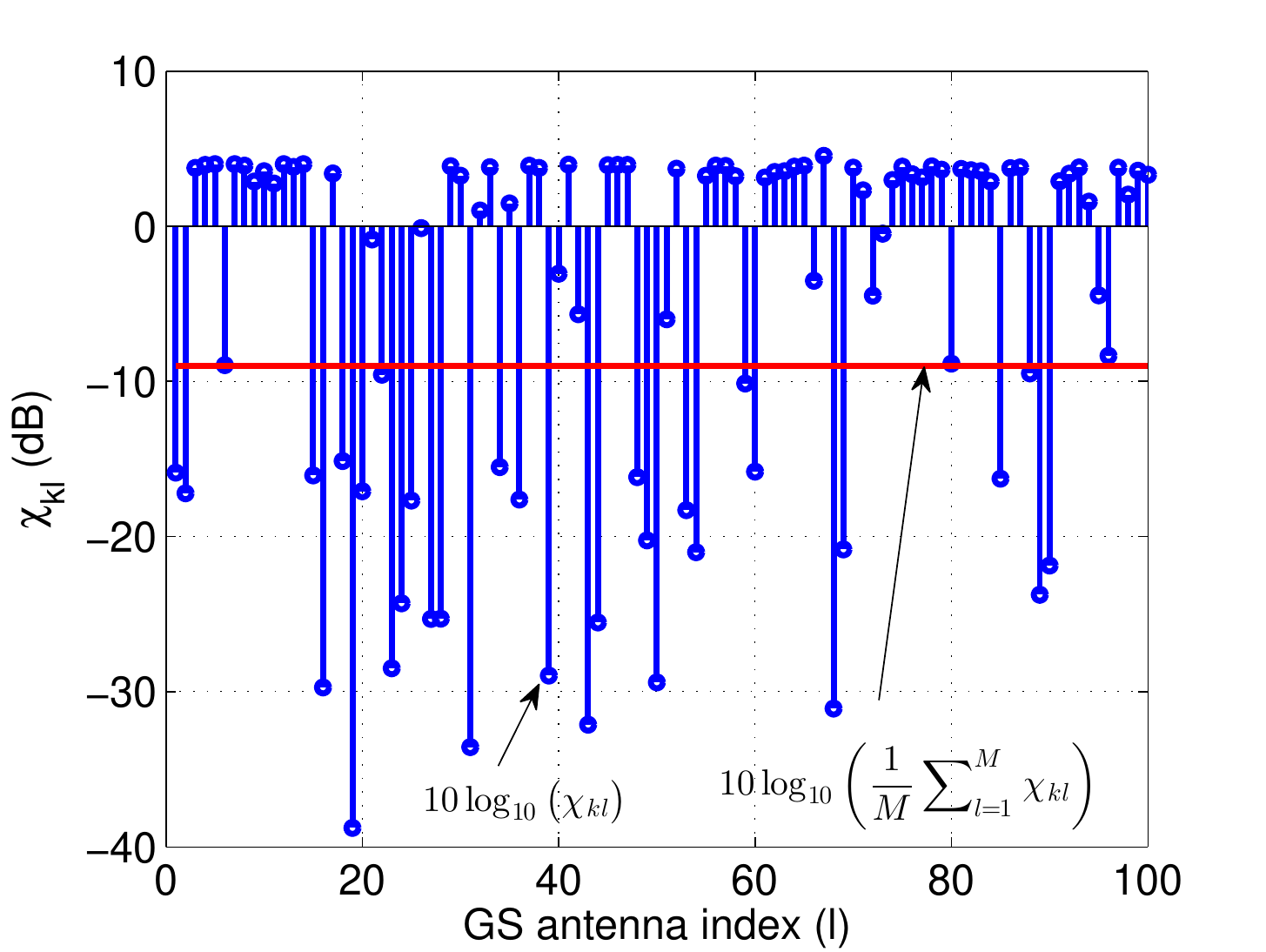}
	\label{PLF_vs_Antenna_Index}}\\
\subfigure[]{\includegraphics[scale=.6]{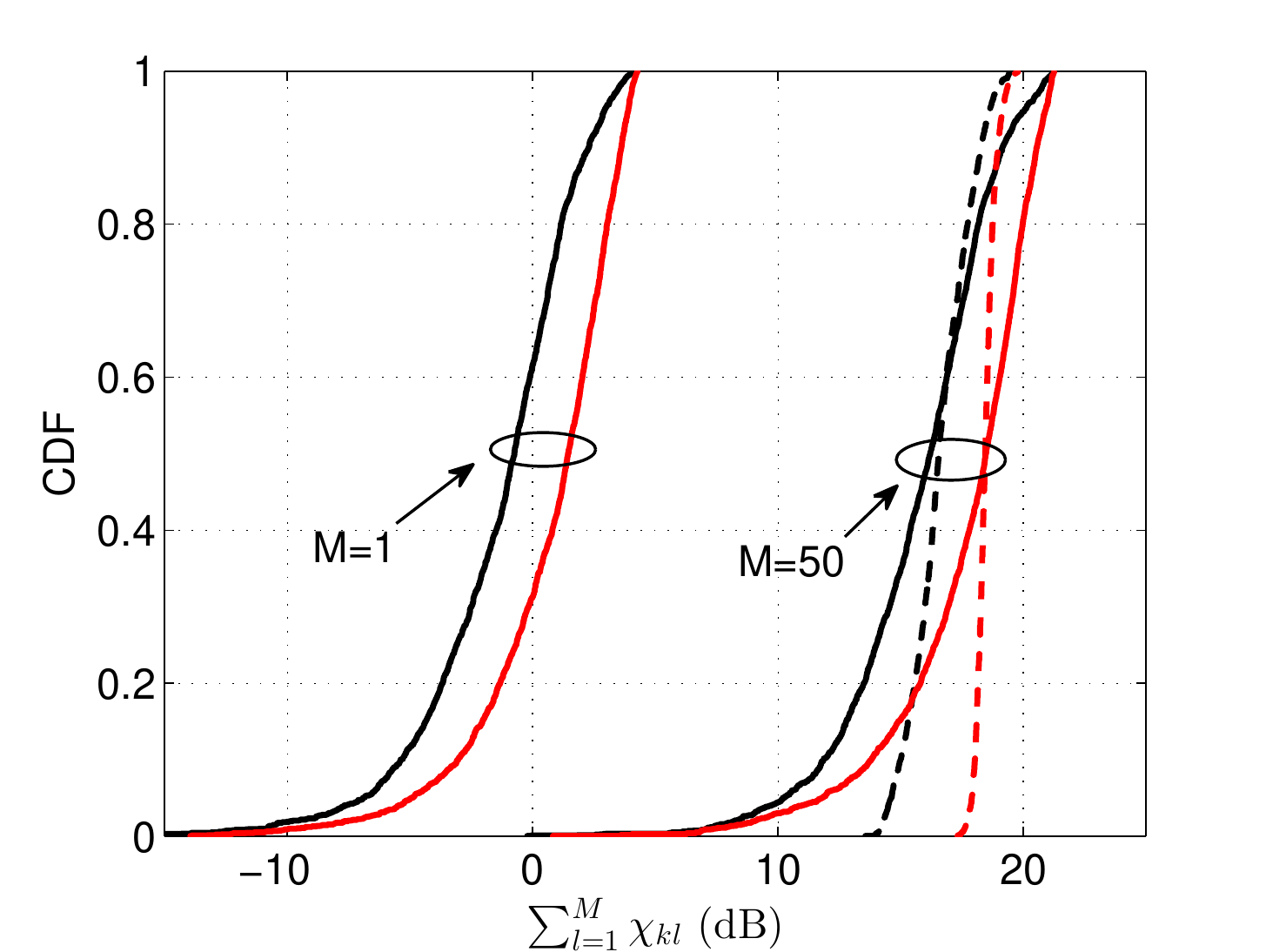}
	\label{CDF_Sum_Gain_Cir}} 
\subfigure[]{\includegraphics[scale=.6]{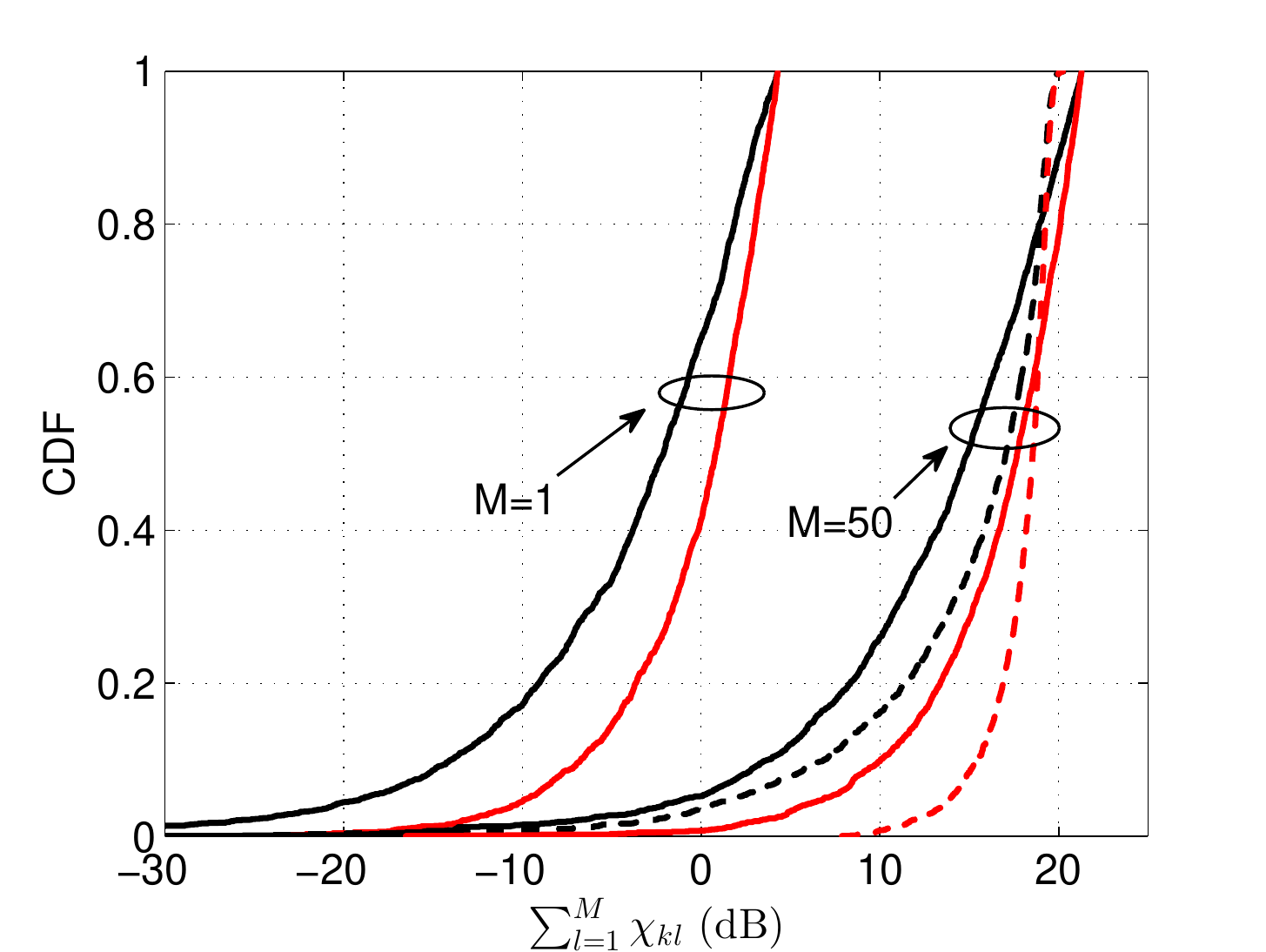} 
	\label{CDF_Sum_Gain_Lin}}%
\caption{(a) Effective gain (including antenna gain and polarization mismatch 
loss) with identically oriented GS array elements for $\lambda = 12\ \text{cm}, 
d_k = 100\ \text{m}, \delta_x = \frac{\lambda}{2} \ \text{m}, d_{\mathrm{len}}= 
\frac{\lambda}{2}\ \text{m}$ with 
$E^{l}_{\theta}=\frac{1}{\sqrt{2}},E^{'k}_{\theta}=\frac{i}{\sqrt{2}}$ and 
$E^{l}_{\psi}=E^{'k}_{\psi}=\frac{1}{\sqrt{2}}$. (b) Effective gain with 
arbitrarily oriented GS array elements for $M_x=100, M_y=1, \theta_k = 
\frac{\pi}{3} , \phi_k=\pi$. (c) CDF of ($\sum_{l=1}^M\chi_{kl}$) for $M_x=50, 
M_y =1$ and $E_\theta^{l}=\frac{1}{\sqrt{2}}, E_\psi^{l}=\frac{i}{\sqrt{2}}, 
E_\theta^{'k}=\frac{1}{\sqrt{2}},E_\theta^{'k}=\frac{-i}{\sqrt{2}}$ (circularly 
polarized antenna elements; solid lines: identically oriented GS elements, 
dashed lines: pseudo-randomly oriented GS elements, red lines: isotropic 
antenna pattern, black lines: dipole antenna pattern). (d) CDF of 
($\sum_{l=1}^M\chi_{kl}$) for $M_x=50, M_y =1$ and 
$E_\theta^{l}=1,E_\psi^{l}=0,E_\theta^{'k}=1,E_\theta^{'k}=0$ (linearly 
polarized antenna elements).}
\end{figure*}

\section{Impact of polarization mismatch and antenna pattern on the link reliability}\label{sec_link_budget_analysis}

Since each UAV performs instantaneous power control to maintain an equal data 
SNR $\rho_u$ according to \eqref{power_ch_inv_kl}, at the GS, the received 
power on the coherence channel bandwidth $B_c$ remains the same for all drones 
(i.e. $\rho_u \ N_0 \ B_c $ (W), where $N_0$ is the noise spectral density 
(i.e. $N_0 = k_B \ T\ 10^{\mathcal{F}/10} \approx 2\times 10^{-20}$ J, where 
$k_B = 1.38\times 10^{-23}$ J/K, $T = 290$ K, and the receiver noise figure 
$\mathcal{F} = 7$ dB). Therefore, if the distance between the GS and the UAV 
location is much larger than the GS array's aperture size (i.e. $d_{kl}\approx 
d_k, \forall l$), for given positions of GS and $k$-th UAV, the required 
transmit power per data symbol is calculated from \eqref{power_ch_inv_kl} as 
\begin{equation*}
p_{uk} = \rho_u \ N_0 \ B_c \left(\frac{4\pi d_k}{\lambda}\right)^2 \ 
\frac{1}{\frac{1}{M}\sum\nolimits_{l=1}^M\chi_{kl}}\ \ \text{(W)}.
\end{equation*} 
Note that the quantity $\chi_{kl}$ consists of polarization mismatch loss 
factors, and transmit and receive antenna patterns. The required pilot power is 
obtained using \eqref{pilot_power} as 
\begin{equation*}
p_p = \rho_p \ N_0 \ B_c \left(\frac{4\pi d_\mathrm{wc}}{\lambda}\right)^2 \ 
\frac{1}{\chi_\mathrm{wc}} \ \ \text{(W)}.
\end{equation*} 
Then the instantaneous uplink power (in W) required by the $k$-th UAV over bandwidth of $B$ Hz is
\begin{align}
\label{inst_power_link_budget}
p_{uk,\mathrm{Tot}} 
& = \frac{B}{B_c} \left(\Lambda\ p_{uk} + 
\left(\frac{K}{T_{\mathrm{len}}}\right) p_p\right)\nonumber\\
& = B \ N_0\ \left(\frac{4\pi}{\lambda}\right)^2\\
&\hspace{.2cm}\times\left(\Lambda \ \rho_u  \ d_k^2 \ 
\frac{1}{\frac{1}{M}\sum\limits_{l=1}^M\chi_{kl}} + 
\left(\frac{K}{T_{\mathrm{len}}}\right) \ \rho_p \ d_\mathrm{wc}^2 \ 
\frac{1}{\chi_\mathrm{wc}} \right).\nonumber
\end{align}

In this section, to show the effect of the polarization mismatch loss and antenna patterns we analyze the effective gain $\sum\nolimits_{l=1}^M\chi_{kl}$ for the randomly and uniformly
distributed UAV positions within a spherical shell with inner radius 
$R_\mathrm{min}=20$~m and outer radius $R=500$ m. The uniformly distributed 
UAV locations inside a spherical volume are obtained using the procedure as 
detailed in \cite[p.130]{Knuth:1997}. We consider an ULA with antenna spacing 
$\delta_x=\frac{\lambda}{2}=6.25$ cm ($f_c=2.4$ GHz). The roll, pitch, and yaw 
angles both at the GS and at the UAV are uniformly distributed in the interval 
$[-\frac{\pi}{2},\frac{\pi}{2}]$, $[-\frac{\pi}{2},\frac{\pi}{2}]$, and 
$[0,\frac{\pi}{2}]$, respectively.

When all GS array elements are identically oriented (i.e. 
$\boldsymbol{\mathrm{R}}_{kl}^{\times,+}=\boldsymbol{\mathrm{I}}_3$ and 
$\chi_{kl}\approx \chi_k, \forall l$), with an arbitrarily chosen roll, 
pitch, and yaw angles, Figure \ref{Effective_Gain_3D_Journal_} shows the 
effective gain ($\chi_{kl}$) for varying elevation and azimuth angles with 
$E^{l}_{\theta}=~\frac{1}{\sqrt{2}},E^{l}_{\psi}=\frac{i}{\sqrt{2}}$, 
$E^{'k}_{\theta}=E^{'k}_{\psi}=\frac{1}{\sqrt{2}}$. It can be observed that the 
gain is very low at certain orientation angles (below $-50$ dB in some cases). 
This implies that if all the GS antenna array elements are identically 
oriented, most likely the signal will be lost for certain positions and 
orientation of the UAV.

For example, consider the following parameters: $K=20$, $\rho_u=10$ dB, 
$\rho_p=10$ dB, $\tau_{\mathrm{dl}}=\frac{T_{\mathrm{len}}}{8}$, $B=20$ MHz, 
$B_c~=~3$~MHz, $f_c=2.4$ GHz, and $v_\mathrm{max}=20$ m/s. If $d_{k}=400$ m, 
since $d_\mathrm{wc} = R$, the total transmit power required by the UAV is 
$p_{uk,\mathrm{Tot}} =  \frac{9.2\times 
10^{-3}}{\frac{1}{M}\sum\nolimits_{l=1}^M\chi_{kl}}+\frac{2.3\times 
10^{-5}}{\chi_\mathrm{wc}}$ W. Now the required uplink power depends on the 
relative orientation of the GS and UAV antenna elements. When all GS elements 
are identically oriented, as a result of low gain as shown in Figure 
\ref{Effective_Gain_3D_Journal_}, the required uplink power will be very high. 
For example, if we consider $\frac{1}{M}\sum\nolimits_{l=1}^M\chi_{kl}=-40$~dB 
and $\chi_\mathrm{wc}=-50$~dB, the required transmit power is 
$p_{uk,\mathrm{Tot}} \approx 94$~W. Since the UAV's power supply is limited in 
practice, the outage probability (as defined in \eqref{outage_prob}) will be 
high. This situation can be avoided by arbitrarily orienting the GS array 
elements.

When all GS array elements are arbitrarily oriented, 
Figure~\ref{PLF_vs_Antenna_Index} shows the effective gain experienced by an
individual antenna element for $M_x=100$, $M_y =1$, $\theta_k =\frac{\pi}{3}$, and $\phi_k=\pi$. It can be observed that not all elements experience low gain. As a result, the nulls as shown in Figure \ref{Effective_Gain_3D_Journal_} can be canceled out due to polarization diversity. For example, from Figure \ref{PLF_vs_Antenna_Index}, it can be observed that the average gain (i.e. $\frac{1}{M}\sum_{l=1}^M\chi_{kl}$) is approximately equal to $-8$ dB.

Figure \ref{CDF_Sum_Gain_Cir} shows the CDF of ($\sum_{l=1}^M\chi_{kl}$) with 
identical and pseudo-randomly oriented GS elements for $M_x=~1,\ 50,\ M_y =1$, 
and $E_\theta^{l}=~\frac{1}{\sqrt{2}}, E_\psi^{l}=~\frac{i}{\sqrt{2}}, 
E_\theta^{'k}=~\frac{1}{\sqrt{2}}$, $E_\theta^{'k}=~\frac{-i}{\sqrt{2}}$ 
(circularly polarized cross-dipoles both at the GS and at the UAV). The solid 
and dashed lines denote the gain values with identical and pseudo-randomly 
oriented GS elements, respectively. The black lines represent the gain values 
with an omni-directional antenna pattern (only in the azimuthal direction) 
while the red lines represent the gain values power with a hypothetical 
isotropic antenna pattern (i.e. the gain is assumed to be $1$ for all 
$\theta_k$ and $\phi_k$). Similarly, Figure~ \ref{CDF_Sum_Gain_Lin} shows the 
CDF of ($\sum_{l=1}^M\chi_{kl}$) for $M_x=50, M_y =1$ and 
$E_\theta^{l}=1,E_\psi^{l}=0,E_\theta^{'k}=1,E_\theta^{'k}=0$ (linearly 
polarized dipoles both at the GS and at the UAV).

{From Figures \ref{CDF_Sum_Gain_Cir} and \ref{CDF_Sum_Gain_Lin} we make the following observations:
\begin{itemize}
\item By increasing the number of antennas from $1$ to $50$, the gain is 
increased by a factor of $50$ ($\approx 17$ dB).

\item Employing linearly polarized antennas either at the GS or UAV results in 
lower gain. Irrespective of the orientation of the GS array elements, almost 
always the gain varies between $-30$ and $21$ dB. 

\item Circularly polarized cross-dipoles perform far better than the linearly 
polarized dipoles. Consider a threshold value of $10$ dB gain. When GS elements 
are identically oriented (solid lines), the probability of experiencing gain 
below $10$ dB (This corresponds to $\frac{1}{M}\sum_{l=1}^M\chi_{kl} \approx 
-7$ dB) is $0.045$ for circularly polarized cross-dipoles and $0.26$ for 
linearly polarized dipoles. This is due to the nulls as observed in Figure 
\ref{Effective_Gain_3D_Journal_}. On the other hand, when the GS elements are 
pseudo-randomly oriented (dashed lines), the probability is zero and $0.16$ for 
circular polarized and linearly polarized dipoles, respectively. 

\item With circularly polarized cross-dipoles, the value of $\chi_\mathrm{wc}$ 
is around $-17$ dB for identical orientation and $-3.5$~dB for arbitrary 
orientation. This will significantly reduce the uplink power. For example, if 
we consider $\frac{1}{M}\sum\nolimits_{l=1}^M\chi_{kl}=-12$ dB and 
$\chi_\mathrm{wc}=-20$ dB, the required transmit power is $p_{uk,\mathrm{Tot}} 
= 0.15$ W. This means that, using arbitrarily oriented GS elements, it is 
possible to achieve $100\%$ coverage with very low uplink transmit power. As it 
has been seen earlier, this is not possible with identically oriented array 
elements.

\item Since the cross-dipole provides quasi-isotropic gain pattern, the gain difference with the isotropic antenna pattern is only around $3$ dB. By adding third dipole, the difference in gain can be further reduced.
\end{itemize}}

The above results clearly suggest that by using simple cross-dipole antenna elements with circular polarization both at the GS (with arbitrary orientation) and at the UAV one can achieve the link reliability requirements of the UAV networks. 

\begin{figure*}
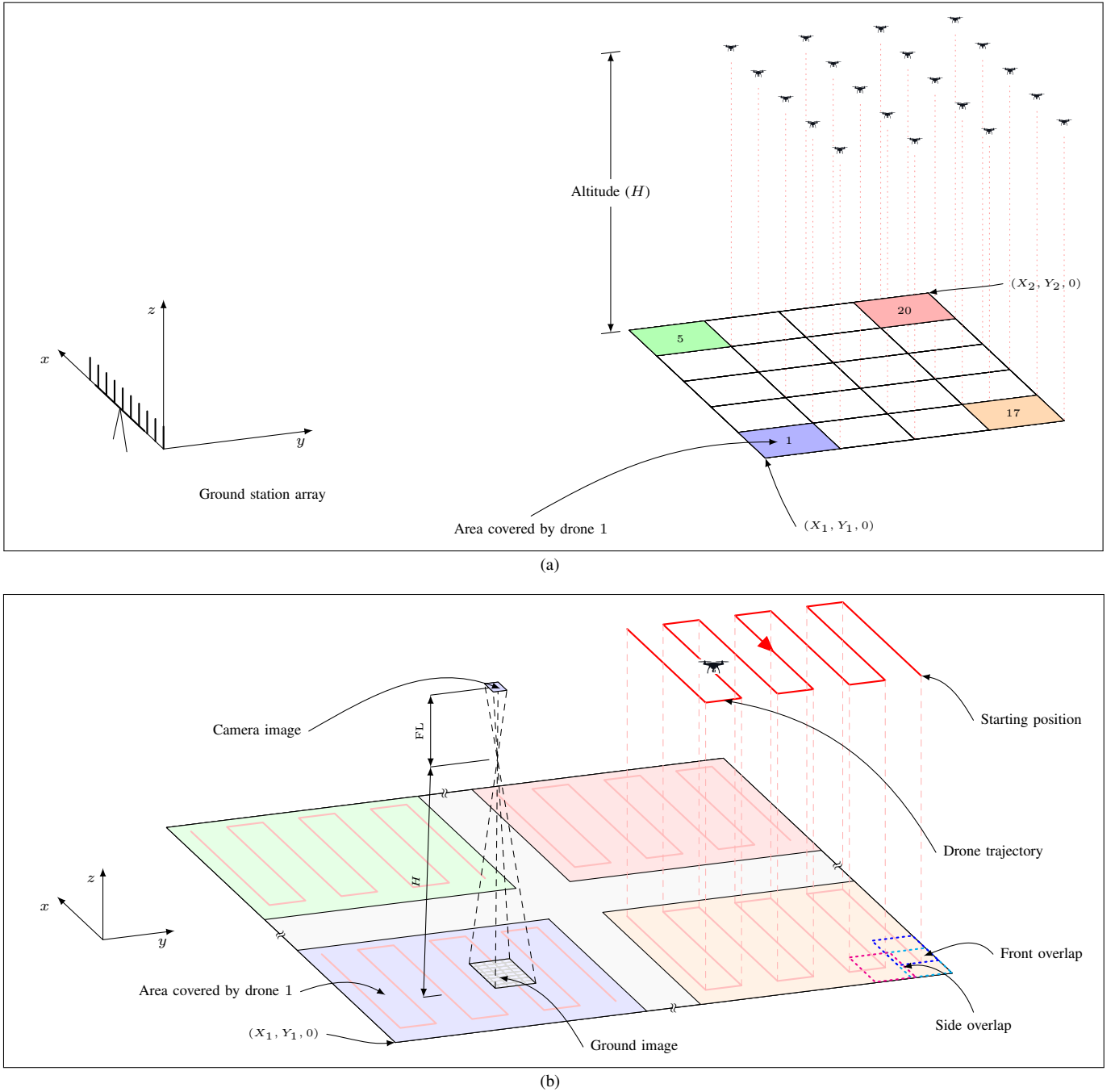

\centering
\subfigure[]{\fbox{\input{projection_simple_wide_journal_2016}}\label{drone_scenario}}\\
\subfigure[]{\fbox{\input{projection_simple_journal_2016}}\label{drone_scenario_1}}\\
\caption{(a) Drone surveillance scenario. (b) Enlarged version of Figure 
\ref{drone_scenario} with camera's geometry and drone's trajectory (The line 
colors correspond to the colored region as shown in \ref{drone_scenario}, Blue: 
Drone $1$, Green: Drone $5$, Orange: Drone $17$, Red: Drone $20$).}
\end{figure*}

\begin{figure*}
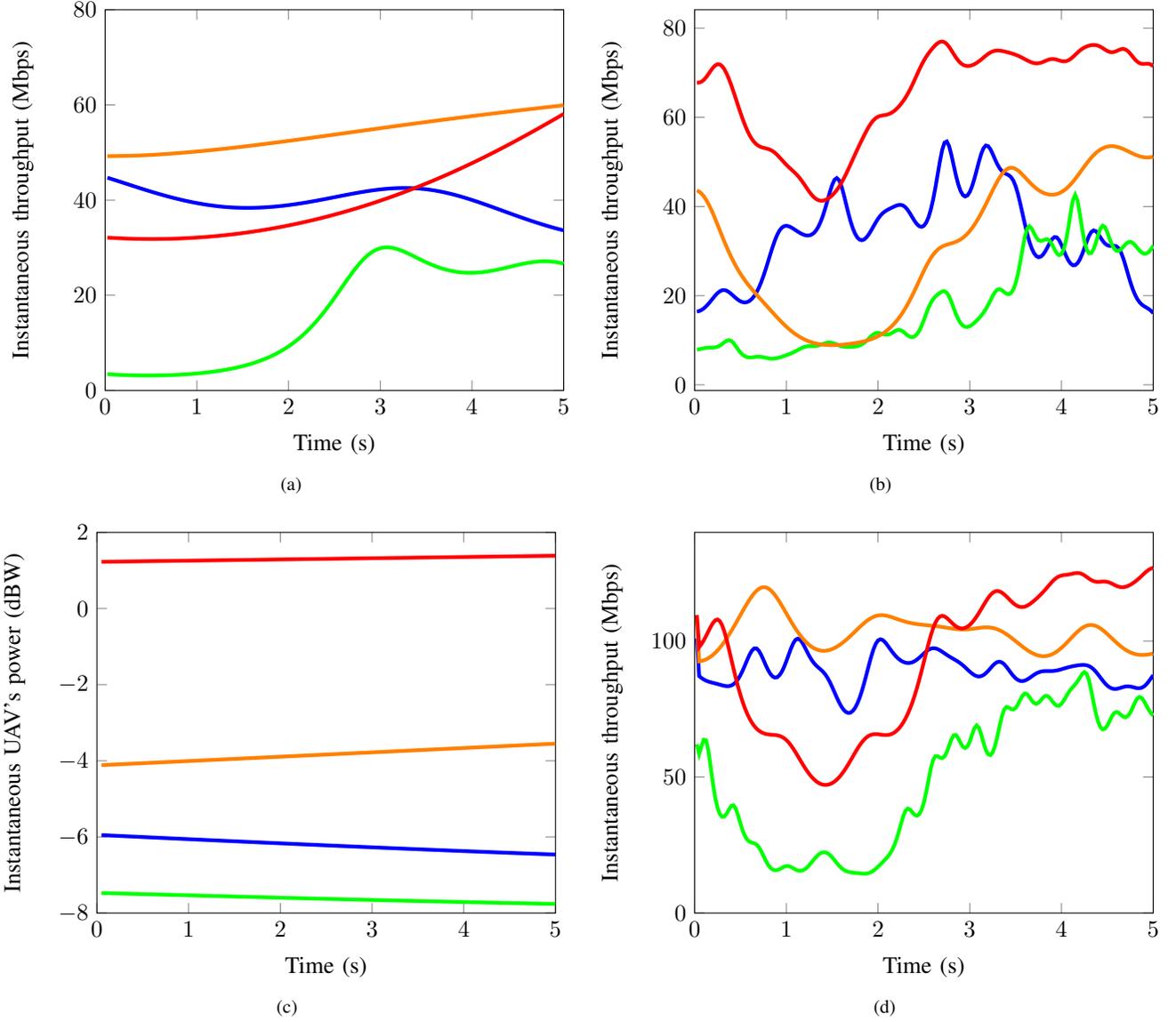

\centering
\subfigure[]{\input{myfile_inst_cap1}
\label{inst_cap1}}  
\subfigure[]{\input{myfile_inst_cap2}
\label{inst_cap2}} \\
\subfigure[]{\input{myfile_power1}
\label{inst_power}} 
\subfigure[]{\input{myfile_inst_cap1_M1000}
\label{inst_cap1_M1000}}
\caption{ (a) Instantaneous throughput achieved by the four corner drones for 
$\delta_x=\frac{\lambda}{2}$ and $M_x=100$. (b) Instantaneous throughput 
achieved for $\delta_x=5\lambda$ and $M_x=100$. (c) Instantaneous transmit 
power of the four corner drones. (d) Instantaneous throughput achieved for 
$\delta_x=\frac{\lambda}{2}$ and $M_x=1000$ (Note that the number of antennas 
in increased by a factor of $10$ and the uplink power is reduced by a factor of 
$10$.)
}
\end{figure*}

\section{Surveillance use case}
Consider a scenario with $K=20$ drones each equipped with a camera (with 
resolution of $r_{py}\times r_{px}$ [pixel], $r_{py}>r_{px}$) scanning a 
particular geographical region with an area $A$ and transmitting images or 
streaming videos to the GS located at the origin (see Figure 
\ref{drone_scenario}). An ULA with $M_x=100$ elements is located along the 
$x$-axis. The GS array elements are arbitrarily oriented and the UAV's dipoles 
are oriented along the $x$- and $z$-axes. Let $X_1=-1000$, $X_2=2000$, 
$Y_1=2000$ and $Y_2=6000$. The total area is $A=3$~km~$\times 4$~km and the 
area covered by each drone is 
$A_{\mathrm{drone}}=\frac{A}{K}=600$~m~$\times 1000$~m. For simplicity of the 
analysis, we assume that, at certain altitude $H$ m, all $20$ drones move 
simultaneously at constant speed $v$ m/s along their trajectory as shown in 
Figure \ref{drone_scenario_1}. Let the starting position of the $k$-th drone be 
$\left(X_1+(i-1)\frac{(X_2-X_1)}{5},Y_1+j\frac{(Y_2-Y_1)}{4},H\right)$, $k= 
5(j-1)+i$, $i=1,...,5,\ j=1,...,4$.

When $H=100$ m, $M_x=100$, $M_y=1$, $f_c=2.4$~GHz, $B = 20$~MHz, $B_c = 3$~MHz, 
$\rho_u = 10$~dB, $\rho_p = 20$~dB, $v=30$~m/s, and 
$E_\theta^{l}=\frac{i}{\sqrt{2}}$, $E_\psi^{l}=\frac{i}{\sqrt{2}}$, 
$E_\theta^{'k}=\frac{i}{\sqrt{2}}$, $E_\theta^{'k}=\frac{i}{\sqrt{2}}$,
 Figures \ref{inst_cap1} and \ref{inst_cap2} show the instantaneous 
throughput achieved by the four corner drones (with index $1, 5, 
17$, and $20$ each drone covers colored regions in Figure \ref{drone_scenario}) 
with antenna spacing $\delta_x=\frac{\lambda}{2}=6.25$ cm and 
$\delta_x=5\lambda=0.625$ m, respectively\footnotemark. Due to the movement of the drones, 
the cumulative interference from the other $K-1=19$ drones fluctuates as the 
azimuth and elevation angles vary with time which results in varying 
instantaneous throughput. With increased antenna spacing $\delta_x=5\lambda$, 
the capacity fluctuates more frequently due to increased resolving capability 
of the GS array. The instantaneous transmit powers of the drones obtained from 
\eqref{inst_power_link_budget} are shown in Figure \ref{inst_power}. The pilot 
power is chosen based on the worst-case values of the distance ($6325$~m at 
$(X_2,Y_2,H)$) and $\chi_\mathrm{wc}=-10$~dB. Further, with 
$\delta_x=\frac{\lambda}{2}$, when increasing the number of GS antennas by a 
factor of $10$ (i.e. $M_x$ is changed from $100$ to $1000$) and reducing  the 
transmit power values (as shown in Figure \ref{inst_power}) by a factor of 
$10$, the instantaneous throughput performance is shown in Figure 
\ref{inst_cap1_M1000}.
\footnotetext[5]{From the received signal as given in \eqref{received_signal_k}, the instantaneous uplink throughput achieved by the $k$-th UAV can be expressed as $S_{k}^{\mathrm{inst}}=\Lambda B \log_2\left(1+\frac{p_{uk}|\hat{\boldsymbol{g}}_k^H\boldsymbol{g}_k|^2}{\sum\nolimits_{j=1,j\neq k}^{K}p_{uj}|\hat{\boldsymbol{g}}_k^H\boldsymbol{g}_j|^2+|\hat{\boldsymbol{g}}_k|^2}\right)$ (bits/s).}

\footnotetext[6]{GSD is the distance between the centers of two neighbouring pixels measured on the ground while  $\mathrm{PS}$ is the distance between two pixels measured on the camera's sensor. Depending on the mission, the target GSD varies from a few centimeters to a few meters \cite{drone_mapping_handbook,rosnell2012}.}

\begin{table*}[t]
\parbox{.49\linewidth}{
\centering
\caption{Data rate requirements for image transmission ($K=20$,
  $r_{py}\times r_{px}=2664\times 1496$, $b=24$, $\mathrm{OL}_y=70\%$)}
\label{data_size_comparison_image}
    \begin{tabular}{|l|l|l|l|}    
    \hline
	    \textbf{GSD$\rightarrow$} & 2 cm & 5 cm & 20 cm \\ 
    \hline
		\textbf{Drone speed $v$ $\rightarrow$} & 20 m/s & 30 m/s & 30 m/s \\
	\hline
    $Q_{\mathrm{image}}$ (Uncompressed) & 120 Mbps \!\!\!\! & 72 Mbps & 18 
    Mbps\\ 
    \hline
	$Q_{\mathrm{image}}^{\mathrm{sum}}$ (Uncompressed)& 
	2.4 Gbps & 1.44 Gbps\!\!\! & 360 Mbps\!\!\!\! \\ 
	\hline
	$M_{\mathrm{req}}$ (Uncompressed) & 2195  & 313 & 20 \\ 
	\hline
	$M_{\mathrm{req}}$ (Compressed, $\mathrm{CR}=2:1$) & 187  & 61 & 9 \\ 
	\hline
	\end{tabular}
}
\hfill 
\hspace{.02\linewidth} 
\parbox{.48\linewidth}{
\centering
\caption{Data rate requirements for compressed ($\mathrm{CR}=200:1$) video transmission}\label{data_size_comparison_video}
    \begin{tabular}{ | l | l  | l | }    \hline
    \textbf{ $\boldsymbol{r_{py}\times r_{px}}\rightarrow$} & $4096\times 2160$ & $2664\times 1496$ \\ \hline
    $Q_{\mathrm{video}}$ (60 FPS) & 64 Mbps & 29 Mbps \\ \hline
		    $Q_{\mathrm{video}}^{\mathrm{sum}}$ (60 FPS) & 1.28 Gbps & 580 Mbps \\ \hline
		    $M_{\mathrm{req}}$ (60 FPS) & 221  & 41 \\ \hline
		    $M_{\mathrm{req}}$ (30 FPS) & 49  & 15 \\ \hline
	  \end{tabular}%
}
\end{table*}

In surveillance missions, the main requirement is to scan the region of 
interest with certain spatial resolution or ground sampling distance (GSD). As 
illustrated in Figure \ref{drone_scenario_1}, for a given target GSD and the 
dimensions of the camera sensor, the drones have to fly at an altitude of $H = 
\frac{\mathrm{GSD}\times \mathrm{FL}}{\mathrm{PS}}$ m (where 
$\textrm`\mathrm{PS}\textrm'$ is the camera sensor's pixel size and 
$\textrm`\mathrm{FL}\textrm'$ is the focal length\footnotemark). The area 
covered by a camera image is 
$A_{\mathrm{image}}~=~r_{px}\cdot~r_{py}\cdot\mathrm{GSD}^2$. If 
$\mathrm{FL}~=~5\times~10^{-3}$~m and $\mathrm{PS}~=~2.3\times~10^{-6}$~m 
\cite[Ch.4]{drone_mapping_handbook}, for $\mathrm{GSD}=2\times 
10^{-2}$, $2\times 10^{-1}$, and $1$ m, the  altitudes are approximately equal 
to $44, 435$, and $2174$ m, respectively. 
When $r_{py}\times r_{px}=2664\times 1496$, the corresponding area covered by a camera image is approximately $53.3\times  30$, $533\times 300$, and $2664\times 1496$ m$^2$, respectively. Let $b$ be the number of bits per pixel generated by the camera's sensor and $\mathrm{CR}$ be the compression ratio. Then the number of bits generated by an image is $D_{\mathrm{image}}=\frac{r_{px}\cdot r_{py} \cdot b}{ \mathrm{CR}}$. We assume that the camera is oriented with the sensor width (long dimension) parallel to the flight direction. Let $\textrm`\mathrm{OL}_y\textrm'$ and $\textrm`\mathrm{OL}_x\textrm'$ be the front and side image overlap, respectively, required by the mission (see Figure \ref{drone_scenario_1}). The time  between two consecutive images is $t=\frac{r_{py}\cdot \mathrm{GSD} \cdot (1-\mathrm{OL}_y)}{v}$, where $v$ is the drone speed.  Then, for image transmission, the data rate required per drone is  
\begin{equation}\label{Q_image}
Q_{\mathrm{image}}=\frac{D_{\mathrm{image}}}{t}=\frac{r_{px}\cdot b \cdot v}{ \mathrm{GSD} \cdot  \mathrm{CR} \cdot (1-\mathrm{OL}_y)}\  \text{(bits/sec)}.
\end{equation}
The sum throughput required by $K$ drones is  
$Q_{\mathrm{image}}^{\mathrm{sum}}~=~K\times~Q_{\mathrm{image}}$ (bits/sec). 
For video transmission, the sum throughput requirement is 
\begin{equation}\label{Q_video}
Q_\mathrm{video}^{\mathrm{sum}}=K\times\frac{r_{px}\cdot r_{py} \cdot b \cdot 
\mathrm{FPS} }{\mathrm{CR}} \ \text{(bits/sec)},\end{equation}
 where  $\mathrm{FPS}$  is the number of frames per second.

Table \ref{data_size_comparison_image} shows the data
  rate requirements (from \eqref{Q_image}) for image transmission with different target GSDs
  and drone speeds when $K=20$, $r_{py}\times r_{px}=2664\times 1496$, 
  $b=24$~bits/pixel, and $\mathrm{OL}_y=70\%$. Similarly, the data rate 
  requirements for compressed video transmission with different camera 
  resolutions
  and frame rates are shown in Table \ref{data_size_comparison_video}. Massive 
  MIMO technology can meet this high-throughput, reliable simultaneous 
  communication requirement of UAV swarm by employing the number of antennas 
  ($M_{\mathrm{req}}$) as given in Tables \ref{data_size_comparison_image} and 
  \ref{data_size_comparison_video}. The lower bound derived as given in 
  \eqref{ergodic_lower_bound_gk} is the average value of the instantaneous 
  capacities shown in Figures \ref{inst_cap1} and \ref{inst_cap2}. As discussed 
  earlier, since we apply channel inversion power control, for the considered 
  trajectory of the drones in this use case, the ergodic rate expression will 
  be the same as \eqref{ergodic_plane_wave_mrc} except the quantity $\Omega$. 
  Since $\Omega$ decays with antenna spacing, it can be made close to zero by 
  appropriately selecting the antenna spacing: $\delta_x$ and $\delta_y$. 
  Therefore, by assuming $\Omega=0$, the values of number of GS antennas 
  ($M_{\mathrm{req}}$) as shown in Tables \ref{data_size_comparison_image} and 
  \ref{data_size_comparison_video} are calculated using \eqref{pre_log_factor}, 
  \eqref{M_req}, and \eqref{Q_image} for the following parameters: 
  $\rho_u=10$~dB, $\rho_p=20$~dB,  $B=20$~MHz, $B_c=3$~MHz, $f_c=2.4$~GHz, and 
  $\tau_{\mathrm{dl}}=\frac{T_{\mathrm{len}}}{8}$. 
	
  The numbers of antennas are calculated assuming 
  transmission with a constant bit rate; hence  if the drone speed is
  $20$~m/s, the time it takes to transmit each  image is
  $0.8$, $2$, and $8$ seconds, respectively. However, in certain
  missions (e.g. search and rescue operations), faster transmission
  might be required. That can be accomplished by increasing   the
  number of antennas beyond the values in Tables \ref{data_size_comparison_image} and 
  \ref{data_size_comparison_video}. Further, as it can be seen from Figure 
  \ref{inst_cap1}, the instantaneous throughput stays at longer time duration 
  due to the similar elevation and azimuth angles of multiple drones. To 
  characterize such events, we have investigated the outage capacity 
  performance of Massive MIMO in LoS conditions in \cite{prabhu-spawc2017}. We 
  have shown that with large number of GS antennas, the outage capacity can be 
  increased as the fluctuations in the interference power becomes negligible 
  due to an interference hardening effect. 	

{In this use case, the total time to complete the mission is $T_{\mathrm{mission}}=\frac{A}{K \cdot v \cdot r_{px}\cdot \mathrm{GSD} \cdot (1-\mathrm{OL}_x)}$ seconds, i.e., $K$ times faster than the single drone case. For $\mathrm{GSD}=5$ cm, $v=30$ m/s, $r_{px}\times r_{py} = 2664 \times 1496$, and $\mathrm{OL}_x=60\%$, the mission can be completed within $6$ minutes and $16$ seconds. On the other hand, a single drone would require more than two hours to complete the mission.
}

\section{Discussions}

The proposed Massive MIMO based communication framework could be used for several multi-drone applications that require high throughput communication with high reliability. For example, consider the mission of  collecting data from a remote region that is unreachable by humans (e.g. during a search-and-rescue operation or during a natural disaster). The antenna array need not necessarily be located on the ground: One can even think of an aerial network that consists of a master drone (hovering at certain altitude) equipped with an antenna array and a swarm of small UAVs (each equipped with a single antenna) surrounding the master drone. 

In practice, the position of the UAVs can have an arbitrary distribution within 
any arbitrary three-dimensional region. Furthermore, the drones' mobility 
patterns can be either known a priori, or not. In this work we have selected 
the random-within-a-sphere mobility model as it facilitates the derivation of 
certain capacity results in closed form. However, we do not expect that 
different mobility models would yield significantly different final performance 
results, as long as the UAVs move fast enough so that each codeword spans over 
many realizations of the inner products of the spatial signatures. For other 
distributions of the drone positions, the rate expressions would be similar, 
but the quantity $\Omega$ will be slightly different. This quantity can be 
interpreted as the correlation between the spatial signatures of the $k$-th and 
the $j$-th UAV. Since $\Omega$ is a function of the antenna spacing, 
optimization of the array geometry is possible. In particular,  since the 
spatial resolvability of an array increases with the element spacing, even if 
the drones are spatially close to each other, by increasing the antenna spacing 
one can reduce the  correlation between their spatial signatures.

Moreover, from \eqref{pre_log_factor} it can be observed that the drone speed 
has only a slight impact on the pre-log factor and hence on the data rate. For 
example, consider $\tau_{\mathrm{dl}} =\frac{T_{\mathrm{len}}}{8}$, 
$B_c=2$~MHz, $f_c = 5$~GHz and $K=100$. If we change the UAV speed from
$v_{\mathrm{max}}=0$ m/s to $v_{\mathrm{max}}=30$ m/s, the change of the pre-log
factor is very small, i.e. $0.875-0.825= 0.05$.  In \eqref{asym_high_snr_mrc}, 
the pre-log factor is chosen based on the maximum drone speed -- yielding  a 
worst-case design. However, due to the LoS propagation, it may be possible to 
reduce the frequency of the pilot transmissions. Specifically, if a UAV moves 
along a trajectory that is known or can be estimated, its channel response may 
be predicted. This can further boost the payload data rate achieved per drone.

Massive MIMO offers both multiplexing gains and a range extension due to the 
array gain. In the surveillance use case described above, the distances between 
the GS and the four corner drones are on the order of several kilometers. Even 
with  a single drone,  existing wireless standards cannot support the required 
range  and throughput, and Massive MIMO may be the only feasible solution. In 
particular, Massive MIMO technology appears to be much preferable and promising 
as compared to multi-hop solutions, which are known to suffer from  serious 
drawbacks in terms of reliability and latency \cite{pinto2017}.

\section{Conclusions}

Massive MIMO has  tremendous potential in the scenario of communication between
a ground station and a swarm of drones:
\begin{itemize}
	\item The multiplexing ability of the array offers huge transmission 
	capacity. See Tables~\ref{data_size_comparison_image} and 
	\ref{data_size_comparison_video} for an example of a drone-swarm 
	surveillance case study, offering sum throughput of $2.4$~Gbps
	(with $2194$ antennas) and $1.44$~Gbps (with $313$ antennas) in $20$~MHz
	bandwidth to $20$ drones.
	\item The array gain of $M$ offers a range extension of a factor $\sqrt{M}$ 
	in line-of-sight. Therefore, the use of Massive MIMO may reduce the need for
	multi-hop solutions. Alternatively, the array gain can be used to reduce the
	UAV's transmit power by a factor $M$.
	\item Maximal-ratio processing with very low complexity and per-antenna
	distributed processing is sufficient to obtain very good performance. 
	Additional gains are obtainable through zero-forcing as pointed out in 
	Section	\ref{rate_opt_ant_spacing}. However, a more detailed study of 
	zero-forcing has to be regarded as future work.
	\item With (pseudo-) randomly oriented ground station antennas, diversity
	against polarization mismatches is naturally obtained. This facilitates the 
	use of simple antenna elements, such as	cross-dipoles, reducing the 
	concerns of	antenna pattern designs.
	\item If the drones are uniformly distributed inside of a spherical shell, 
	then the optimal antenna spacing is an integer multiple of half a 
	wavelength for a linear array (exactly) and for a rectangular array 
	(approximately). 
\end{itemize}
We hope that the community will reduce our ideas in this paper to practice in
the future and perform field trials.

\appendices


\section{APPENDIX A: 3D Rotation Model}\label{section_uav_orientation}
A counterclockwise rotation around the
reference coordinate axes is described by 1) Roll
($\alpha_x \in [-\frac{\pi}{2},\frac{\pi}{2}]$): 
angle of rotation around the $x$-axis 2)
Pitch ($\alpha_y\in [-\frac{\pi}{2},\frac{\pi}{2}]$): 
angle of rotation around $y$-axis
3) Yaw ($\alpha_z\in [0,2\pi]$): 
angle of rotation around the $z$-axis. 
For example, the transformed unit direction vectors due to a rotation
around the $x$-axis is     
$(\hat{\boldsymbol{x}}^{'}\ \hat{\boldsymbol{y}}^{'}\ \hat{\boldsymbol{z}}^{'})=
    (\hat{\boldsymbol{x}}\ \hat{\boldsymbol{y}}\ \hat{\boldsymbol{z}})
    \boldsymbol{\mathrm{R}}_{x}(\alpha_x)$,
where the $3 \times 3$ matrix
$\boldsymbol{\mathrm{R}}_{x}(\alpha_x)$ is given by 
$\boldsymbol{\mathrm{R}}_{x}(\alpha_x)=$ 
$\left(\begin{matrix}
  1 & 0 & 0\\ 0 & \cos{\alpha_x} & -\sin{\alpha_x}\\
  0 & \sin{\alpha_x} & \cos{\alpha_x}
\end{matrix}\right)$.
Similarly, the rotation matrices around $y$- and $z$-axes are given by 
$\boldsymbol{\mathrm{R}}_{y}(\alpha_y)=$ 
$\left(\begin{matrix}
  \cos{\alpha_y} & 0 & \sin{\alpha_y}\\
  0 & 1 & 0\\ -\sin{\alpha_y} & 0 & \cos{\alpha_y}
\end{matrix}\right)$ 
and 
$\boldsymbol{\mathrm{R}}_{z}(\alpha_z)~=~\left(\begin{matrix}
  \cos{\alpha_z} & -\sin{\alpha_z} & 0 \\
  \sin{\alpha_z} & \cos{\alpha_z} & 0 \\ 0 & 0 & 1
\end{matrix}\right)$, respectively. In practice, the rotation of the UAVs may 
take place at around any of the three axes irrespective of the current state of 
the rotation. In that case, the elements of the rotation matrix depend on
the order that the axes are rotated. For example, the rotation matrix
that rotates in the order around $z$-, $y$-, and $x$-axes is obtained
by $\boldsymbol{\mathrm{R}}(\alpha_z,\alpha_y,\alpha_x) =$ $
    \boldsymbol{\mathrm{R}}_{x}(\alpha_x)
    \boldsymbol{\mathrm{R}}_{y}(\alpha_y)
    \boldsymbol{\mathrm{R}}_{z}(\alpha_z)$.


\section{APPENDIX B: Proof of Equation \eqref{ergodic_lower_bound_gk}}\label{ergodic_lower_bound_gk_appendix}
\subsubsection{Calculation of $\mathrm{var}(a_2)$}
The variance of the second term of \eqref{effective_noise} is
\begin{equation}\label{var_b}
\mathrm{var}(a_2)=\sum\nolimits_{j=1,j\neq k}^{K}\mathbb{E}\{p_{uj}p_{uk}|\hat{\boldsymbol{g}}_k^H \boldsymbol{g}_j|^2\}.
\end{equation}
Let $\boldsymbol{w}_k$ be the $k$-th column of $\boldsymbol{W}$. Then, $\mathbb{E}\{p_{uj}p_{uk}|\hat{\boldsymbol{g}}_k^H \boldsymbol{g}_j|^2\}$ in \eqref{var_b} can be obtained as
\begin{align}\label{var_b_j}
 \mathbb{E}
 &\{p_{uj}p_{uk}|\hat{\boldsymbol{g}}_k^H \boldsymbol{g}_j|^2\}\nonumber\\
 &= \mathbb{E}\{p_{uj}p_{uk}\hat{\boldsymbol{g}}_k^H \boldsymbol{g}_j 
 \boldsymbol{g}_j^H \hat{\boldsymbol{g}}_k\}\nonumber\\
 &= \mathbb{E}\bigg\{ p_{uj}p_{uk}\bigg(\boldsymbol{g}_k 
 +\frac{1}{\sqrt{p_p}}\boldsymbol{w}_k\bigg)^H \boldsymbol{g}_j 
 \boldsymbol{g}_j^H 
 \bigg(\boldsymbol{g}_k+\frac{1}{\sqrt{p_p}}\boldsymbol{w}_k\bigg)\bigg\} 
 \nonumber \\
 &= \mathbb{E}\big\{p_{uj}p_{uk}\boldsymbol{g}_k^H\boldsymbol{g}_j
 \boldsymbol{g}_j^H\boldsymbol{g}_k\big\}\nonumber\\
 &\hspace{.5cm}+\frac{1}{\sqrt{p_p}}\mathbb{E}
 \big\{\boldsymbol{g}_k^H\}\mathbb{E}\big\{p_{uj}p_{uk}\boldsymbol{g}_j
 \boldsymbol{g}_j^H\}\mathbb{E}\big\{\boldsymbol{w}_k\big\}\\
 &\hspace{.5cm}+\frac{1}{\sqrt{p_p}}\mathbb{E}\big\{\boldsymbol{w}_k^H\}
 \mathbb{E}\big\{p_{uj}p_{uk}\boldsymbol{g}_j\boldsymbol{g}_j^H\}
 \mathbb{E}\big\{\boldsymbol{g}_k\big\}\nonumber\\
 &\hspace{.5cm}+\frac{1}{p_p}\mathbb{E}\big\{\boldsymbol{w}_k^H
 \boldsymbol{g}_j\boldsymbol{g}_j^H\boldsymbol{w}_k\big\}\nonumber \\
 &= \mathbb{E}\left\{p_{uj}p_{uk}|\boldsymbol{g}_k^H
 \boldsymbol{g}_j|^2\right\}+\frac{1}{p_p}
 \mathbb{E}\left\{\!p_{uj}p_{uk}\norm{\boldsymbol{g}_j}^2\!\right\}\nonumber\\
 &= \mathbb{E}\left\{p_{uj}p_{uk}|\boldsymbol{g}_k^H
 \boldsymbol{g}_j|^2\right\}\!+\!\frac{M\rho_u}{p_p}
 \mathbb{E}\left\{\frac{1}{\beta_k\chi_k}\right\},\nonumber
\end{align}
where we used the facts that $\boldsymbol{g}_k$ and $\boldsymbol{w}_k$ are independent, $\mathbb{E}\big\{\boldsymbol{w}_k\big\}=\boldsymbol{0}$, and $\norm{\boldsymbol{g}_j}^2=M\beta_j\chi_j$.

After substituting \eqref{pilot_power} and \eqref{var_b_j} into \eqref{var_b}, we get
\begin{equation}
\begin{array}{r@{}l}
\label{var_b_final}
\mathrm{var}(a_2)= &{} \displaystyle 
\frac{(K-1)M\rho_u}{\rho_p}\chi_\mathrm{wc}
\left(\frac{\lambda}{4 \pi d_\mathrm{wc}}\right)^2
\mathbb{E}\bigg\{\frac{1}{\beta_k\chi_k}\bigg\}\\
&{} +\sum\nolimits_{j=1,j\neq k}^{K}
\mathbb{E}\big\{p_{uj}p_{uk}|\boldsymbol{g}_k^H\boldsymbol{g}_j|^2\big\}.
\end{array}
\end{equation}

\subsubsection{Calculation of $\mathrm{var}(a_1)$}
The variance of the first term of \eqref{effective_noise} is
\begin{equation}
\begin{array}{r@{}l}
\label{denom_first_term}
\mathrm{var}(a_1) &{} = 
\mathbb{E}\big\{|p_{uk}\hat{\boldsymbol{g}}_k^H\boldsymbol{g}_k-
\mathbb{E}\{p_{uk}\hat{\boldsymbol{g}}_k^H\boldsymbol{g}_k\}|^2\big\} \\
&{} =\mathbb{E}\big\{p_{uk}^2|\hat{\boldsymbol{g}}_k^H\boldsymbol{g}_k|^2\big\} 
-|\mathbb{E}\{p_{uk}\hat{\boldsymbol{g}}_k^H\boldsymbol{g}_k\}|^2.
\end{array}
\end{equation}

From \eqref{var_b_j}, by applying \eqref{pilot_power} and \eqref{power_ch_inv}, we get
\begin{align}\label{Expectation_mod_ak2}
\mathbb{E}\{p_{uk}^2|\hat{\boldsymbol{g}}_k^H \boldsymbol{g}_k|^2\}& = 
\mathbb{E}\big\{p_{uk}^2\norm{\boldsymbol{g}_k}^4\}+\frac{1}{p_p}\mathbb{E}\big\{p_{uk}^2\norm{\boldsymbol{g}_k}^2\big\}\nonumber\\&=M^2\rho_u^2+\frac{M\rho_u}{p_p}\mathbb{E}\bigg\{\frac{1}{\beta_k\chi_k}\bigg\}
\end{align}
and 
\begin{align}\label{Expectation_ak}
\mathbb{E}\{p_{uk}\hat{\boldsymbol{g}}_k^H \boldsymbol{g}_k\} &= \mathbb{E}\bigg\{p_{uk}\left(\boldsymbol{g}_k+\frac{1}{\sqrt{p_p}}\boldsymbol{w}_k\right)^H \boldsymbol{g}_k\bigg\} \nonumber \\
&=\mathbb{E}\big\{p_{uk}\norm{\boldsymbol{g}_k}^2\}+\frac{1}{\sqrt{p_p}}\mathbb{E}\big\{\boldsymbol{w}_k^H \big\} \mathbb{E}\big\{p_{uk}\boldsymbol{g}_k\big\}\nonumber\\
&=\mathbb{E}\big\{p_{uk}\norm{\boldsymbol{g}_k}^2\}\\
& = M\rho_u.\nonumber
\end{align}

After substituting \eqref{Expectation_mod_ak2} and \eqref{Expectation_ak} into \eqref{denom_first_term}, we get
\begin{equation}\label{var_a}
\mathrm{var}(a_1) = \frac{M\rho_u}{\rho_p}\chi_\mathrm{wc}\left(\frac{\lambda}{4 \pi d_\mathrm{wc}}\right)^2\mathbb{E}\bigg\{\frac{1}{\beta_k\chi_k}\bigg\}.
\end{equation}

\subsubsection{Calculation of $\mathrm{var}(a_3)$}
The variance of the third term of \eqref{effective_noise} is 
\begin{equation}
\begin{array}{r@{}l}
\label{var_c}
\mathrm{var}(a_3)
&{} = \mathbb{E}\{p_{uk} \hat{\boldsymbol{g}}_k^H \boldsymbol{n} 
\boldsymbol{n}^H \hat{\boldsymbol{g}}_k\}\\
&{} = \mathbb{E}\big\{p_{uk}\boldsymbol{g}_k^H\boldsymbol{n}
\boldsymbol{n}^H\boldsymbol{g}_k\big\}\\
&{} \displaystyle \hspace{.5cm} 
+\frac{1}{\sqrt{p_p}}\mathbb{E}\big\{p_{uk}\boldsymbol{g}_k^H\}
\mathbb{E}\{\boldsymbol{n}\boldsymbol{n}^H\}\mathbb{E}\{\boldsymbol{w}_k\big\}\\
&{} \displaystyle \hspace{.5cm} 
+\frac{1}{\sqrt{p_p}}\mathbb{E}\big\{\boldsymbol{w}_k^H\}
\mathbb{E}\{\boldsymbol{n}\boldsymbol{n}^H\}\mathbb{E}\{p_{uk}
\boldsymbol{g}_k\big\}\\
&{} \displaystyle \hspace{.5cm} + 
\frac{1}{p_p}\mathbb{E}\big\{p_{uk}\boldsymbol{w}_k^H
\boldsymbol{n}\boldsymbol{n}^H\boldsymbol{w}_k\!\big\}\\ 
&{} \displaystyle = M\rho_u+\frac{M}{\rho_p}\chi_\mathrm{wc}
\left(\frac{\lambda}{4 \pi d_\mathrm{wc}}\right)^2
\mathbb{E}\bigg\{\frac{1}{\beta_k\chi_k}\bigg\}. 
\end{array}
\end{equation}
After substituting \eqref{var_b_final}, \eqref{Expectation_ak}, \eqref{var_a}, and \eqref{var_c} into \eqref{ergodic_lower_bound_hgo}, we get \eqref{ergodic_lower_bound_gk}.

\section{APPENDIX C: Proof of equations \eqref{mean_higher_order_term} and \eqref{mean_sinc}}\label{proof_mean_sinc_higher_order_term}
\subsubsection{Proof of Equation \eqref{mean_higher_order_term}}
By letting 
\[
b_{ll'} = \frac{\pi}{\lambda}
\big(((p-1)^2-(p'-1)^2)\delta_x^2 + ((q-1)^2-(q'-1)^2)\delta_y^2\big),
\] 
Equation \eqref{mean_higher_order_term} can be written as
\begin{align}\label{mean_high_app}
\mathbb{E}
	& \bigg\{e^{i\frac{\pi}{\lambda}\frac{1}{d_k}
	\big[((p-1)^2-(p'-1)^2)\delta_x^2+((q-1)^2-(q'-1)^2)\delta_y^2\big]}\bigg\}
	\nonumber\\
	& = \mathbb{E}\left\{e^{i\frac{b_{ll'}}{d_k}}\right\} \\
	& = \mathbb{E}\left\{\cos\left(b_{ll'}/d_k\right)\right\} + 
	i\mathbb{E}\left\{\sin\left(b_{ll'}/d_k\right)\right\}.\nonumber
\end{align}
Here
\begin{align}
\label{mean_cos_a_r}
  & \mathbb{E}\big\{\cos\big(b_{ll'}/d_k\big)\big\} = 
  \int_{R_{\mathrm{min}}}^{R}\cos\left(\frac{b_{ll'}}{r}\right)f_{d_k}(r)\ 
  dr\nonumber\\
  & = \int_{R_{\mathrm{min}}}^{R}\cos\left(\frac{b_{ll'}}{r}\right)
  \frac{3r^2}{R^3-{R_{\mathrm{min}}^3}} \ dr\nonumber\\
  & = \frac{1}{R^3-{R_{\mathrm{min}}^3}}
  \bigg(\!r^3\cos\left(\frac{b_{ll'}}{r}\right)\bigg|_{R_\mathrm{min}}^R 
  -\int_{R_\mathrm{min}}^R\! r^3 
  d\cos\left(\frac{b_{ll'}}{r}\right)\!\!\bigg)\nonumber\\
  & = \frac{1}{R^3-{R_{\mathrm{min}}^3}}
  \bigg(r^3\cos\left(\frac{b_{ll'}}{r}\right)\bigg|_{R_\mathrm{min}}^R
  \!\!\!\!\! -b_{ll'}\!\!\int_{R_\mathrm{min}}^R\!\! r 
  \sin\left(\frac{b_{ll'}}{r}\right) dr\!\bigg)\nonumber\\
  & = \frac{1}{R^3-{R_{\mathrm{min}}^3}}
  \Bigg(r^3\cos\left(\frac{b_{ll'}}{r}\right)
  \bigg|_{R_\mathrm{min}}^R\\
  & \hspace{.5cm} 
  -\frac{b_{ll'}}{2}\bigg(r^2\sin\left(\frac{b_{ll'}}{r}\right)
  \bigg|_{R_\mathrm{min}}^R\!\!+b_{ll'}\int_{R_\mathrm{min}}^R 
  \cos\left(\frac{b_{ll'}}{r}\right) \ dr\bigg)\Bigg)\nonumber\\
  & = \frac{1}{R^3-{R_{\mathrm{min}}^3}}
  \Bigg(r^3\cos\left(\frac{b_{ll'}}{r}\right)\bigg|_{R_\mathrm{min}}^R 
  \!\!-\frac{b_{ll'}}{2}r^2\sin\left(\frac{b_{ll'}}{r}\right)
  \!\bigg|_{R_\mathrm{min}}^R\nonumber\\
  & \hspace{.5cm} 
  -\frac{b_{ll'}^2}{2}\bigg(r\cos\left(\frac{b_{ll'}}{r}\right)
  \bigg|_{R_\mathrm{min}}^R-b_{ll'}\int_{R_\mathrm{min}}^R 
  \frac{\sin\left(\frac{b_{ll'}}{r}\right)}{r} \ dr\bigg)\Bigg)\nonumber\\
  & = \frac{1}{2(R^3-{R_{\mathrm{min}}^3)}}
  \Bigg((2r^2-b_{ll'}^2)r\cos\left(\frac{b_{ll'}}{r}\right)
  \bigg|_{R_\mathrm{min}}^R\nonumber\\
  & \hspace{.5cm} 
  -b_{ll'}r^2\sin\left(\frac{b_{ll'}}{r}\right)
  \bigg|_{R_\mathrm{min}}^R-b_{ll'}^3\int_{R_\mathrm{min}}^R 
  \frac{\sin\left(\frac{b_{ll'}}{r}\right)}{r} \ dr\Bigg).\nonumber
\end{align}
The last integral in \eqref{mean_cos_a_r} can be calculated as
\begin{align}
\label{Si_integral}
\int_{R_\mathrm{min}}^R \frac{\sin\left(\frac{b_{ll'}}{r}\right)}{r} \ dr 
& = -\int_{\big(b_{ll'}/R_\mathrm{min}\big)}^{\frac{b_{ll'}}{R}} 
\frac{\sin(t)}{t} \ dt\nonumber\\
& = -\mathrm{Si}\big(b_{ll'}/r\big)\bigg|_{R_\mathrm{min}}^R\\
& = \mathrm{Si}\big(b_{ll'}/R\big)-\mathrm{Si}\big(b_{ll'}/R_\mathrm{min}\big),
\nonumber
\end{align}
where $\mathrm{Si}(x)=\int_0^x\frac{\sin t}{t} dt$. By substituting \eqref{Si_integral} into \eqref{mean_cos_a_r}, we get 
\begin{align*}
\mathbb{E}
& \big\{\cos\big(b_{ll'}/d_k\big)\big\}\nonumber\\
& = \frac{1}{2(R^3-{R_{\mathrm{min}}^3)}}\nonumber\\
&\hspace{.4cm}\times\bigg(\big(2R^2-b_{ll'}^2\big)
R\cos\big(b_{ll'}/R\big)-b_{ll'}R^2\sin\big(b_{ll'}/R\big)\nonumber\\
&\hspace{0.85cm}+b_{ll'}^3\mathrm{Si}\big(b_{ll'}/R\big) 
\!-\!\big(2R_\mathrm{min}^2\!-\!b_{ll'}^2\big)
R_\mathrm{min}\cos\big(b_{ll'}/R_\mathrm{min}\big)\nonumber\\
&\hspace{0.85cm}+b_{ll'}R_\mathrm{min}^2\sin\big(b_{ll'}/R_\mathrm{min}\big)-
b_{ll'}^3\mathrm{Si}\big(b_{ll'}/R_\mathrm{min}\big)\bigg).
\end{align*}
Similarly, 
\begin{align*}
\mathbb{E}
& \big\{\sin\big(b_{ll'}/d_k\big)\big\}\nonumber\\
& = \frac{1}{2(R^3-{R_{\mathrm{min}}^3)}}\nonumber\\ 
&\hspace{.4cm}\times\bigg(\big(2R^2-b_{ll'}^2\big)R
\sin\big(b_{ll'}/R\big)+b_{ll'}R^2\cos\big(b_{ll'}/R\big)\nonumber\\
&\hspace{0.85cm}+b_{ll'}^3\mathrm{Ci}\big(b_{ll'}/R\big) 
\!-\!\big(2R_\mathrm{min}^2\!-\!b_{ll'}^2\big)R_\mathrm{min}
\sin\big(b_{ll'}/R_\mathrm{min}\big)\nonumber\\
&\hspace{0.85cm}-b_{ll'}R_\mathrm{min}^2\cos\big(b_{ll'}/R_\mathrm{min}\big)
-b_{ll'}^3\mathrm{Ci}\big(b_{ll'}/R_\mathrm{min}\big)\bigg),
\end{align*}
where $\mathrm{Ci}(x)=-\int_x^\infty\frac{\cos t}{t} dt$. Now by letting 
$\mathbb{C}(b_{ll'})~=~\mathbb{E}\big\{\cos\big(b_{ll'}/d_k\big)\big\}$ and 
$\mathbb{D}(b_{ll'})=\mathbb{E}\big\{\sin\big(b_{ll'}/d_k\big)\big\}$, the 
expectation in \eqref{mean_high_app} is expressed as given in 
\eqref{mean_higher_order_term}.

\subsubsection{Proof of Equation \eqref{mean_sinc}}
\begin{align*}
&\mathbb{E}\left\{e^{i\frac{2\pi}{\lambda}
	((p-p')\delta_x\sin{\theta_k}\cos{\phi_k}+
	(q-q')\delta_y\sin{\theta_k}\sin{\phi_k})}\right\}\nonumber\\
& = \int_{0}^{2\pi}\int_{0}^{\pi}e^{i\frac{2\pi}{\lambda}
	((p-p')\delta_x\sin{\theta}\cos{\phi}+
	(q-q')\delta_y\sin{\theta}\sin{\phi})}\nonumber\\
&\hspace{3cm}\times f_{\theta_k}(\theta)f_{\phi_k}(\phi)\ 
d\theta d\phi\nonumber \\
&\!=\!\!\frac{1}{4\pi}\!\!\int_{0}^{2\pi}\!\!\!\!\!\int_{0}^{\pi}\!\!\!\!e^{i\frac{2\pi}{\lambda}((p-p')\delta_x\sin{\theta}\cos{\phi}+(q-q')\delta_y\sin{\theta}\sin{\phi})}\!
 \sin{\theta}\ d\theta d\phi.
\end{align*}
By letting $\alpha_1=\frac{(p-p')\delta_x}{\lambda}$ and $\alpha_2=\frac{(q-q')\delta_y}{\lambda}$, the above integral can be rewritten as 
\begin{align*}
\frac{1}{4\pi}\int_{0}^{2\pi}\int_{0}^{\pi}
e^{i2\pi(\alpha_1\sin{x}\cos{y}+\alpha_2\sin{x}\sin{y})}\sin{x}\ dx dy.
\end{align*}

First we observe that $\alpha_1\cos y+\alpha_2\sin y = \sqrt{\alpha_1^2+\alpha_2^2} \left(\frac{\alpha_1}{\sqrt{\alpha_1^2+\alpha_2^2}}\cos y+\frac{\alpha_2}{\sqrt{\alpha_1^2+\alpha_2^2}}\sin y\right)$ and we can find an angle $\xi\in [0,2\pi)$, such that $\cos \xi=\frac{\alpha_1}{\sqrt{\alpha_1^2+\alpha_2^2}}$ and $\sin \xi=\frac{\alpha_2}{\sqrt{\alpha_1^2+\alpha_2^2}}$. Thus we can simplify the expression as $\alpha_1\cos y+\alpha_2\sin y = \sqrt{\alpha_1^2+\alpha_2^2}\ \big(\cos \xi\cos y+\sin \xi\sin y\big)=\sqrt{\alpha_1^2+\alpha_2^2}\ \cos (y-\xi)$, 
which leads to 
\begin{align}
\label{four}
\frac{1}{4\pi} 
& \int_{0}^{2\pi}\int_{0}^{\pi} e^{i2\pi\sin{x}(\alpha_1\cos{y} +
	\alpha_2\sin{y})}\sin{x}\ dx dy\nonumber\\
& = \frac{1}{4\pi}\int_{0}^{2\pi}\int_{0}^{\pi}e^{i2\pi\alpha
  \sin{x}\cos{(y-\xi)}}\sin{x}\ dx dy\\
& = \frac{1}{4\pi}\int_{0}^{2\pi}\int_{0}^{\pi}e^{i2\pi\alpha
  \sin{x}\cos{y}}\sin{x}\ dx dy,\nonumber
\end{align}
where $\alpha=\sqrt{\alpha_1^2+\alpha_2^2}$. By applying the following 
symmetries of the trigonometric functions $\cos y~=~\cos(2\pi-y)$, 
$\sin x~=~\sin(\pi-x)$ and $\cos y~=~-\cos(\pi-y)$ we consecutively calculate
\begin{align}
\label{sim1}
\frac{1}{4\pi}
&  \int_{0}^{2\pi}  \int_{0}^{\pi} e^{i2\pi\alpha \sin x \cos y} 
   \sin x \ dx dy\nonumber\\
& = \frac{1}{2\pi}\int_{0}^{\pi} \int_{0}^{\pi} e^{i2\pi\alpha \sin x 
	\cos y} \sin x \ dx dy\nonumber\\
& = \frac{1}{\pi}\int_{0}^{\pi} \int_{0}^{\frac{\pi}{2}} 
e^{i2\pi\alpha \sin x \cos y} \sin x \ dx dy \\ 
& = \frac{1}{\pi}\bigg(\int_{0}^{\frac{\pi}{2}} \int_{0}^{\frac{\pi}{2}}
e^{i2\pi\alpha \sin x \cos y} \sin x \ dx dy \nonumber\\
&\hspace{1cm} + \int_{0}^{\frac{\pi}{2}}  \int_{0}^{\frac{\pi}{2}} 
e^{-i2\pi\alpha \sin x \cos y} \sin x \ dx dy\bigg).\nonumber
\end{align}
By defining the function 
\[
h(\alpha) = \int_{0}^{\frac{\pi}{2}} \int_{0}^{\frac{\pi}{2}} 
e^{i2\pi \alpha \sin x \cos y} \sin x \ dx dy
\] 
and using (\ref{sim1}), we get 
\begin{align}
\label{seven}
\frac{1}{4\pi} \int_{0}^{2\pi} \int_{0}^{\pi} e^{i2\pi\alpha \sin x
  \cos y} \sin x \ dx dy 
& = \frac{1}{\pi}\big( h(\alpha) + h^*(\alpha) \big)\nonumber\\
& = \frac{2}{\pi}\Re\{h(\alpha)\}.
\end{align}
We calculate the Fourier transform of $h(\alpha)$ as 
\begin{align}
\label{fourier}
 H(f)
 & = \mathcal{F}\{h(\alpha)\} = \int_{-\infty}^{\infty} h(\alpha) 
 e^{-i2\pi\alpha f} d\alpha \nonumber\\
 & = \int_{-\infty}^{\infty} \int_{0}^{\frac{\pi}{2}}
    \int_{0}^{\frac{\pi}{2}} e^{i2\pi \alpha \sin x \cos y} \sin x \ dx dy
	\ e^{-i2\pi \alpha f} \ d\alpha \nonumber\\
 & = \int_{0}^{\frac{\pi}{2}} \int_{0}^{\frac{\pi}{2}}
    \bigg(\int_{-\infty}^{\infty} e^{-i2\pi\alpha(f-\sin x \cos y)}
    d\alpha\bigg)\sin x \ dx dy \nonumber \\ 
 & = \int_{0}^{\frac{\pi}{2}} \int_{0}^{\frac{\pi}{2}}
     \delta(f - \sin x \cos y) \sin x \ dx dy\\
 & =\!\!
\begin{cases}
  \int_{0}^{\frac{\pi}{2}}\int_{0}^{\frac{\pi}{2}} \delta(f-\sin x
  \cos y) \sin x\ dx dy, &\!\!\!\! \text{if}\ f \in [0,1], \\
  0, &\!\!\!\! \text{if}\ f \not\in [0,1].
\end{cases}\nonumber
\end{align}
Here we used the facts that $\sin x \cos y \in [0,1]$ when 
$x,y~\in~\big[0,\frac{\pi}{2}\big]$ and that
\begin{equation}
\int_A f(x)\ \delta(g(x)) dx = \sum\nolimits_i \frac{f(x_i)}{|g^{'}(x_i)|},
\label{deltaintegral}
\end{equation}
where $x_i\in A \subseteq \mathbb{R}$ are all roots of $g(x)$ (i.e. $g(x_i)=0$) 
in the set $A$. If $g(x)$ has no roots in $A$, then the integral in 
(\ref{deltaintegral}) is equal to zero. We proceed with the calculation of the 
inner integral in (\ref{fourier}). We observe that in this case we have $f\in 
[0,1],\sin x \in [0,1]$ and $\cos y \in [0,1]$. Using (\ref{deltaintegral}) we 
can conclude that
\begin{align}
\label{innerintegral}
\int_{0}^{\frac{\pi}{2}} 
&  \delta(f - \sin x \cos y) \sin x\ dx\nonumber\\
& = 
\begin{cases}
  \frac{\sin(\arcsin (\frac{f}{\cos y}))}{|-\cos(\arcsin(\frac{f}{\cos
      y})) \cos y|}, & \text{if}\ f\in [0,\cos y], \\
  0, & \text{if}\ f \in (\cos y,1]
\end{cases} \\
& =
\begin{cases}
  \frac{f}{\cos y \sqrt{\cos^2 y - f^2}}, & \text{if}\ y\in[0,\arccos f], \\
  0, & \text{if}\ y \in (\arccos f,\frac{\pi}{2}].
\end{cases}\nonumber
\end{align}
Now we can calculate the double integral in (\ref{fourier}) by applying several variable changes. Inserting (\ref{innerintegral}) in (\ref{fourier}) we obtain  
\begin{align}
  \label{pihalf}
  & \int_{0}^{\frac{\pi}{2}}\int_{0}^{\frac{\pi}{2}}  \delta(f-\sin x \cos y) 
  \sin x\ dx dy\nonumber\\
  & = 
  \int_{0}^{\arccos f} \frac{f}{\cos y \sqrt{\cos^2 y-f^2}}\ dy
  \ \ / y = \arcsin t /  \nonumber\\
  & = \int_{0}^{\sqrt{1-f^2}} \!\!\!\!\frac{f}{(1-t^2) \sqrt{1-t^2-f^2}}\ dt
  \ \ / t \!=\! (p-p')\sqrt{1-f^2}\! / \nonumber\\ 
  & = \int_{0}^{1} \frac{f}{(1-(p-p')^2(1-f^2)) \sqrt{1-(p-p')^2}}\ du
  \nonumber \\ 
  &\hspace{2cm} \ / p-p' = \cos t / \\ 
  & = \int_{0}^{\frac{\pi}{2}} \frac{f}{\sin^2t+f^2\cos^2t}\ dt\nonumber\\
  & = \int_{0}^{\frac{\pi}{2}} \frac{2f}{(f^2+1)+(f^2-1)\cos 2t}\ dt
  \ \ / t= \frac{\pi-2u}{4} / \nonumber\\ 
  & = \int_{-\frac{\pi}{2}}^{\frac{\pi}{2}} 
  \frac{f}{(f^2+1)+(f^2-1)\sin (p-p')}\ du\nonumber\\
  & = \frac{f}{2f}\arcsin\bigg(\frac{(f^2+1)\sin (p-p') + (f^2-1)}
  {(f^2+1)+(f^2-1)\sin(p-p')}\bigg)
  \bigg|_{-\frac{\pi}{2}}^{\frac{\pi}{2}}\nonumber\\
  & = \frac{\pi}{2}. \nonumber 
\end{align}
Combining (\ref{fourier}) and (\ref{pihalf}) we get 
\begin{equation}
  H(f) = \frac{\pi}{2}\mathrm{rect\bigg(f-\frac{1}{2}\bigg)} =
  \begin{cases}
    \frac{\pi}{2}, & \text{if}\ f\in[0,1], \\
    0,  & \text{if}\ f \not\in [0,1],
  \end{cases}\nonumber
\end{equation} where the rectangular pulse function is defined as 
\begin{equation}\mathrm{rect(t)} = \begin{cases}
1, & \text{if}\ t\in\left[-\frac{1}{2},\frac{1}{2}\right],\\
0, & \text{if}\ t\not\in\left[-\frac{1}{2},\frac{1}{2}\right].
\end{cases}\nonumber
\end{equation} 
The function $h(\alpha)$ can be derived by taking the inverse Fourier transform $H(f)$ as follows
\begin{equation}
\begin{array}{r@{}l}
\label{halpha}
  h(\alpha) &{} =
  \mathcal{F}^{-1}\{H(f)\} = \frac{\pi}{2}\mathcal{F}^{-1}
  \bigg\{\mathrm{rect\bigg(f-\frac{1}{2}\bigg)}\bigg\}\\ 
  &{} \displaystyle = \frac{\pi}{2}e^{i2\pi\frac{1}{2}\alpha}
  \mathcal{F}^{-1}\{\mathrm{rect(f)}\} =
  \frac{\pi}{2}e^{i\pi\alpha}\mathrm{sinc}(\alpha).
\end{array}
\end{equation}
Using \eqref{four}, \eqref{seven}, and \eqref{halpha}, the original integral can 
be obtained as
\begin{equation}
\begin{array}{r@{}l}
  \displaystyle \frac{1}{4\pi} 
  &{} \displaystyle \int_{0}^{2\pi}\int_{0}^{\pi} 
  e^{i2\pi\sin{x}(\alpha_1\cos{y}+\alpha_2\sin{y})}\sin{x}\ dx dy\\ 
  &{} \displaystyle =
  \frac{2}{\pi}\Re\{h(\alpha)\}   =
  \frac{2}{\pi}\frac{\pi}{2}\Re\{e^{i\pi\alpha}\}\mathrm{sinc}(\alpha)\\ 
  &{} \displaystyle =
  \cos(\pi\alpha)\ \mathrm{sinc}(\alpha) =
  \frac{\cos(\pi\alpha)\sin(\pi\alpha)}{\pi\alpha}\\
  &{} \displaystyle =
  \frac{\sin(2\pi\alpha)}{2\pi\alpha} =
  \mathrm{sinc}(2\alpha).
\end{array}
\end{equation}

Finally, by substituting $\alpha_1$, $\alpha_2$, and $\alpha$, we get
\begin{equation*}
\begin{array}{r@{}l}
 \mathbb{E} & 
 \Big\{e^{i\frac{2\pi}{\lambda}((p-p')\delta_x\sin{\theta_k}\cos{\phi_k}+
 	(q-q')\delta_y\sin{\theta_k}\sin{\phi_k})}\Big\}\\
 &{} = \mathrm{sinc}\bigg(\frac{2}{\lambda}\sqrt{(p-p')^2\delta_x^2+
 	(q-q')^2\delta_y^2}\bigg).
\end{array}
\end{equation*}


\vspace{5cm}
\begin{IEEEbiography}[{\includegraphics[width=1.1in,height=1.1in,clip,keepaspectratio]{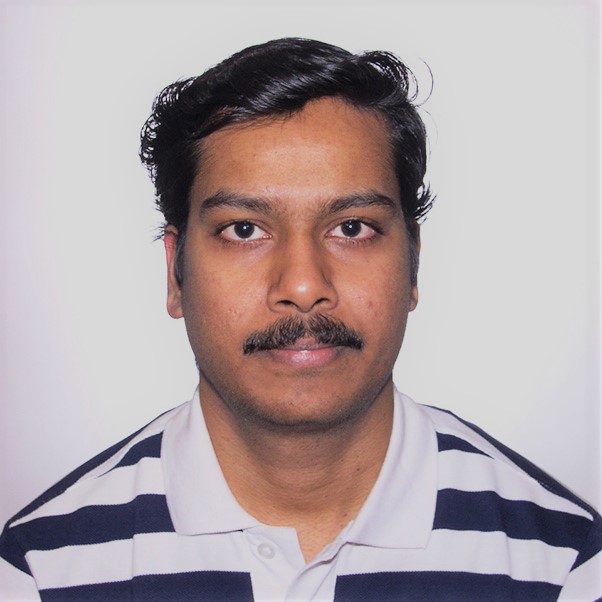}}]{Prabhu
 Chandhar} received the Ph.D. Degree from Indian Institute of Technology 
Kharagpur, West Bengal, India. From Nov. 2015 to Oct. 2017, he was a 
Postdoctoral Researcher at the Division of Communication Systems, Link\"oping 
University (LiU), Link\"oping, Sweden. From Aug. 2009 to July 2010, he worked 
as a Senior Research Fellow at Vodafone IIT KGP Centre of Excellence in 
Telecommunications (VICET), IIT Kharagpur, India. His research interests are 
within the fields of signal processing and communication theory.
\end{IEEEbiography}


\begin{IEEEbiography}[{\includegraphics[width=1.33in,height=1.33in,clip,keepaspectratio]{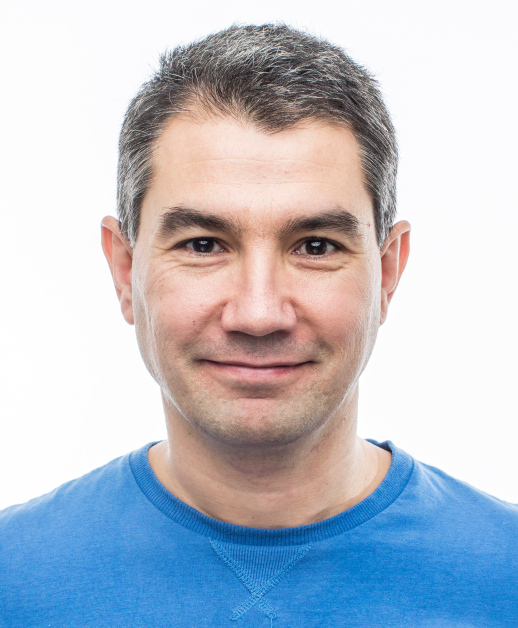}}]{Danyo
 Danev} received his MSc in mathematics from Sofia University in
1996 and his PhD in electrical engineering from Link\"oping University
in 2001. In 2005 he obtained Docent title in Data Transmission. He is
currently Associate Professor at Link\"oping University, Sweden.  His
research interests are within the fields of coding, information and
communication theory. He has authored or co-authored 2 book chapters,
17 journal papers and more than 30 conference papers on these
topics. He is currently teaching a number of communication
engineering and mathematics courses. Since 2012 he is board member
of the IEEE Sweden VT/COM/IT Chapter.
\end{IEEEbiography}


\begin{IEEEbiography}[{\includegraphics[width=1.16in,height=1.16in,clip,keepaspectratio]{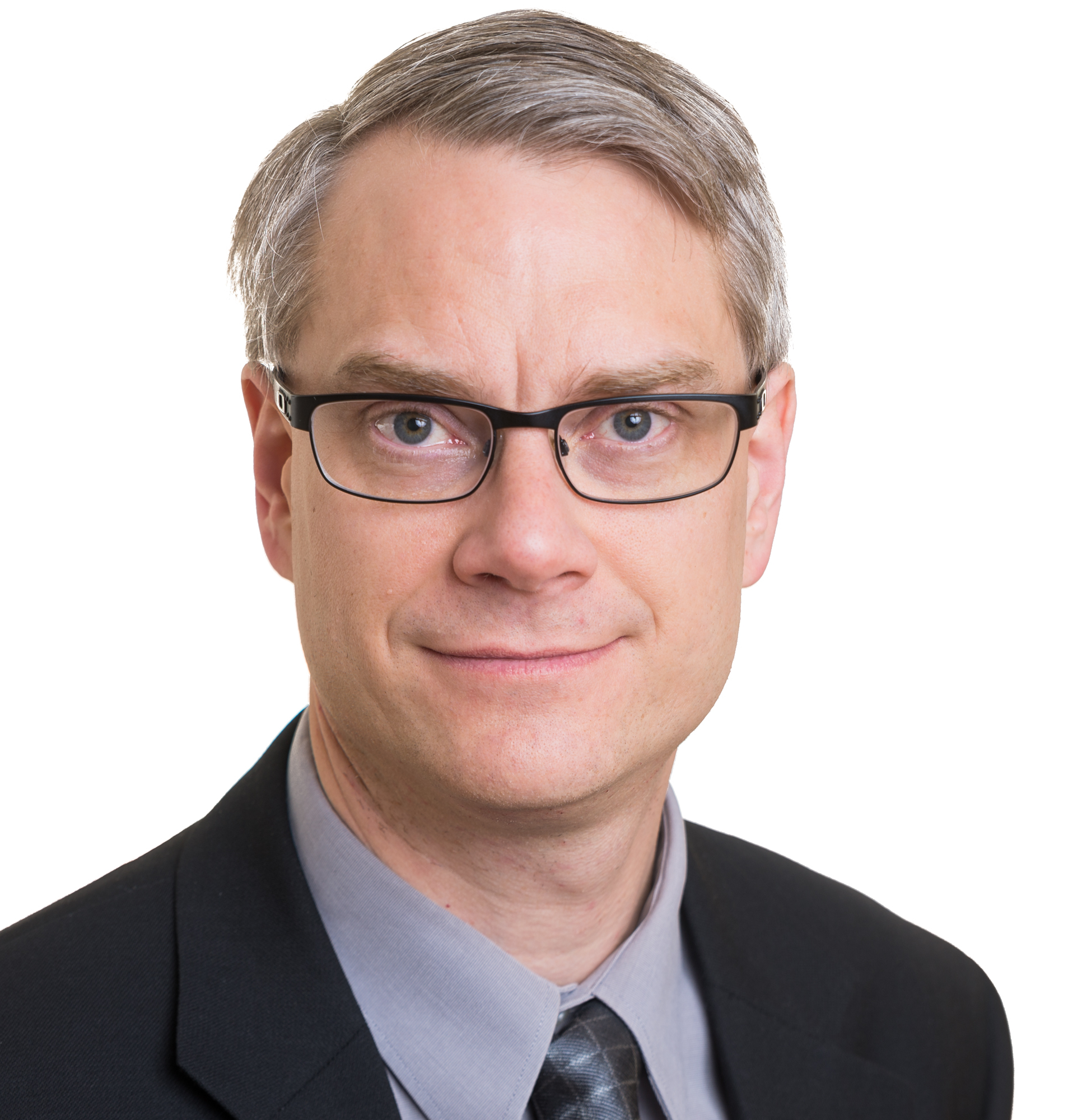}}]{Erik
 G. Larsson} received the Ph.D. degree from Uppsala University,
Uppsala, Sweden, in 2002.

He is currently Professor of Communication Systems at Link\"oping
University (LiU) in Link\"oping, Sweden. He was with the Royal
Institute of Technology (KTH) in Stockholm, Sweden, the University of
Florida, USA, the George Washington University, USA, and Ericsson
Research, Sweden.  In 2015 he was a Visiting Fellow at Princeton
University, USA, for four months.  His main professional interests are
within the areas of wireless communications and signal processing. He
has co-authored some 130 journal papers on these topics, he is
co-author of the two Cambridge University Press textbooks
\emph{Space-Time Block Coding for Wireless Communications} (2003) and
\emph{Fundamentals of Massive MIMO} (2016). He is co-inventor on 16
issued and many pending patents on wireless technology.

He was Associate Editor for, among others, the \emph{IEEE Transactions
  on Communications} (2010-2014) and the \emph{IEEE Transactions on
  Signal Processing} (2006-2010).  From 2015 to 2016 he served as
chair of the IEEE Signal Processing Society SPCOM technical committee,
and in 2017 he is the past chair of this committee.  From 2014 to 2015
he served as chair of the steering committee for the \emph{IEEE
  Wireless Communications Letters}.  He was the General Chair of the
Asilomar Conference on Signals, Systems and Computers in 2015, and its
Technical Chair in 2012.  He is a member of the IEEE Signal Processing
Society Awards Board during 2017--2019.

He received the IEEE Signal Processing Magazine Best Column Award
twice, in 2012 and 2014, the IEEE ComSoc Stephen O. Rice Prize in
Communications Theory in 2015, and the IEEE ComSoc Leonard G. Abraham
Prize in 2017. He is a Fellow of the IEEE.
\end{IEEEbiography}


\begin{thebibliography}{10}
\providecommand{\url}[1]{#1}
\csname url@samestyle\endcsname
\providecommand{\newblock}{\relax}
\providecommand{\bibinfo}[2]{#2}
\providecommand{\BIBentrySTDinterwordspacing}{\spaceskip=0pt\relax}
\providecommand{\BIBentryALTinterwordstretchfactor}{4}
\providecommand{\BIBentryALTinterwordspacing}{\spaceskip=\fontdimen2\font plus
\BIBentryALTinterwordstretchfactor\fontdimen3\font minus
  \fontdimen4\font\relax}
\providecommand{\BIBforeignlanguage}[2]{{%
\expandafter\ifx\csname l@#1\endcsname\relax
\typeout{** WARNING: IEEEtran.bst: No hyphenation pattern has been}%
\typeout{** loaded for the language `#1'. Using the pattern for}%
\typeout{** the default language instead.}%
\else
\language=\csname l@#1\endcsname
\fi
#2}}
\providecommand{\BIBdecl}{\relax}
\BIBdecl

\bibitem{prabhu2016_uav}
P.~Chandhar, D.~Danev, and E.~G. Larsson, ``Massive {MIMO} as enabler for
  communications with drone swarms,'' in \emph{Proc. International Conference
  on Unmanned Aircraft Systems (ICUAS)}, June 2016, pp. 347--354.

\bibitem{prabhu2016_spawc}
------, ``On ergodic rates and optimal array geometry in line-of-sight
  {M}assive {MIMO},'' in \emph{Proc. {IEEE} SPAWC}, July 2016, pp. 1--6.

\bibitem{hayat2016_1}
S.~Hayat, E.~Yanmaz, and R.~Muzaffar, ``Survey on unmanned aerial vehicle
  networks for civil applications: A communications viewpoint,'' \emph{{IEEE}
  Commun. Surveys Tuts.}, vol.~18, no.~4, pp. 2624--2661, Fourth-Quarter 2016.

\bibitem{gupta2016}
L.~Gupta, R.~Jain, and G.~Vaszkun, ``Survey of important issues in {UAV}
  communication networks,'' \emph{{IEEE} Commun. Surveys Tuts.}, vol.~18,
  no.~2, pp. 1123--1152, Second-Quarter 2016.

\bibitem{asadpour2014}
M.~Asadpour, B.~Van~den Bergh, D.~Giustiniano, K.~Hummel, S.~Pollin, and
  B.~Plattner, ``Micro aerial vehicle networks: An experimental analysis of
  challenges and opportunities,'' \emph{{IEEE} Commun. Mag.}, vol.~52, no.~7,
  pp. 141--149, July 2014.

\bibitem{andre2014}
T.~Andre, K.~Hummel, A.~Schoellig, E.~Yanmaz, M.~Asadpour, C.~Bettstetter,
  P.~Grippa, H.~Hellwagner, S.~Sand, and S.~Zhang, ``Application-driven design
  of aerial communication networks,'' \emph{{IEEE} Commun. Mag.}, vol.~52,
  no.~5, pp. 129--137, May 2014.

\bibitem{yanmaz2013}
E.~Yanmaz, R.~Kuschnig, and C.~Bettstetter, ``Achieving air-ground
  communications in 802.11 networks with three-dimensional aerial mobility,''
  in \emph{Proc. {IEEE} INFOCOM}, Apr. 2013, pp. 120--124.

\bibitem{Asadpour2016}
M.~Asadpour, K.~A. Hummel, D.~Giustiniano, and S.~Draskovic, ``Route or carry:
  Motion-driven packet forwarding in micro aerial vehicle networks,''
  \emph{{IEEE} Trans. Mobile Comput.}, vol.~16, no.~3, pp. 843--856, Mar. 2017.

\bibitem{asadpour2013LDD}
M.~Asadpour, D.~Giustiniano, K.~A. Hummel, S.~Heimlicher, and S.~Egli, ``Now or
  later?: Delaying data transfer in time-critical aerial communication,'' in
  \emph{Proc. ACM Conference on Emerging Networking Experiments and
  Technologies CoNEXT}, 2013, pp. 127--132.

\bibitem{hayat2016}
S.~Hayat, E.~Yanmaz, and C.~Bettstetter, ``Experimental analysis of
  multipoint-to-point {UAV} communications with {IEEE} 802.11n and 802.11ac,''
  in \emph{Proc. {IEEE} PIMRC}, Aug. 2015, pp. 1991--1996.

\bibitem{abatti2005}
{Abatti J.}, \emph{{Small Power: The Role of Micro and Small UAVs in the
  Future}}.\hskip 1em plus 0.5em minus 0.4em\relax {Maxwell AFB, AL: Air
  Command and Staff College}, 2005.

\bibitem{olsson2010}
P.~Olsson, J.~Kvarnstrom, P.~Doherty, O.~Burdakov, and K.~Holmberg,
  ``Generating {UAV} communication networks for monitoring and surveillance,''
  in \emph{Proc. International Conference on Control Automation Robotics Vision
  (ICARCV)}, Dec. 2010, pp. 1070--1077.

\bibitem{Marzetta16Book}
{T. L. Marzetta, E. G. Larsson, H. Yang, and H. Q. Ngo}, \emph{{Fundamentals of
  {M}assive {MIMO}}}.\hskip 1em plus 0.5em minus 0.4em\relax {Cambridge
  University Press}, 2016.

\bibitem{marzetta2010}
T.~Marzetta, ``Noncooperative cellular wireless with unlimited numbers of base
  station antennas,'' \emph{{IEEE} Trans. Wireless Commun.}, vol.~9, no.~11,
  pp. 3590--3600, Nov. 2010.

\bibitem{larsson2014}
E.~G. Larsson, O.~Edfors, F.~Tufvesson, and T.~Marzetta, ``Massive {MIMO} for
  next generation wireless systems,'' \emph{{IEEE} Commun. Mag.}, vol.~52,
  no.~2, pp. 186--195, Feb. 2014.

\bibitem{rasool2009}
J.~Rasool, G.~Oien, J.~Hakegard, and T.~Myrvoll, ``On multiuser {MIMO} capacity
  benefits in air-to-ground communication for air traffic management,'' in
  \emph{Proc. ISWCS}, Sept. 2009, pp. 458--462.

\bibitem{jiang2014}
Y.~Jiang, A.~Tiwari, M.~Rachid, and B.~Daneshrad, ``{MIMO} for airborne
  communications [industry perspectives],'' \emph{{IEEE} Wireless Commun.
  Mag.}, vol.~21, no.~5, pp. 4--6, Oct. 2014.

\bibitem{su2013}
W.~Su, J.~D. Matyjas, M.~J. Gans, and S.~Batalama, ``Maximum achievable
  capacity in airborne {MIMO} communications with arbitrary alignments of
  linear transceiver antenna arrays,'' \emph{{IEEE} Trans. Wireless Commun.},
  vol.~12, no.~11, pp. 5584--5593, Nov. 2013.

\bibitem{michailidis2010}
E.~T. Michailidis and A.~G. Kanatas, ``Three-dimensional {HAP}-{MIMO} channels:
  Modeling and analysis of space-time correlation,'' \emph{{IEEE} Trans. Veh.
  Technol.}, vol.~59, no.~5, pp. 2232--2242, June 2010.

\bibitem{sarris2005}
I.~Sarris and A.~Nix, ``Maximum {MIMO} capacity in line-of-sight,'' in
  \emph{Proc. International Conference on Information, Communications and
  Signal Processing}, Dec. 2005, pp. 1236--1240.

\bibitem{bohagen2007}
F.~Bohagen, P.~Orten, and G.~Oien, ``Design of optimal high-rank line-of-sight
  {MIMO} channels,'' \emph{{IEEE} Trans. Wireless Commun.}, vol.~6, no.~4, pp.
  1420--1425, Apr. 2007.

\bibitem{halsig2015}
T.~Halsig and B.~Lankl, ``Array size reduction for high-rank {L}o{S} {MIMO}
  {ULA}s,'' \emph{{IEEE} Microw. Wireless Compon. Lett.}, vol.~4, no.~6, pp.
  649--652, Dec. 2015.

\bibitem{jinhui2013}
J.~Chen, ``When does asymptotic orthogonality exist for very large arrays?'' in
  \emph{Proc. {IEEE} GLOBECOM}, Dec. 2013, pp. 4146--4150.

\bibitem{yeqing2015}
Y.~Hu, Y.~Hong, and J.~Evans, ``Interference in {L}o{S} {M}assive {MIMO} is
  well approximated by a {B}eta-mixture,'' in \emph{Proc. {IEEE} ICC
  Workshops}, June 2015, pp. 1137--1142.

\bibitem{tan2015}
W.~Tan, S.~Jin, J.~Wang, and Y.~Huang, ``Achievable sum-rate analysis for
  {M}assive {MIMO} systems with different array configurations,'' in
  \emph{Proc. {IEEE} WCNC}, Mar. 2015, pp. 316--321.

\bibitem{yang2017}
H.~Yang and T.~L. Marzetta, ``Massive {MIMO} with max-min power control in
  line-of-sight propagation environment,'' \emph{IEEE Transactions on
  Communications}, vol.~65, no.~11, pp. 4685--4693, Nov. 2017.

\bibitem{shafi2006}
M.~Shafi, M.~Zhang, A.~Moustakas, P.~Smith, A.~Molisch, F.~Tufvesson, and
  S.~Simon, ``Polarized {MIMO} channels in {3-D}: {M}odels, measurements and
  mutual information,'' \emph{{IEEE} J. Select. Areas Commun.}, vol.~24, no.~3,
  pp. 514--527, Mar. 2006.

\bibitem{manh_tuan_dao2011}
M.-T. Dao, V.-A. Nguyen, Y.-T. Im, S.-O. Park, and G.~Yoon, ``{3D} polarized
  channel modeling and performance comparison of {MIMO} antenna configurations
  with different polarizations,'' \emph{{IEEE} Trans. Antennas Propag.},
  vol.~59, no.~7, pp. 2672--2682, July 2011.

\bibitem{jaeckel2012}
S.~Jaeckel, K.~Borner, L.~Thiele, and V.~Jungnickel, ``A geometric polarization
  rotation model for the {3-D} spatial channel model,'' \emph{{IEEE} Trans.
  Antennas Propag.}, vol.~60, no.~12, pp. 5966--5977, Dec. 2012.

\bibitem{Balanis2005}
C.~A. Balanis, \emph{Antenna Theory: Analysis and Design}.\hskip 1em plus 0.5em
  minus 0.4em\relax Wiley-Interscience, 2005.

\bibitem{sun2017_airframe}
R.~Sun, D.~W. Matolak, and W.~Rayess, ``Air-ground channel characterization for
  unmanned aircraft systems--{P}art {IV}: Airframe shadowing,'' \emph{{IEEE}
  Trans. Veh. Technol.}, vol.~66, no.~9, pp. 7643--7652, Sept. 2017.

\bibitem{matolak2017_water}
D.~W. Matolak and R.~Sun, ``Air-ground channel characterization for unmanned
  aircraft systems--{P}art {I}: Methods, measurements, and models for
  over-water settings,'' \emph{{IEEE} Trans. Veh. Technol.}, vol.~66, no.~1,
  pp. 26--44, Jan. 2017.

\bibitem{sun2017_hilly}
R.~Sun and D.~W. Matolak, ``Air-ground channel characterization for unmanned
  aircraft systems--{P}art {II}: Hilly and mountainous settings,'' \emph{{IEEE}
  Trans. Veh. Technol.}, vol.~66, no.~3, pp. 1913--1925, Mar. 2017.

\bibitem{matolak2017_urban}
D.~W. Matolak and R.~Sun, ``Air-ground channel characterization for unmanned
  aircraft systems--{P}art {III}: The suburban and near-urban environments,''
  \emph{{IEEE} Trans. Veh. Technol.}, vol.~66, no.~8, pp. 6607--6618, Aug.
  2017.

\bibitem{rappaport96}
{Rappaport T. S.}, \emph{{Wireless Communications Principles and
  Practice}}.\hskip 1em plus 0.5em minus 0.4em\relax {Prentice Hall Inc.},
  1996.

\bibitem{teal2002}
P.~D. Teal, T.~D. Abhayapala, and R.~A. Kennedy, ``Spatial correlation for
  general distributions of scatterers,'' \emph{{IEEE} Signal Process. Lett.},
  vol.~9, no.~10, pp. 305--308, Oct. 2002.

\bibitem{Knuth:1997}
D.~E. Knuth, \emph{The Art of Computer Programming, Volume 2 (3rd Ed.):
  Seminumerical Algorithms}.\hskip 1em plus 0.5em minus 0.4em\relax Boston, MA,
  USA: Addison-Wesley Longman Publishing Co., Inc., 1997.

\bibitem{drone_mapping_handbook}
``Drones and aerial observation: New technologies for property rights, human
  rights, and global development - {A} primer,''
  \url{http://drones.newamerica.org/primer/DronesAndAerialObservation.pdf},
  [{O}nline], Accessed: \today.

\bibitem{rosnell2012}
T.~Rosnell and E.~Honkavaara, ``Point cloud generation from aerial image data
  acquired by a quadrocopter type micro unmanned aerial vehicle and a digital
  still camera,'' \emph{Sensors}, vol.~12, no.~1, pp. 453--480, Jan. 2012.

\bibitem{prabhu-spawc2017}
P.~Chandhar, D.~Danev, and E.~G. Larsson, ``On the outage capacity in {M}assive
  {MIMO} with line-of-sight,'' in \emph{Proc. IEEE SPAWC}, July 2017, pp. 1--6.

\bibitem{pinto2017}
L.~R. Pinto, L.~Almeida, and A.~Rowe, ``Demo abstract: Video streaming in
  multi-hop aerial networks,'' in \emph{Proc. ACM/IEEE International Conference
  on Information Processing in Sensor Networks}, Apr. 2017, pp. 283--284.

\end{thebibliography}
\end{document}